%% file: NGC6334_accepted_arxiv.tex
\documentclass[twocolumn,longauth]{aa} 
\usepackage{natbib}
\usepackage{url}
\PassOptionsToPackage{hyphens}{url}
\usepackage{graphicx}
\usepackage{subfigure}
\usepackage{amsmath}
\usepackage{txfonts}
\usepackage{xcolor}
\usepackage{listings} 
\usepackage{times}
\usepackage{multirow}
\usepackage{float}
\usepackage{epstopdf}
\usepackage{longtable}
\usepackage{soul}
\usepackage[colorlinks=true,linkcolor=red,citecolor=blue, breaklinks=true]{hyperref}

\newcommand{\hii}{\ion{H}{ii}}

\newcommand{\planck}{\textit{Planck}}  
\newcommand{\disperse}{{\tt  DisPerSE}}  

\newcommand{\Artemis}{{ArT\'eMiS}}
\def\NHUNIT{\ifmmode {\rm \,cm^{-2}} \else $\rm \,cm^{-2}$ \fi} 

\def\nhh{\ifmmode N_{\rm H_{2}}\else $N_{\rm H_{2}}$\fi} 
\def\nhhc{\ifmmode N_{\rm H_{2}}^0\else $N_{\rm H_{2}}^0$\fi} 
\def\nhhbg{\ifmmode N_{\rm H_{2}}^{\rm bg}\else $N_{\rm H_{2}}^{\rm bg}$\fi} 
\def\nh{\ifmmode N_{\rm H}\else $N_{\rm H}$\fi}
\def\ml{\ifmmode M_{\rm line}\else $M_{\rm line}$\fi}  
\def\sunpc{\ifmmode \rm M_\odot/\rm pc\else $\rm M_\odot/\rm pc$\fi}  
\def\rout{\ifmmode R_{\rm out}\else $R_{\rm out}$\fi}  
\def\av{\ifmmode A_{\rm V}\else $A_{\rm V}$\fi}   
\def\fwhmdec{\ifmmode FHWM_{\rm dec}\else $FWHM_{\rm dec}$\fi}   
\def\rflat{\ifmmode R_{\rm flat}\else $R_{\rm flat}$\fi}   
\def\kms{\ifmmode {km\,s$^{-1}$}\else km\,s$^{-1}$\fi}  

\def\arcm{\ifmmode {^{\scriptstyle\prime}}
          \else $^{\scriptstyle\prime}$\fi}
\newdimen\sa  \newdimen\sb
\def\parcs{\sa=.07em \sb=.03em
     \ifmmode \hbox{\rlap{.}}^{\scriptstyle\prime\kern -\sb\prime}\hbox{\kern -\sa}
     \else \rlap{.}$^{\scriptstyle\prime\kern -\sb\prime}$\kern -\sa\fi}
\def\parcm{\sa=.08em \sb=.03em
     \ifmmode \hbox{\rlap{.}\kern\sa}^{\scriptstyle\prime}\hbox{\kern-\sb}
     \else \rlap{.}\kern\sa$^{\scriptstyle\prime}$\kern-\sb\fi}

\def\rev{}
\def\revbis{}
\begin{document} 
\input{bistro_authors2.tex}

\title{Dust polarized emission observations of NGC 6334}
\subtitle{BISTRO reveals the details of the complex but organized magnetic field structure of the high-mass star-forming hub-filament network}

  \titlerunning{Dust polarized emission observations of NGC 6334}

\authorrunning{D. Arzoumanian et al.}


\abstract{
{\it Context.}  
Molecular  filaments and hubs have received special attention recently thanks to new studies showing their key role in  star formation. 
While the (column) density and velocity structures of both filaments and hubs have been carefully studied, their magnetic field (B-field) properties are not yet  characterized. Consequently, the role of B-fields in the formation and evolution of  
hub-filament systems  is not well constrained.\\ 
{\it Aims.} We aim to understand the role of the B-field and its interplay with turbulence and gravity in the dynamical evolution of the NGC 6334 filament network that harbours cluster-forming hubs and high-mass star formation. \\ 
{\it Methods.} 
We present new  observations of the dust polarized emission at 850\,$\mu$m towards the $2\,{\rm pc} \times 10$\,pc map of NGC 6334 at a spatial resolution of 0.09\,pc obtained with the JCMT/SCUBA-2/POL-2 as part of the BISTRO survey. 
We study the distribution and dispersion of the polarized intensity ($PI$), the polarization fraction ($PF$), and  the  plane-of-the-sky (POS) B-field angle ($\chi_{\rm B_{POS}}$) towards the whole region, along  the 10\,pc-long ridge, and along the sub-filaments connected to the ridge and the hubs.  
We derive the power spectrum of the intensity and $\chi_{\rm B_{POS}}$ along the ridge crest and compare with results obtained from  simulated filaments. \\ 
{\it Results.} The observations  span  $\sim3$ orders of magnitude in $I$ and $PI$ and $\sim2$ orders of magnitude in $PF$ (from $\sim0.2\%$ to  $\sim20\%$). A large scatter in $PI$ and $PF$ is observed for a given value of $I$. Our analyses show a complex B-field structure  when observed over the whole region ($\sim10$\,pc),  however, at smaller scales ($\sim1$\,pc), $\chi_{\rm B_{POS}}$ varies coherently  along the crests of the filament network. 
The observed power spectrum of $\chi_{\rm B_{POS}}$ can  be well represented with a power law function with a slope $-1.33\pm0.23$, which is  $\sim20\%$ shallower than that of $I$. We find that this result is compatible with the properties of simulated filaments and may indicate the physical processes at play in  the formation and evolution of star-forming filaments.
 Along the sub-filaments, $\chi_{\rm B_{POS}}$  rotates  from  being mostly perpendicular {\rev (or randomly orientated with respect)} to the crests to mostly parallel as the sub-filaments merge with the ridge and hubs. This variation of the B-field structure along the sub-filaments may be tracing local velocity flows of matter in-falling onto the ridge and hubs.
 Our analysis also suggests a variation of the energy balance along the crests of these sub-filaments, from magnetically critical/supercritical at their far ends to magnetically subcritical  near the ridge and hubs. We also detect an increase of $PF$ towards the high-column density ($\nhh\gtrsim10^{23}\NHUNIT$) star cluster-forming hubs that may be explained by the increase of grain alignment efficiency due to stellar radiation from the newborn stars, combined with an ordered B-field structure.  \\
{\it Conclusions.}  These observational results reveal for the first time the characteristics of the small scale (down to $\sim0.1\,$pc) B-field structure of a $10\,$pc-long hub-filament system. 
Our analyses show variations of the polarization properties along the sub-filaments  that may be  tracing  the evolution of their physical properties  during their interaction with the  ridge and hubs.
We also  detect an impact of  feedback from young high-mass stars on the local B-field structure and the polarization properties, that could put constraints on possible models for dust grain alignment and provide important hints on the interplay between the star formation activity and interstellar B-fields. 
}


\maketitle

\section{Introduction}\label{intro} 

In the last decade, observations tracing the  thermal  emission  from  the cold dust (e.g., with $Herschel$ and $Planck$) and  the molecular gas of the   interstellar medium (ISM) unveiled its highly filamentary structure  
 \citep[e.g.,][]{Andre2014,Molinari2010,Wang2015,planck2016-XXXII,Hacar2018}. 
Moreover, in star forming regions,
prestellar cores and protostars are mostly observed  along the sample of filaments that are gravitationally unstable \citep[e.g.,][]{Andre2010,Konyves2015,Hacar2013}. 
In addition, detailed analyses of the properties of these filaments and their cores indicate that a relatively small fraction ($\sim15-20\,\%$) of the mass of the filaments is in the form of cores, 
and the mass distribution of these cores (CMF)  has a shape similar to the initial mass function (IMF) of stars   \citep[e.g.,][]{Andre2010,Konyves2015,Konyves2020}.

While we still have to confirm  whether or not  the full mass spectrum of stars can be formed by the same physical process,  
 these observational results  suggest that  fragmentation within self-gravitating filaments is the dominant mode  for star formation  \citep[at least for the low- to intermediate-mass stars, e.g.,][]{Andre2014,Tafalla2015}. 
 If this is the case, the mechanisms responsible for filament formation and fragmentation should explain the observed physical properties of the cores (and eventually that of the stars), e.g., spacing/clustering, mass functions (CMF/IMF), size distribution, angular momentum. Such a filament paradigm should also explain the global properties of the star formation process on the scales of galaxies, e.g., star formation efficiency and rate, (dense) gas depletion timescales.  It is thus crucial to describe the properties and understand the dynamics of these filaments, which  give the initial conditions of core/star formation and may be a main element to understand the global star formation properties on the scale of galaxies. 
 
Observations suggest that, in star-forming molecular clouds, filaments are rarely isolated, but mostly organized in systems. We can divide these systems into two major groups:\\
 1) The {\it"ridge-dominated"} group, which can be described by  a single  dense self-gravitating "main-filament" or "ridge" with a line mass $M_{\rm line}$ larger than $M_{\rm line,crit}$, the thermal critical value for equilibrium\footnote{These filaments  are thermally supercritical with line masses  ($M_{\rm line}$) larger than the critical equilibrium value for isothermal cylinders \citep[e.g.][]{Inutsuka1997},  $M_{\rm line, crit}^T = 2\, c_{\rm s}^2/G\sim16.5\,$M$_\odot$/pc \citep{Ostriker1964}, where  $c_{\rm s}\sim0.2$\,\kms\ is the sound speed for $T\,\sim\,10$\,K. }. These filaments are gravitationally unstable and are observed to  be forming chains of stars along their crests and are often connected from the side to lower-density "sub-filaments" \citep[e.g.,][]{Hill2011,Schneider2010,Hennemann2012,Palmeirim2013,HKirk2013,Arzoumanian2017}. \\
2) The {\it"hub-dominated"} group, composed by several star-forming (mostly high-density) filaments merging into a high-density  "hub" \citep[e.g.,][]{myers2009,Peretto2013,Peretto2014,Williams2018}. 
These hubs have also been identified  to host young star-clusters, suggesting the important role of  hub-filament configurations in star-cluster formation \citep[e.g.,][]{Schneider2012,Kumar2020}.

The formation of these ridge and hub systems is still under debate.  Most observational work has focused on their (column) densities and velocity structures \citep[e.g.,][]{Hennemann2012,Kirk2013,Peretto2013}, which have revealed velocity gradients along lower-density sub-filaments that may be feeding material to the ridges and/or hubs \citep[e.g.,][]{Schneider2010,Palmeirim2013,Peretto2014}.  One remaining question, however, is the role of magnetic fields (B-fields) in hindering or supporting the flow of material.  The relative role of B-fields with respect to both turbulence and gravity is not well constrained  \citep[e.g.,][]{Crutcher2012,Hennebelle2019}.

B-fields are most often characterized via  polarization measurements, assuming that interstellar dust grains are aspherical and that their major axes align perpendicular to the local B-field orientation \citep[e.g.,][]{Lee1985,Lazarian1997,Hildebrand2000,Andersson2015ARAA,Lazarian2015}. 
Consequently, a fraction of the thermal emission from these dust grains will be linearly polarized relative to the direction of the plane-of-sky (POS) B-field \citep[e.g.,][]{Jones1967,Hildebrand1983,Andersson2015}.  
All-sky dust polarized emission of the ISM observed by $Planck$ at 850$\,\mu$m has revealed  organized B-fields on large scales  \citep[$>1-$10\,pc, see, e.g.,][]{planck2015-XIX}. 
Statistical analysis of the relative orientation between the filaments and the plane-of-the-sky (POS) B-field angle shows that low column density filaments are mostly parallel to the local B-field while high column density  filaments tend to be perpendicular to the B-field lines \citep{planck2016-XXXII}.  
Similar results are also inferred from infrared polarization data \citep[e.g.,][]{Sugitani2011,Palmeirim2013,Cox2016,Soler2016}. 
Comparisons of these observations with numerical simulations suggest that B-fields play a dynamically important role in the formation and evolution of filaments in a mostly sub/trans-Alfv\'enic turbulent ISM \citep[][]{Falceta-Goncalves2008,Falceta-Goncalves2009,Soler2013,planck2016-XXXV}.
There is also an indication that  the  orientation of the B-field  towards dense filaments changes from being nearly perpendicular in the surrounding cloud to more parallel in the filament interior \citep{planck2016-XXXIII}. 
 However, the low resolution of $Planck$ data ($\sim5'-10'$ or $\sim0.2-0.6$\,pc at distance $\lesssim500$\,pc or $>0.5$\,pc at distances $>1$\,kpc) 
 is  insufficient  to resolve the B-field structure at  the characteristic   $\sim0.1$\,pc transverse size of molecular filaments \citep[][]{Arzoumanian2011,Arzoumanian2019}, which is also the scale at which filament fragmentation and core formation occurs \citep[][]{Tafalla2015,Kainulainen2017,Shimajiri2019}. Consequently, the geometry of the B-field within filaments and its effects on fragmentation and star formation are mostly unknown.\\
 
To gain insight into the B-field structure along dense filaments and improve our understanding of the role of the magnetic field in the star formation process,  we analyze  new 850\,$\mu$m data obtained towards  the NGC 6334 star-forming filamentary region observed as part of the 
B-field In STar-forming Region Observations (BISTRO) using 
SCUBA-2/POL-2 installed on  the James Clerk Maxwell Telescope (JCMT).  
While the original  BISTRO survey was designed to cover a variety of nearby star-forming regions, with a focus on the Gould Belt molecular clouds \citep{Ward-Thompson2017}, 
the NGC 6334 field is part of BISTRO-2, a follow-up BISTRO survey, which is an extension aiming at observing mostly 
high-mass star-forming regions.\\

NGC 6334 is a high-mass star-forming complex that lies within the Galactic plane at a relatively nearby distance of $1.3\pm0.3$\,kpc \citep{Chibueze2014}.
This region has been the target of multiple studies at
different wavelengths \citep[see][for an extensive review]{Persi2008}. %
At optical  wavelengths, NGC 6334 is seen as a grouping of well-documented \hii\ regions  also known as the "Cat's Paw" \citep[GUM61, GUM 62, GUM 63, GUM 64, \hii\ 351.2+0.5, and GM1-24,][]{Persi2010,Russeil2016}. 
In between these \hii\ bubbles, along the Northeast$-$Southwest direction lies a 10\,pc-long  filamentary cloud that is very bright at (sub)millimeter wavelengths \citep{Kraemer1999,Matthews2008,Russeil2013,Zernickel2013,Tige2017}. This filamentary cloud is dominated by both a dense ridge threaded by sub-filaments, and by two hub-like structures   towards its Northeast end (cf. Fig.\,\ref{IQUmaps}). 
 This ridge itself  is actively forming  high-mass stars as revealed  by compact radio emission, ultra-compact \hii\ regions, maser sources, and molecular outflows identified along or next to its crest
 \citep{Sandell2000,McCutcheon2000,Munoz2007,Qiu2011}.
The NGC 6334 filament has a line mass  $\ml \sim1000\,$\sunpc$\, (> M_{\rm line,crit})$  
with column densities $\nhh\gtrsim10^{23}\,\NHUNIT$ over most of its 10\,pc long crest  \citep{Andre2016} and it is fragmented into a series of relatively massive cores with a mean mass $\sim10\,$M$_\odot$ 
  \citep[e.g.,][]{Shimajiri2019}. 
The filament inner width is observed to be of the order of 0.1\,pc  \citep{Andre2016},  compatible with the findings derived from statistical analysis of dust continuum $Herschel$ observations of nearby and less massive filaments  \citep[][]{Arzoumanian2011,Arzoumanian2019,Koch2015}. This similarity suggests  that the formation process of stars from low to high-mass stars may be similar, and gravitational fragmentation of $\sim0.1$\,pc-wide  filaments may be also the main mode of intermediate- to high-mass star formation  \citep{Shimajiri2019,Andre2019}.\\ 
 
{\rev  The magnetic field structure of NGC 6334 has been investigated  at large scales ($\sim$10\,pc)  using  starlight polarization  and at small scales ($\lesssim$0.1\,pc) towards the dense cores using interferometric observations of dust polarized emission with the SMA  \citep{Li2006,Zhang2014,LiHb2015,Juarez2017}.  
By combining and interpolating some of the latter measurements, 
\citet{LiHb2015}   suggested that the orientation of the B-field does not change significantly over the various scales, pointing to the  important dynamical role played by the B-field  in the formation of NGC 6334.  
 }
In this paper, we study the B-field structure and the polarization properties towards the NGC 6334 high-mass star-forming region {\rev at scales ranging from about 0.1\,pc to 10\,pc}, {\rev using the dust polarized emission at $850\,\mu$m observed as part of the BISTRO survey. We also present the B-field structure observed at lower resolution by \planck\  around the NGC 6334 region up to scales of $\sim40\,$pc. 
We further} compare the characteristics derived from  this high-mass star-forming region with those inferred from the analysis of molecular clouds forming lower mass stars. 
This paper is organized as follows: 
In Sect.\,\ref{obs}, we describe the observations and the data reduction. In Sect.\,\ref{ana1}, we present the spatial distribution of the observed polarized emission and scatter plots relating the different polarization parameters. In Sect.\,\ref{PropCrests}, we analyse the polarization and physical properties of different crests identified in the filament network. In Sect.\,\ref{powspecSection}, we present a power spectrum  analysis of the observed emission (total intensity and B-field structure) along the filament crest.  In Sect.\,\ref{disc}, we discuss the possible physical origin of the observed polarization properties and the B-field structure of the high-mass star-forming hub-filament system. We give a summary of the analysis and results in Sect.\,\ref{Summary}.

\section{Observations}\label{obs}
\subsection{SCUBA-2/POL-2 850\,$\mu$m BISTRO observations}

\begin{figure*}[!h]
   \centering
     \resizebox{19.cm}{!}{
\includegraphics[angle=0]{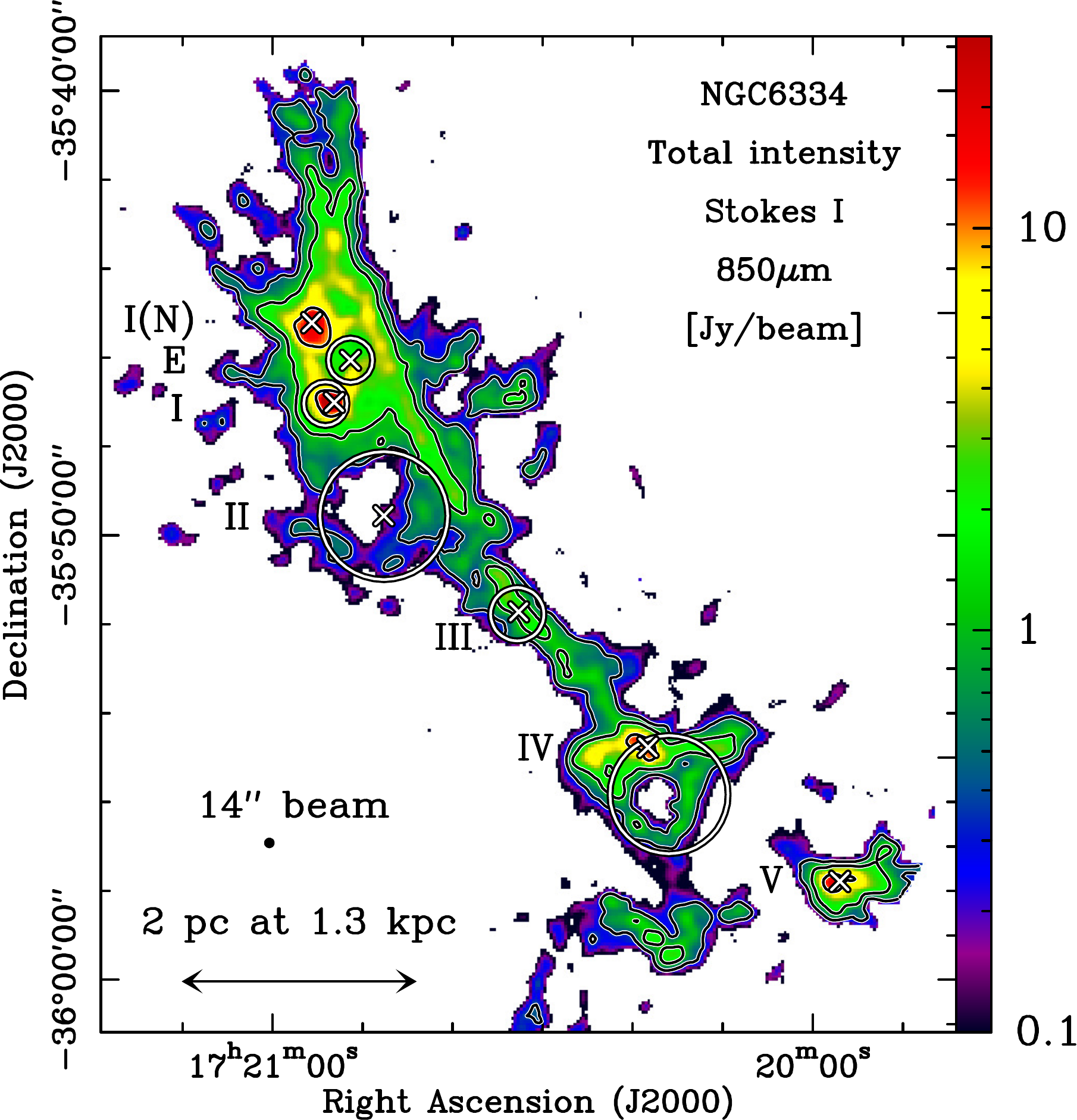}
\includegraphics[angle=0]{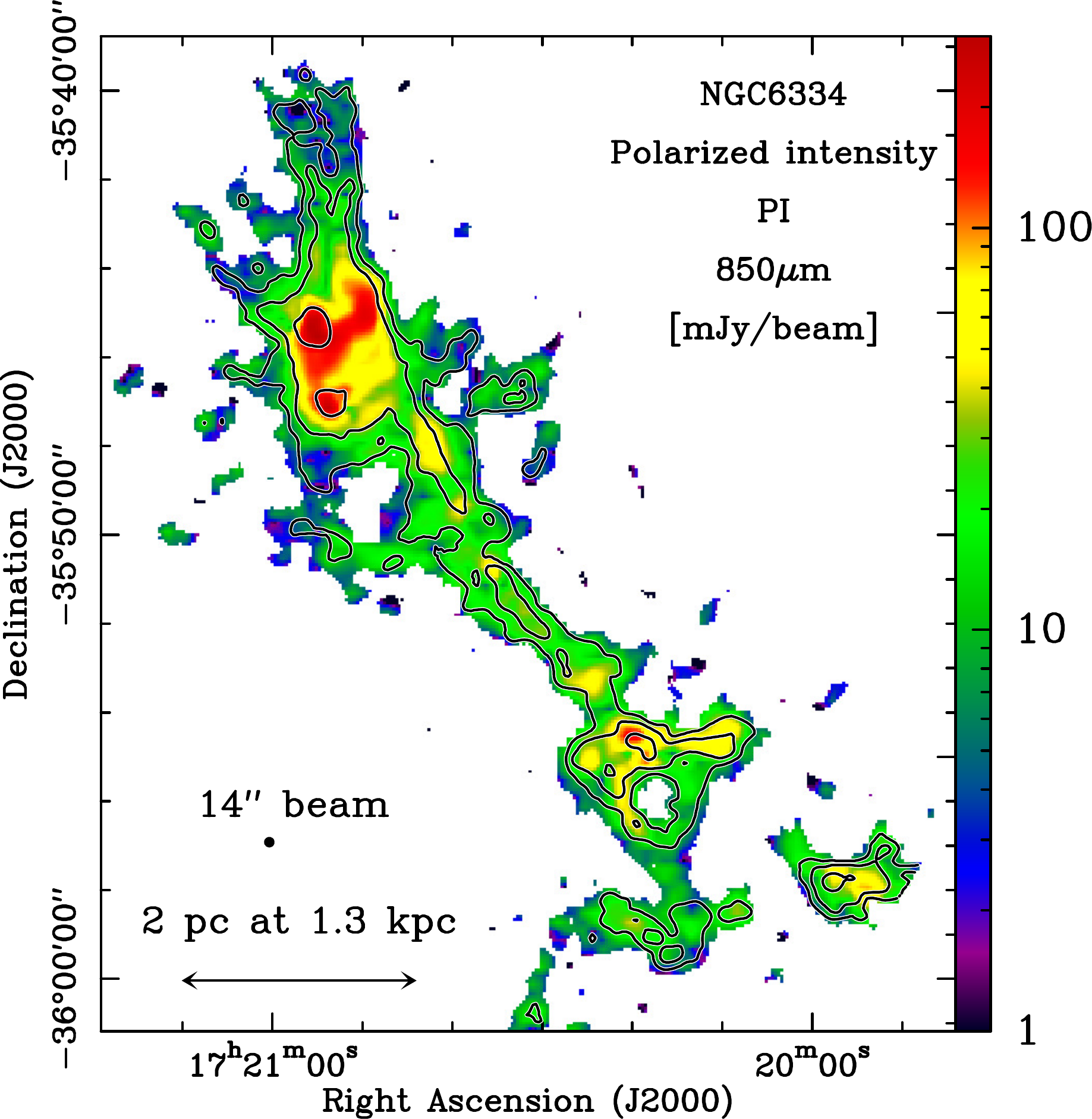}}
   \resizebox{19.cm}{!}{
\includegraphics[angle=0]{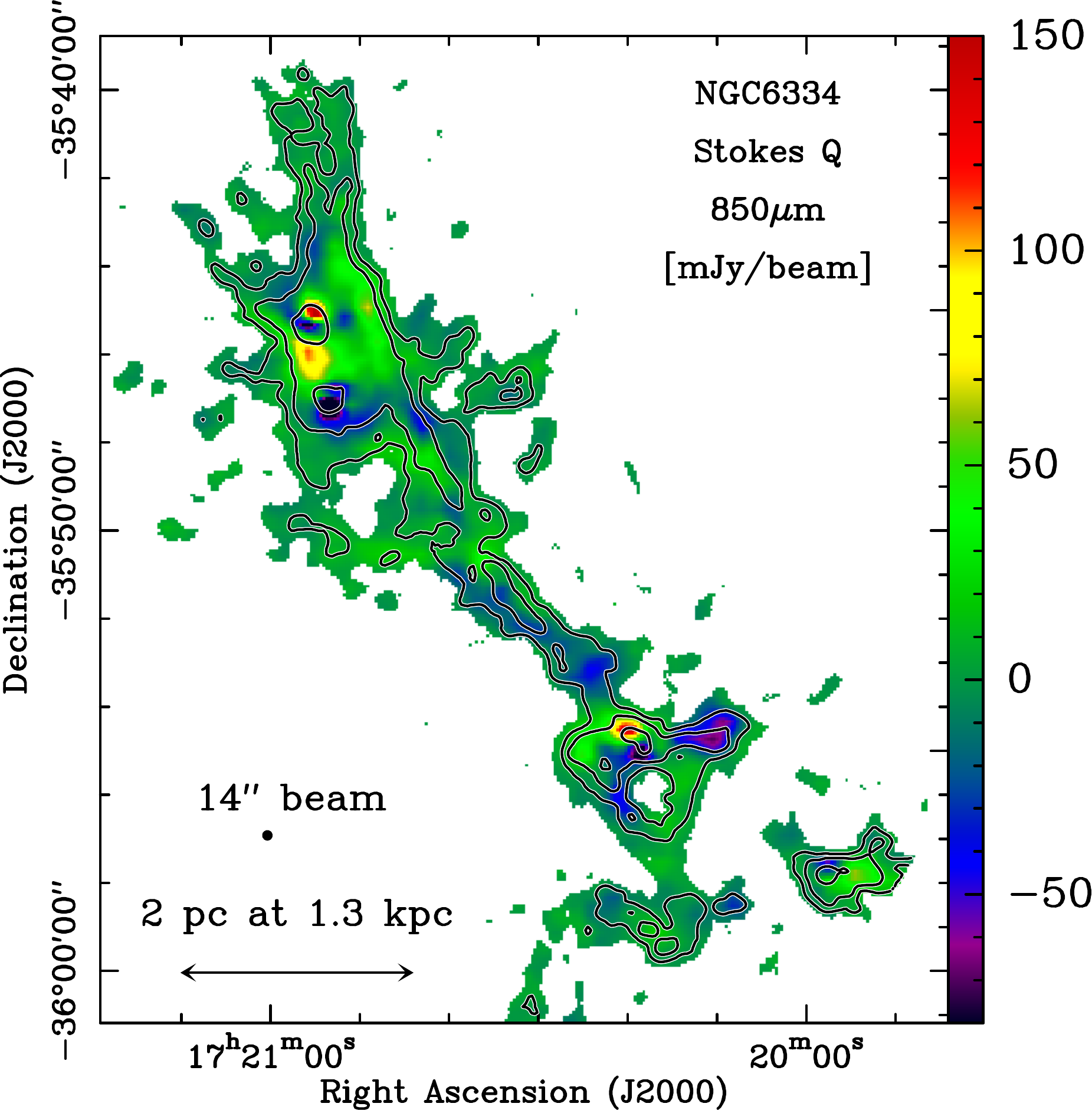}
\includegraphics[angle=0]{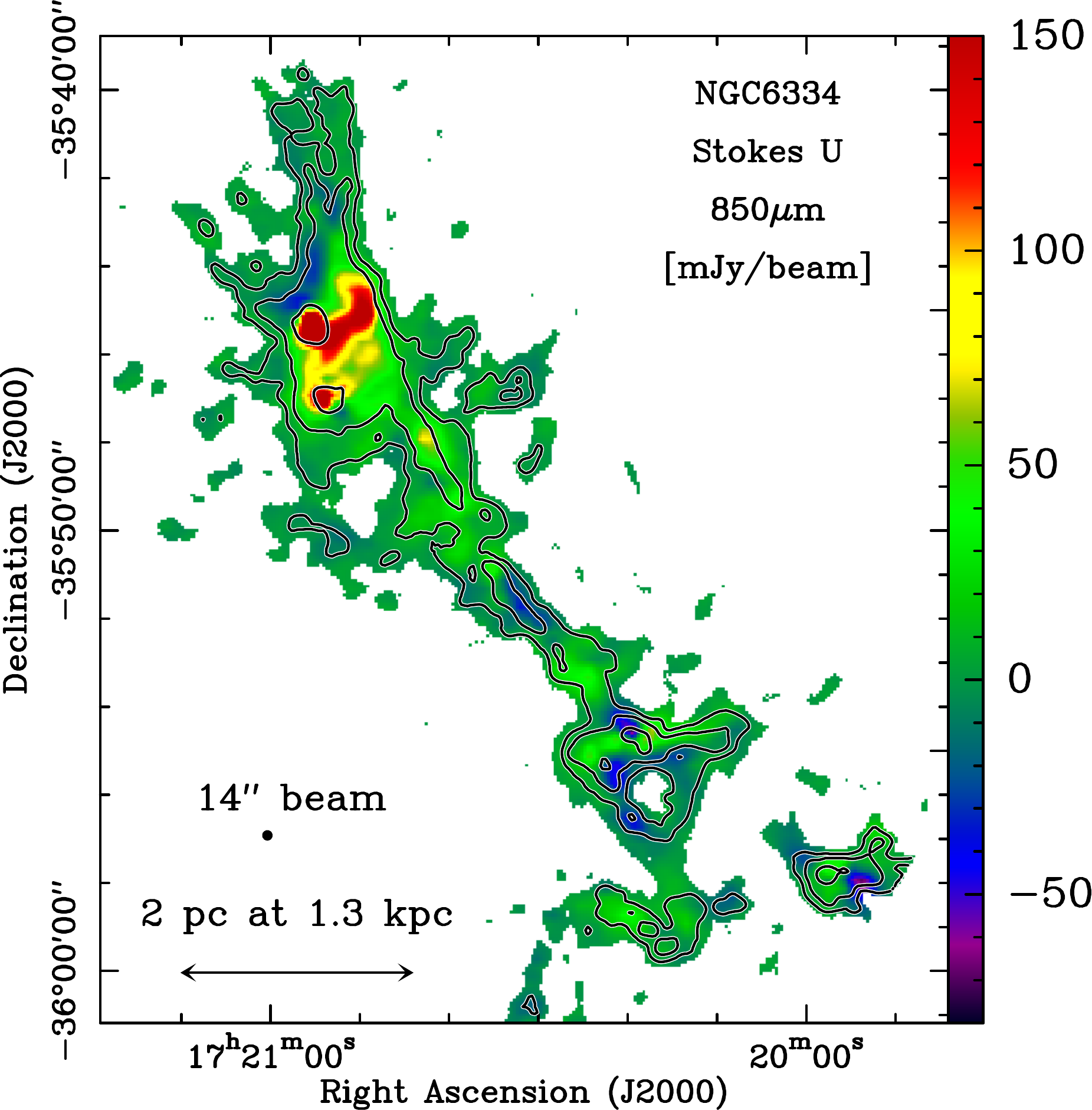}
}
\vspace{-0.3cm}
\color{black}
  \caption{ Stokes parameter maps of the total $I$ (top-left) and the polarized $Q$ and $U$ (bottom-left and  -right, respectively) thermal dust emission at 850\,$\mu$m  observed with the JCMT SCUBA-2/POL-2  towards NGC 6334  as part of the BISTRO survey.   The top right hand side panel shows the map of the (debiased) polarized emission, $PI$.  %
  The half-power-beam-width (HPBW) resolution of these maps is $14\arcsec$.
  For all maps, the plotted emission corresponds to  {\it SNR}$(I)>25$.
 The contours, which  are the same for all  plots, correspond to $I=0.4, 1.4,$ and $8\,$Jy\,beam$^{-1}$ or $\nhh\sim(2, 6,\, $and$\, 37) \times10^{22}\NHUNIT$. The lowest contour of $I=0.4\,$Jy\,beam$^{-1}$ is equivalent to {\it SNR}$(I)\sim 300$ and {\it SNR}$(PI)\sim 10$.
 The white crosses on the $I$ map  indicate the seven high-mass star-forming regions numbered from I to V from North to South in addition to I(N) and E in the North of the field. The white circles show the positions and sizes of UC\hii\ and \hii\  regions \citep[see][and references therein]{Persi2008}.  
}          
  \label{IQUmaps}
    \end{figure*}

The  total and polarized dust thermal continuum emission towards NGC 6334 
was observed  using SCUBA-2/POL-2  \citep{Bastien2011,Holland2013,Friberg2016} installed on the JCMT between August 2017 and April 2019 (project code M17BL011). 
The observations were carried out under dry weather conditions with the atmospheric opacity at 225\,GHz ranging between 0.03 and 0.07.
The observations were done with the standard SCUBA-2/POL-2 DAISY mapping mode with a constant scanning speed of $4\arcsec$\,s$^{-1}$ (POL2-DAISY)  and  a data sampling rate of 8\,Hz. Two maps of  about $12\arcmin$ in diameter each were taken towards the North and South of the elongated filament to cover the entire 10\,pc-long  structure. A  total of 30\,hours on-source were needed to observe the two fields (with 20 exposures per field). The North and South fields were centred   on   (17:20:50.011, $-$35:45:34.33) and  (17:20:19.902, $-$35:54:30.45) in (R.A. J2000, Dec. J2000), respectively (see Fig.\,\ref{Err_maps} for the limits of the two observed DAISY sub-fields). The two sub-fields were then combined to create the mosaicked 
$\sim20\arcmin\times20\arcmin$ Stokes parameter maps (Fig.\,\ref{IQUmaps}).
The flux conversion factor (FCF) of POL-2 at 850\,$\mu$m is taken to be 725\,Jy\,pW$^{-1}$\,beam$^{-1}$ for the three Stokes $I$, $Q$, and $U$ parameters \citep{Friberg2016}. 
The spatial distributions of the Stokes $I$, $Q$, and $U$  parameters are derived from their time-series measurements  using the {\it pol2map}\footnote{\url{http://starlink.eao.hawaii.edu/docs/sc22.htx/sc22.html}}  data reduction pipeline (including {\it skyloop}) that is based on  the 
{\it Starlink} routine {\it makemap} \citep{Chapin2013,Currie2014} and   optimized for the SCUBA-2/POL-2 data  \citep[c.f.,][for details on the data reduction process]{Coude2019}. 

The JCMT has a diameter of 15\,m and achieves an effective half-power-beam-width (HPBW) angular resolution of  $14\arcsec$ at 850\,$\mu$m \citep{Dempsey2013,Friberg2016}. 
The data are projected onto grid maps with pixel sizes of $4\arcsec$ %
and $12\arcsec$ ($\sim$ HPBW), both at the spatial resolution of  $14\arcsec$ (HPBW of the JCMT at 850\,$\mu$m).

Figure\,\ref{IQUmaps} presents the maps of the three Stokes parameters of NGC 6334. The Stokes $I$ map is very similar to what was observed by the previous generation bolometer on the JCMT \citep[e.g.,][]{Matthews2008} and the APEX/\Artemis\ map at 350\,$\mu$m \citep[at an angular resolution of $8\arcsec$,][]{Andre2016}. 
  The Stokes $Q$ and $U$ parameter maps  provide  a first  insight into the spatial structure of the  linearly polarized emission of the cold dust grains. The spatial distribution of these two parameters is different and %
  some abrupt changes  from positive to negative can be seen. These variations are a first indication of the change of the POS B-field structure.

\begin{figure*}[!h]
   \centering
      \resizebox{19.5cm}{!}{ \hspace{-.5cm}
   \includegraphics[angle=0]{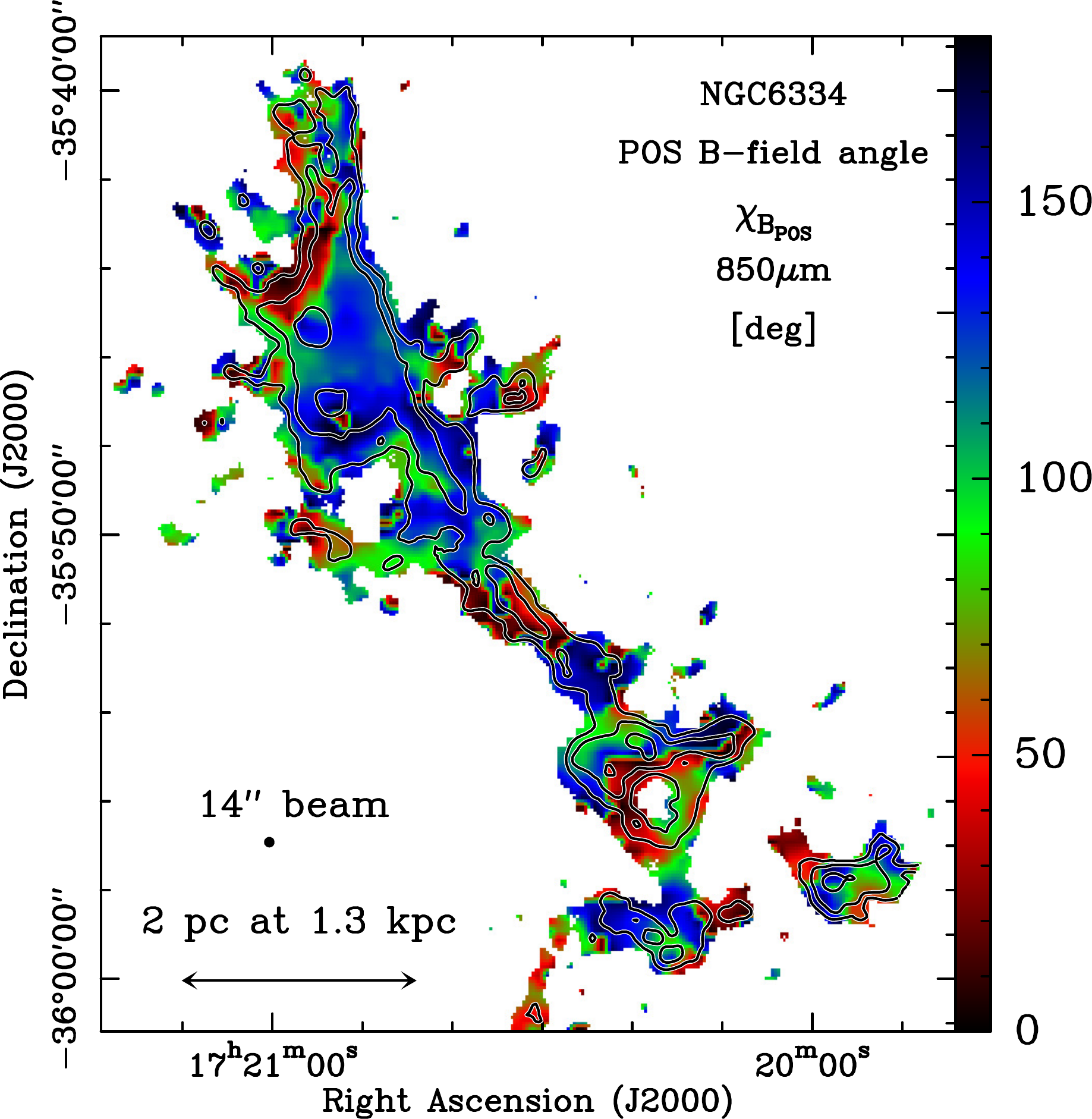}
\includegraphics[angle=0]{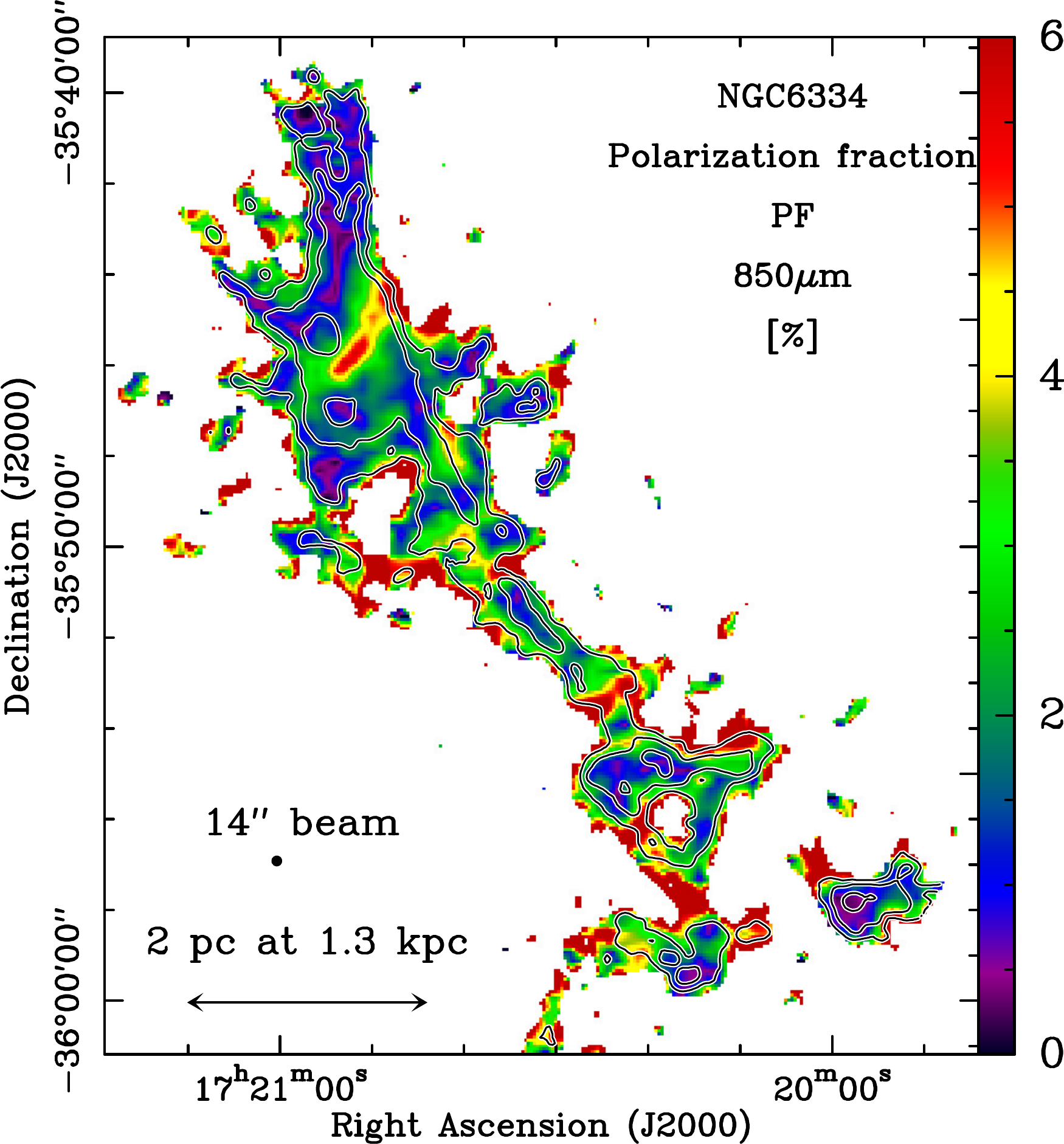}
}\vspace{-.1cm}
  \caption{ 
   Maps of the POS B-field angle $\chi_{B_{\rm POS}}$ ({\it Left}) and of the polarization fraction $PF$ ({\it Right}). 
  The HPBW resolution of the maps is $14\arcsec$. The plotted emission correspond to {\it SNR}$(I)>25$. %
    The contours are $I=0.4, 1.4,$ and $8\,$Jy\,beam$^{-1}$, as in Fig.\,\ref{IQUmaps}. 
}          
  \label{PAPF}
    \end{figure*}
    
\subsection{Polarization parameters}\label{polParam}

The polarization parameters are calculated from the combination of the Stokes $I$, $Q$, and $U$ parameters 
using the following relations
  \begin{gather*}
 PI_{\rm obs}= \sqrt{Q^2+U^2}\\
 PF_{\rm obs}=PI_{\rm obs}/I\\
  \psi=0.5\,\arctan(U,Q),
 \end{gather*}
 where $PI_{\rm obs}$, $PF_{\rm obs}$, and $\psi$ are the observed polarized intensity, the observed polarization fraction, and the polarization angle, respectively. 
 The polarization angle is calculated  in the IAU convention, i.e., North to East in the equatorial coordinate system. The POS magnetic field ($B_{\rm POS}$) orientation is obtained by adding $90^\circ$ to the polarization angle 
 \begin{equation}
 \chi_{B_{\rm POS}}= \psi+90^\circ.\label{psi}
  \end{equation}
 Given the noise present in the observed Stokes parameters, $PI_{\rm obs}$ and $PF_{\rm obs}$ are biased positively. We debias  $PI_{\rm obs}$ and $PF_{\rm obs}$ taking into account the uncertainties on $Q$ and $U$, $\delta Q$ and $\delta U$, respectively, using the relations \citep[e.g.,][]{Vaillancourt2006}: 
  \begin{gather}
 PI= \sqrt{Q^2+U^2-0.5(\delta Q^2+\delta U^2)}\label{PI}\\
   PF=PI/I.\label{PF}
     \end{gather}
We calculate the uncertainties on the polarization parameters as:
 \begin{gather*}
\delta PI=\sqrt{(Q\delta Q)^2+(U\delta U)^2}/PI\\ 
\delta PF=PF\, \sqrt{(\delta PI/PI)^2+(\delta I/I)^2} \\
 \delta \psi=0.5 \sqrt{(U\delta Q)^2+(Q\delta U)^2}/PI^2.
  \end{gather*}

 We estimate the signal-to-noise-ratio ({\it SNR}) for each of the quantities and use the following notation in the paper:  {\it SNR(value)=value/$\delta$value}, where $value$ is $I$, $PI$, or $PF$. 
Appendix\,\ref{App1} presents the distribution of the uncertainties of the different parameters over the observed region (Fig.\,\ref{Err_maps}). The scatter plots and histograms of Figs.\,\ref{Err_scatter}, \ref{IQUerr_histo},  and \ref{Err_histo} show the uncertainties of the Stokes parameters and of the polarization properties derived from the BISTRO data. 

  The relations  we used to debias $PI$ and $PF$ (Eqs.\,\ref{PI}\,and\,\ref{PF}) have been shown to be reliable for 
    {\it SNR(PI)}\,$\gtrsim3$ \citep[][]{Vaillancourt2006,Plaszczynski2014,Montier2015,Hull2015}. 
    We thus limit our quantitative analysis for data points with {\it SNR}$(PI)>3$.  
    
\begin{figure*}[!h]
   \centering
   \resizebox{19.cm}{!}{
   \includegraphics[angle=0]{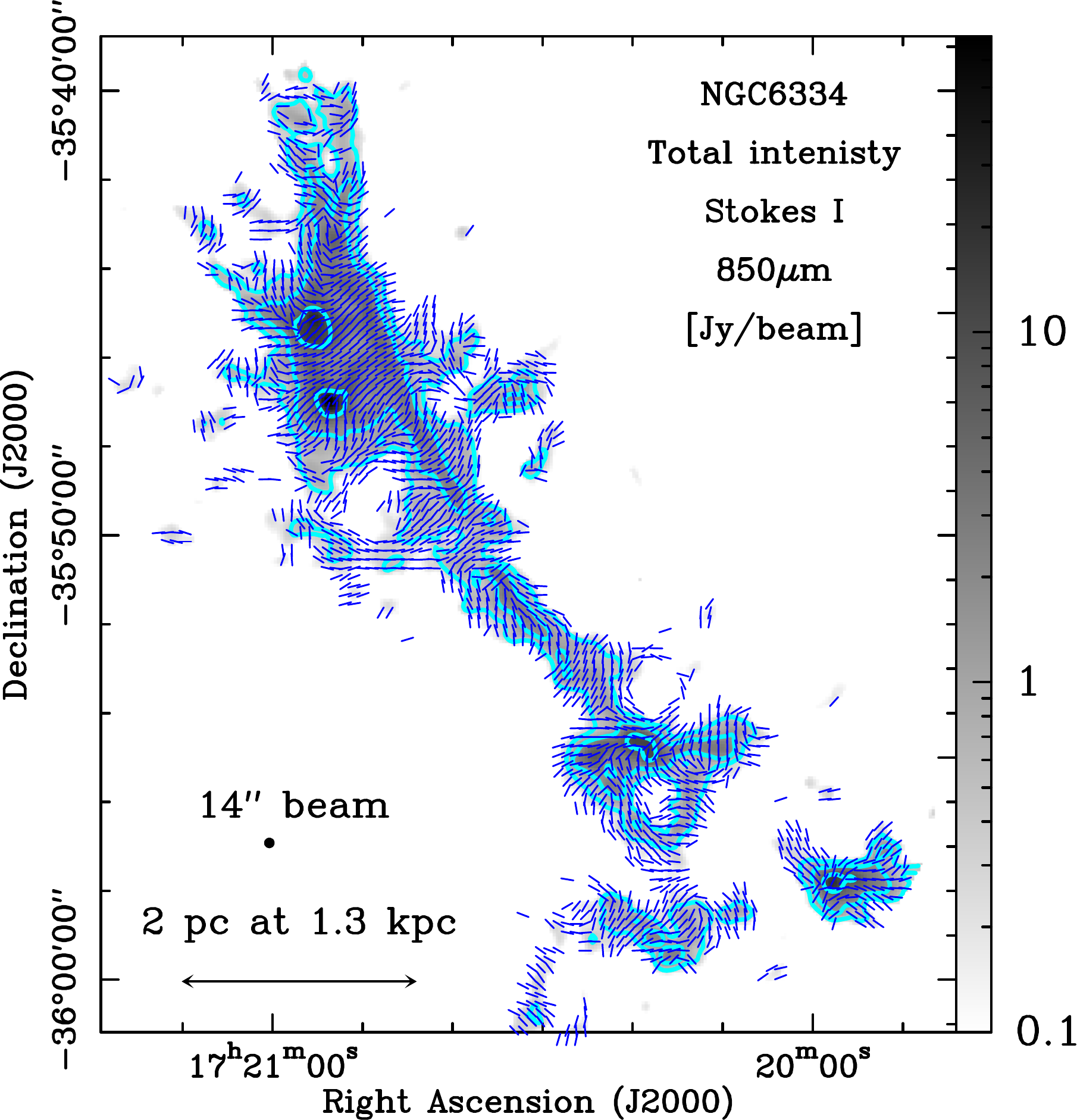}
  }\vspace{-.1cm}
  \caption{ 
Total intensity Stokes $I$ map (same as Fig.\,\ref{IQUmaps}-top left). The blue short lines (all with the same length) show the orientation of the POS B-field angle ($\chi_{B_{\rm POS}}$) for {\it SNR}$(I)>25$ and {\it SNR}$(PI)>3$. 
The contours are $I=0.4, 1.4,$ and $8\,$Jy\,beam$^{-1}$, as in Fig.\,\ref{IQUmaps}. 
}          
  \label{PI_PA1}
    \end{figure*}

\begin{figure*}[!h]
   \centering
   \resizebox{19.cm}{!}{ \includegraphics[angle=0]{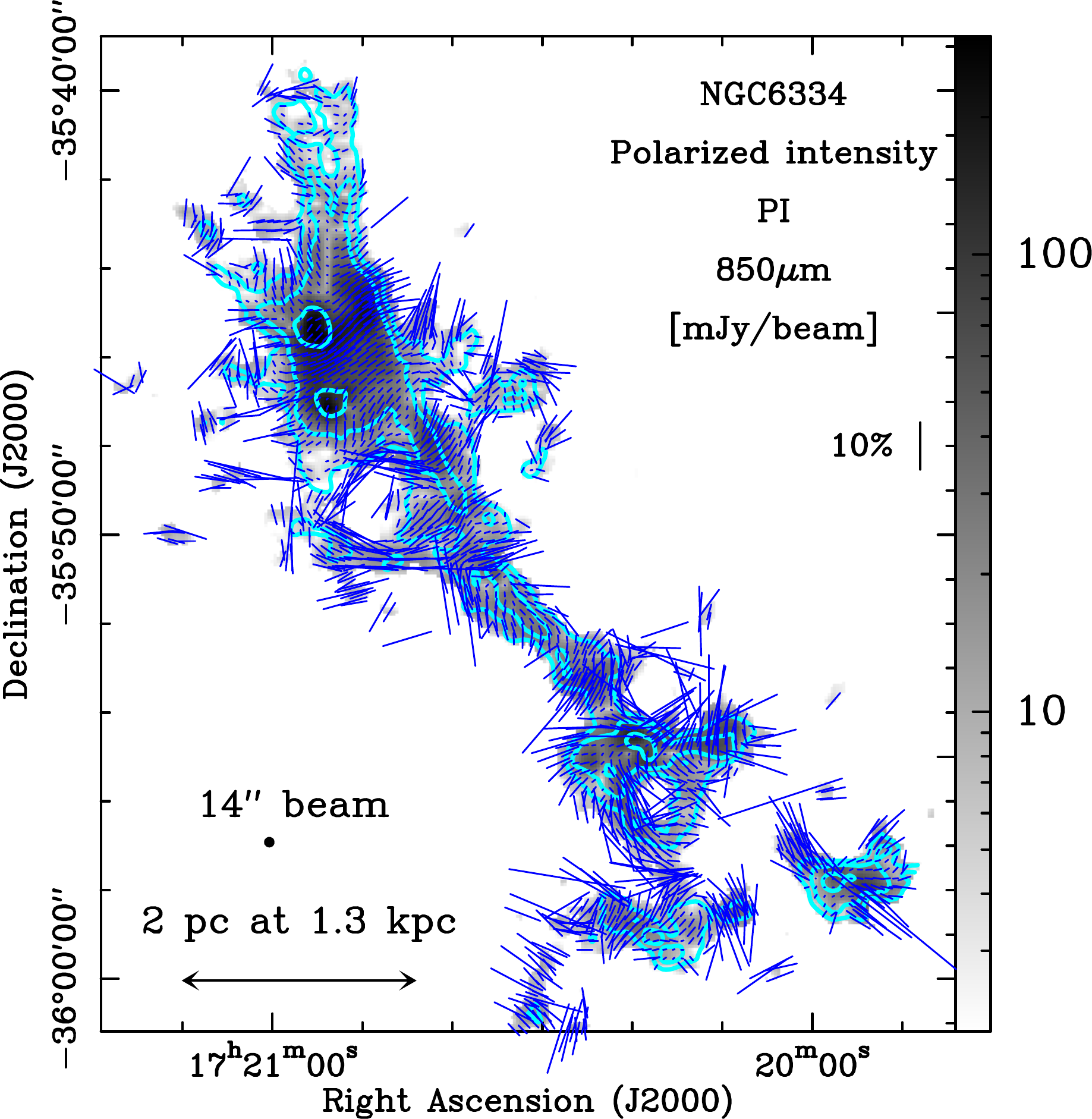}
}\vspace{-.1cm}
  \caption{ 
Polarized intensity ($PI$) map  (c.f., Fig.\,\ref{IQUmaps}-top right). 
  The blue short lines show the orientation of the POS B-field angle (c.f., Fig.\,\ref{PI_PA1}) and their lengths are proportional to the polarization fraction.
        A line showing a polarization fraction of $10\%$ is indicated on the  plot.
      The contours are $I=0.4, 1.4,$ and $8\,$Jy\,beam$^{-1}$, as in Fig.\,\ref{IQUmaps}. 
}          
  \label{PI_PA2}
    \end{figure*}

Figures\,\ref{IQUmaps}, \ref{PAPF}, \ref{PI_PA1}, and  \ref{PI_PA2} show the maps of the derived polarization properties. 
NGC 6334  is very bright at 850\,$\mu$m in both total and polarized emission. 
Most of the emission (corresponding to gas column density $\nhh\gtrsim1\times10^{21}\NHUNIT$, see also below)
 is detected with 
 {\it SNR}$(I)\gg25$ and {\it SNR}$(PI)\gg3$ (see Figs.\,\ref{SNRcontours} and \ref{IQUerr_histo}).
 For our analysis,  {\rev when not mentioned otherwise, } we select  data points with {\it SNR}$(I)>25$, which encompass $96\%$ of the emission with {\it SNR}$(PI)>3$ and correspond to  $\sim1300$ independent beams towards the observed $2\,{\rm pc} \times 10$\,pc field.

\subsection{Column density and comparison with $Herschel$ data}\label{Herschel}

 We estimate the  column density (\nhh)  from the total intensity  Stokes $I$ values at 850\,$\mu$m  
  with 
 the relation $\nhh=I_{850}/(B_{850}[T]\kappa_{850}\mu_{\rm H_2}m_{\rm H})$, where $B_{850}$ is the Planck function, $T=20\,$K is the mean dust temperature 
 of NGC 6334 derived from $Herschel$ data
 {\rev \citep[see,][and Appendix\,\ref{App2a} for a discussion on this adopted value]{Russeil2013},}
 $\kappa_{850}=0.0182 \,{\rm cm}^2$/g is the dust opacity per unit mass of dust + gas at 850\,$\mu$m    \citep[e.g.,][]{Ossenkopf1994},  
  $\mu_{\rm H_2}=2.8$  is the mean molecular weight per hydrogen molecule \citep[e.g.,][]{Kauffmann2008}, and $m_{\rm H}$ is the mass of a hydrogen atom. 
   
In Appendix\,\ref{App2b}, we compare the column density map derived from  BISTRO 850\,$\mu$m data with the 
column density map derived from 
SPIRE+\Artemis\ 350\,$\mu$m at $T=20\,$K \citep[presented by][at a spatial resolution of 8\arcsec]{Andre2016}
smoothed to the same resolution of the BISTRO 14\arcsec\ data.  
As can be seen  in Fig.\,\ref{BiArtMaps}, for column densities $\gtrsim3\times10^{22}\,\NHUNIT$  both maps agree within 
a factor $<2$.
 Furthermore, when we subtract a "Galactic emission"  of $3\times10^{22}\,\NHUNIT$ \citep{Andre2016} from the SPIRE+\Artemis\  column density map, this latter map matches very well  
with the column density map derived from BISTRO observations at 850\,$\mu$m (within the dispersion of the emission towards the observed region). 
This "Galactic emission"   of $3\times10^{22}\,\NHUNIT$ (equivalent to 124\,MJy/sr) may correspond to {\rev a combination of   extended emission physically linked to the NGC 6334 molecular complex and} large-scale foreground or background emission observed towards  NGC 6334, which is itself located within $0.6$\,deg of the Galactic Plane. 
As is usual with ground-based submillimeter (submm) continuum observations, the SCUBA2-POL2  observations carried out with the JCMT are affected by correlated sky noise over the entire observed field. 
Given that  the data reduction subtracts this correlated noise, the data are significantly affected by missing flux \citep[e.g.,][]{Sadavoy2013}
at scales $\gtrsim6\arcmin$ larger than the field-of-view of the DAISY map  with uniform noise coverage. %

Comparing with the emission observed with the $Herschel$ space telescope, which is in principle sensitive to the large scale emission, 
 we  conclude that the BISTRO data of NGC 6334 
 recover most of the (total intensity) flux of the dense, thin, and elongated molecular filamentary structure on which we focus our analysis in the following (c.f.,  Appendix\,\ref{App2b}). 

 \begin{figure*}[!h]
   \centering
     \resizebox{14.cm}{!}{ 
\includegraphics[angle=0]{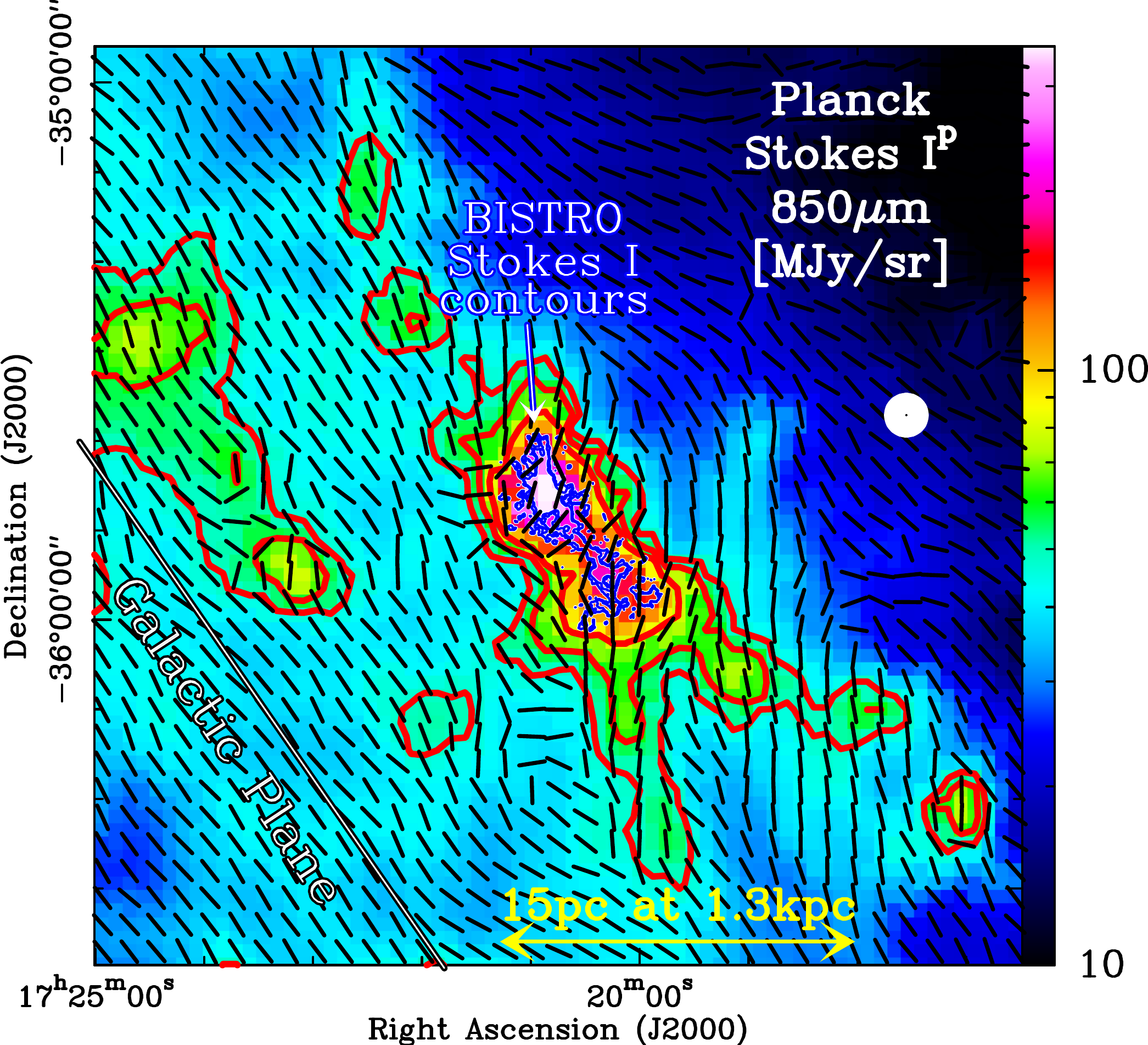}}   
  \caption{ 
  $Planck$ total intensity $I^{\rm p}$ map of the 850$\,\mu$m emission  towards NGC\,6334. %
  The red contours show the $Planck$ intensity at $I^{\rm p}\sim$ 48, 60, and 80\,MJy/sr (at $5\arcmin$ resolution) and the blue contour the BISTRO intensity 0.4\,Jy/14\arcsec-beam or $75\,$MJy/sr (at $14\arcsec$ resolution).
  The black lines are the POS B-field  angles derived from $Planck$ data at the resolution of $5\arcmin$. The  $Planck$ ($5\arcmin$) and BISTRO ($14\arcsec$) beams are shown on the right hand side of the plot as  concentric filled white and black circles, respectively. The Galactic Plane is also indicated. %
}          
  \label{BiPlanckMaps}
    \end{figure*}

  \begin{figure}[!h]
   \centering
     \resizebox{8.cm}{!}{ 
\includegraphics[angle=0]{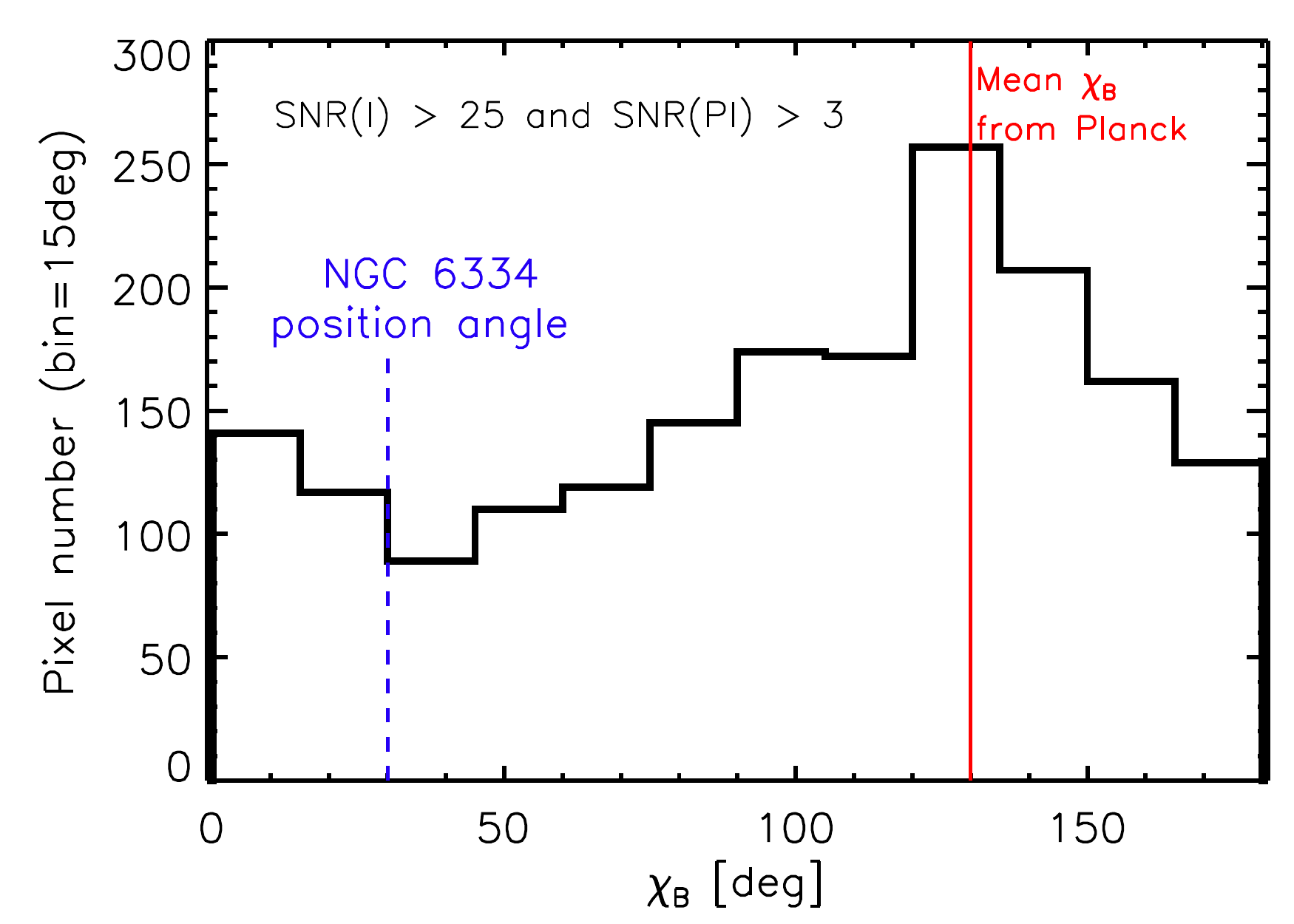}}
  \resizebox{8.cm}{!}{\includegraphics[angle=0]{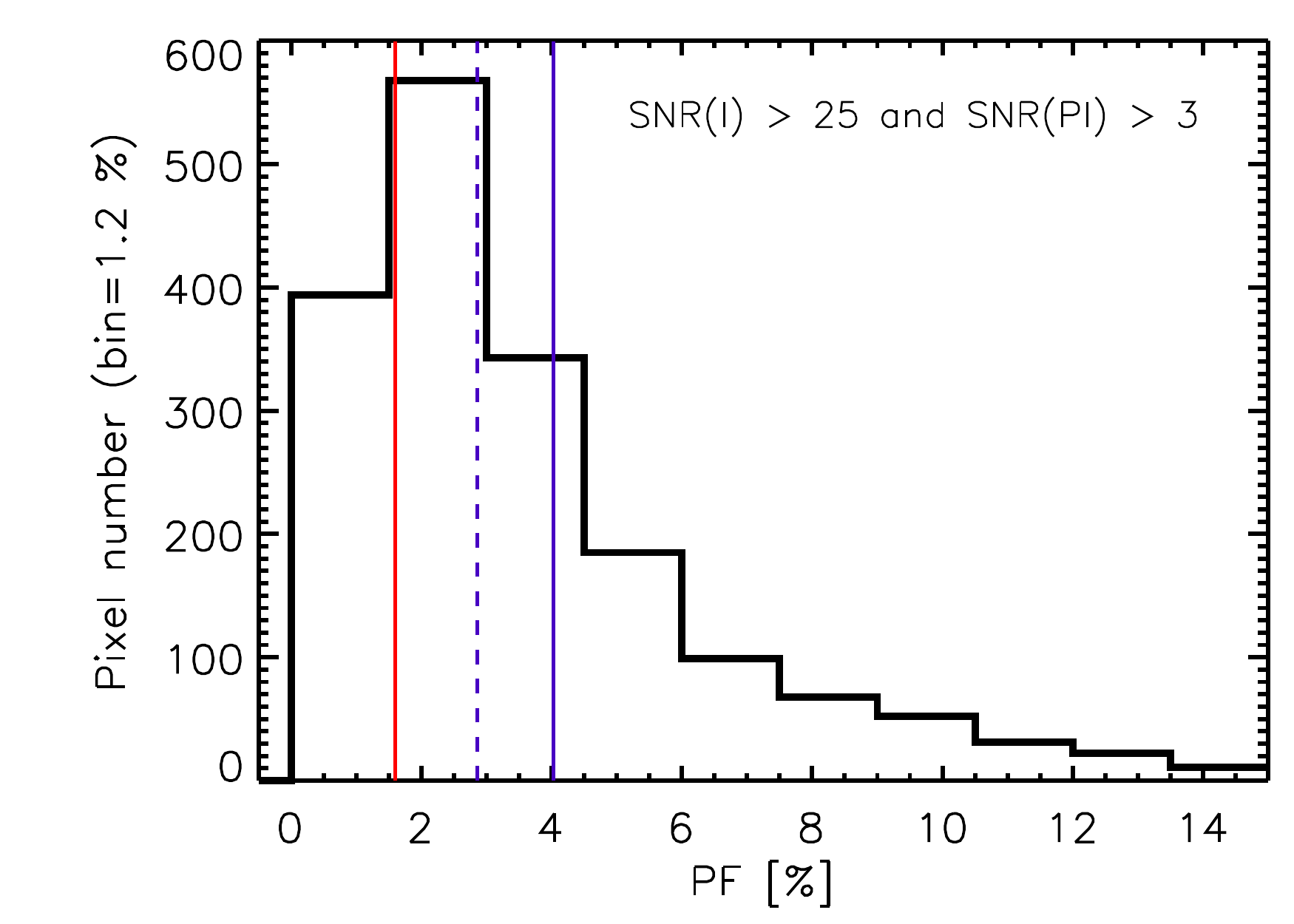}
}\vspace{-0.2cm}
  \caption{{\it Top:} Distribution of $\chi_{B_{\rm POS}}$. %
  The blue dashed  line at $30^\circ$ indicates the mean position angle ($\theta_{\rm fil}$) of NGC 6334. The red solid line at $130^\circ$ shows the mean $\chi_{B_{\rm POS}}^{\rm p}$  derived from $Planck$ data (at the resolution of $5\arcmin$)  towards the field observed with BISTRO (cf., Fig.\,\ref{BiPlanckMaps}). 
 {\it Bottom:} Distribution of  the polarization fraction $PF$. %
  The solid and dashed blue lines indicate the mean and median values of $PF$ at $14\arcsec$, respectively.  The solid red line shows the mean $PF^{\rm p}$ value  derived from $Planck$ data at $5\arcmin$ resolution towards the BISTRO field. 
}          
  \label{histoPlot}
    \end{figure}

    \subsection{Comparison with $Planck$ data}\label{planck}

As for the total intensity (see Sect.\,\ref{Herschel}), the polarized intensity observed by the JCMT/SCUBA2-POL2 may be affected by missing flux at large scales.     
To estimate this missing flux, we use the 850$\,\mu$m $Planck$\footnote{$Planck$ archival data have been retrieved from \url{http://www.esa.int/Planck}} polarization data at a resolution of $5\arcmin$. 
From the Stokes $I^{\rm p}$, $Q^{\rm p}$, and $U^{\rm p}$ $Planck$ maps, we derive the polarization properties  $\chi_{B_{\rm POS}}^{\rm p}$, $PI^{\rm p}$, and $PF^{\rm p}$ using Eqs.
\,\ref{psi},\,\ref{PI},\,and\,\ref{PF}, respectively. Figure\,\ref{BiPlanckMaps} shows a large-scale map around NGC 6334 at a resolution of $5\arcmin$. 
NGC 6334 lies within $0.6^\circ$ of the Galactic plane and the surrounding POS B-field is mostly observed parallel to this plane. 
 Towards the NGC 6334 filamentary molecular cloud, however, the B-field changes orientation to become mostly perpendicular to the elongated filament. 
 This was previously reported by a study using starlight polarization \citep{Li2006}. 
  At the NGC 6334 distance of 1.3\,kpc, the $Planck$ data with a resolution of  $5\arcmin$ ($\sim2$\,pc at 1.3\,kpc) cannot resolve the B-field  structure of  NGC 6334 with a transverse size $\ll2$\,pc. This small-scale  $B_{\rm POS}$ structure is now revealed by BISTRO observations at $14\arcsec$ (see Figs.\,\ref{PI_PA1} and\,\ref{PI_PA2}).

To estimate the missing flux of the BISTRO data, we compare the total and  polarized intensities of the BISTRO and $Planck$ maps at the same  5\arcmin\ resolution and towards the same area. %
The mean values derived from $Planck$ are $I^{\rm p}\sim206\,$MJy/sr and $PI^{\rm p}\sim3.4\,$MJy/sr. The mean values derived from $Planck$ data in the surroundings of the BISTRO map are $I^{\rm p,Galactic}\sim100\,$MJy/sr and $PI^{\rm p,Galactic}\sim1.5\,$MJy/sr (corresponding to the closest red contour to the blue contour in Fig.\,\ref{BiPlanckMaps}). The mean values derived from BISTRO data smoothed to the 5\arcmin\ resolution  are $I^{5\arcmin}\sim106\,$MJy/sr and $PI^{5\arcmin}\sim1\,$MJy/sr. 
The $I^{5\arcmin}$ value is equal to the "Galactic emission" subtracted $Planck$ value of $I^{\rm p}-I^{\rm p,Galactic}\sim106\,$MJy/sr, where $I^{\rm p,Galactic}\sim100\,$MJy/sr corresponds to a column density value  $\sim2.4\times10^{22}$\NHUNIT. The $PI^{5\arcmin}$ value is  compatible within a factor of 2 with  the "Galactic emission" subtracted $Planck$ value of   $PI^{\rm p}-PI^{\rm p,Galactic}\sim1.9\,$MJy/sr.  

This comparison with the $Planck$ data suggests that the BISTRO data recover most of the total and polarized intensity of the NGC 6334 filaments and are essentially free from large scale emission corresponding to 
{\rev extended emission physically linked to the NGC 6334 molecular cloud in addition to} 
line-of-sight  foreground/background emission.  Since, here,  we are interested in studying the polarization properties of this cold and dense gas component towards  NGC 6334, %
BISTRO   high-angular-resolution and high-sensitivity data are very powerful  for such an analysis. If we were to analyse $Planck$ data of NGC 6334, we would have had to remove the contribution to the emission of the surrounding  ISM following the method described in, e.g., \citet[][]{planck2016-XXXIII}.

 \section{Statistical characterization of the whole field}\label{ana1}

 \subsection{Histogram and dispersion of the B-field angles}\label{histoPA}

  Figure\,\ref{histoPlot}-top shows the histogram of the plane-of-the-sky B-field angles $\chi_{B_{\rm POS}}$ towards the whole field for ${\it SNR(I)}>25$ and ${\it SNR(PI)}>3$. 
 The distribution of   $\chi_{B_{\rm POS}}$ spans all the possible values (from $0^\circ$ to $180^\circ$). 
     There is also a peak at an angle of $\sim130^\circ$, which corresponds to the mean $\chi_{B_{\rm POS}}^{\rm p}$  observed by $Planck$ (at a resolution of 5\arcmin) towards the same region (Fig.\,\ref{BiPlanckMaps}).
 The mean POS orientation of the filamentary structure of NGC 6334 elongated Northeast-Southwest is measured to be $\theta_{\rm fil}\sim30^\circ$ (East of North).  
{\rev Thus, while there is a large observed scatter of $\chi_{B_{\rm POS}}$ towards NGC 6334, about $40\%$ of the  $\chi_{B_{\rm POS}}$ are within $\pm25^\circ$ of being perpendicular to the filament.}

    \begin{figure}[!h]
   \centering
 \hspace{-.2cm} 
 \resizebox{9.cm}{!}{ 
\includegraphics[angle=0]{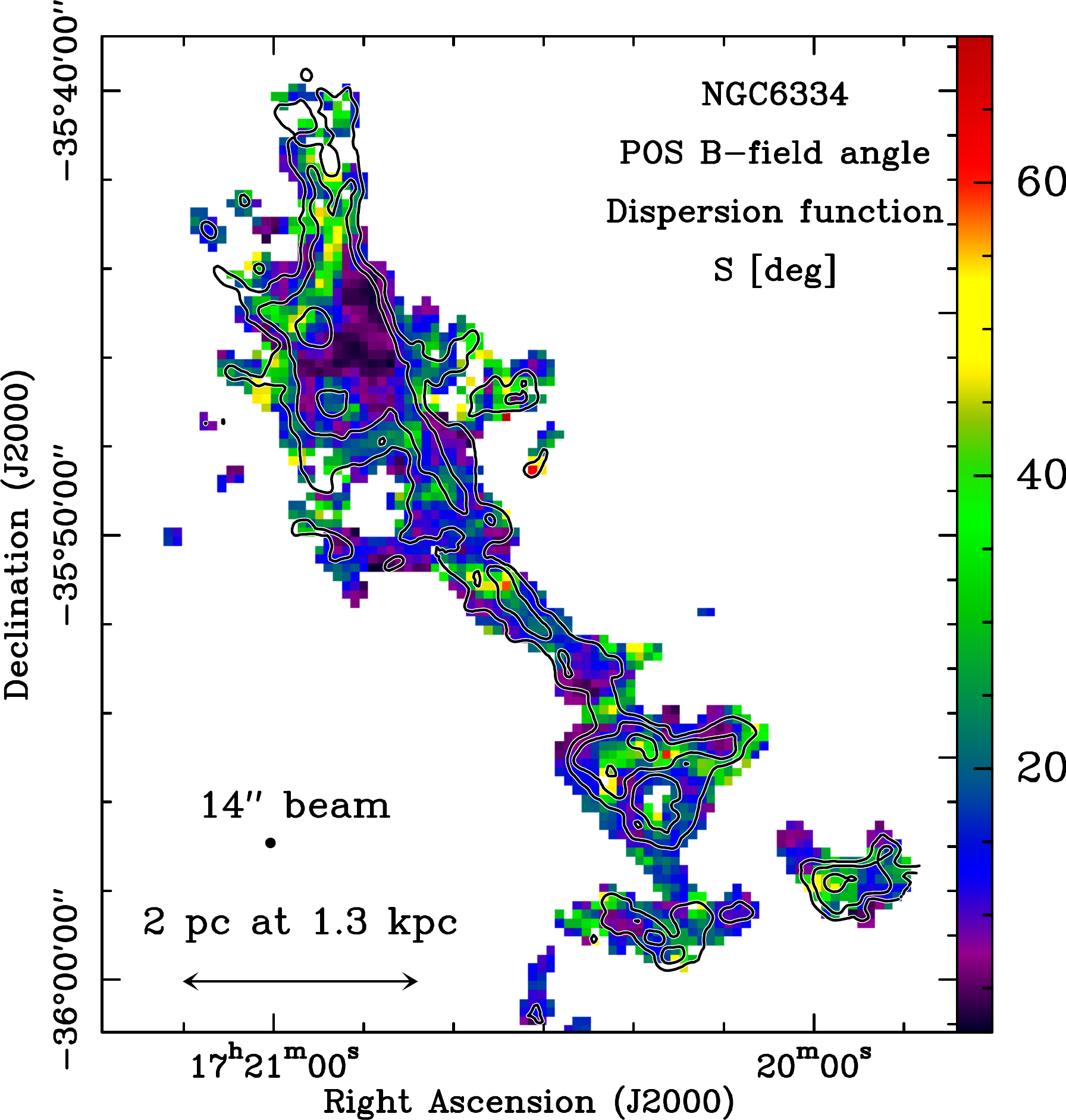}
}\vspace{-.1cm}
  \caption{ Map of $S$ (dispersion function of  $\chi_{B_{\rm POS}}$, see Eq.\,\ref{eqS}) for {\it SNR}$(I)>25$ and {\it SNR}$(PI)>3$. 
 The contours are $I=0.4, 1.4,$ and $8\,$Jy\,beam$^{-1}$, as in Fig.\,\ref{IQUmaps}. 
}          
  \label{SrmsPA}
    \end{figure}

     All the $\chi_{B_{\rm POS}}$ measurements for ${\it SNR(I)}>25$ and ${\it SNR(PI)}>3$ have statistical uncertainties smaller than $10^\circ$  with  typical uncertainties of $\sim7^\circ$, $\sim4^\circ$, and $\sim1^\circ$ for  3\,$<\,${\it SNR(PI)}$\,<$\,5,   5\,$<\,${\it SNR(PI)}$\,<$\,10, and {\it SNR(PI)}$>$10, respectively (see Fig.\,\ref{Err_scatter}-right).
 The observed scatter in $\chi_{B_{\rm POS}}$ is much larger than the intrinsic uncertainties on the measurements and thus traces the variation of the B-field structure over the whole NGC 6334 region. 
 
 Figure\,\ref{PI_PA1}  displays the spatial  distribution of $\chi_{B_{\rm POS}}$ over the whole field indicating that the observed statistical scatter does not result from a completely random field. Indeed, the $\chi_{B_{\rm POS}}$ values seem to be uniform in patches. This morphology suggests a locally ordered magnetic field with a varying $B_{\rm POS}$ structure over larger scales.    
    To assess this "ordered variation" of the B-field, we calculate the dispersion function of  $\chi_{B_{\rm POS}}$ given by
\begin{equation}    
S(x)=\sqrt{\frac{1}{N}\sum_{i=1}^{N}\left[\chi_{B_{\rm POS}}(x)-\chi_{B_{\rm POS}}(x_i)\right]^2},\label{eqS}
\end{equation}  
where $x$ is the pixel position,  $x_i$ is the neighbouring pixel to $x$,  
   and $N$ is the number of $x_i$ pixels surrounding the  pixel $x$. 
 Here we are interested in the measure of the local dispersion of $\chi_{B_{\rm POS}}$. We calculate $S$ for the  nearest four and next-nearest four neighbours surrounding a given pixel on a regular grid, i.e., $N=8$ corresponding to a box of $3\times3$ pixels or $36\arcsec\times36\arcsec$. 
For each pixel position, the $|\chi_{B_{\rm POS}}(x)-\chi_{B_{\rm POS}}(x_i)|\le90^\circ$ condition is imposed on the right hand side of Eq.\,\ref{eqS}
and the supplementary angle ($180^\circ-|\chi_{B_{\rm POS}}(x)-\chi_{B_{\rm POS}}(x_i)|$) is taken otherwise. 

The map of $S$ is shown in Fig.\,\ref{SrmsPA} where patches of low $\chi_{B_{\rm POS}}$ dispersions (i.e., ordered B-field structure) are separated by large values of $S$.

  \begin{figure}[!h]
   \centering
     \resizebox{8.5cm}{!}{ 
\includegraphics[angle=0]{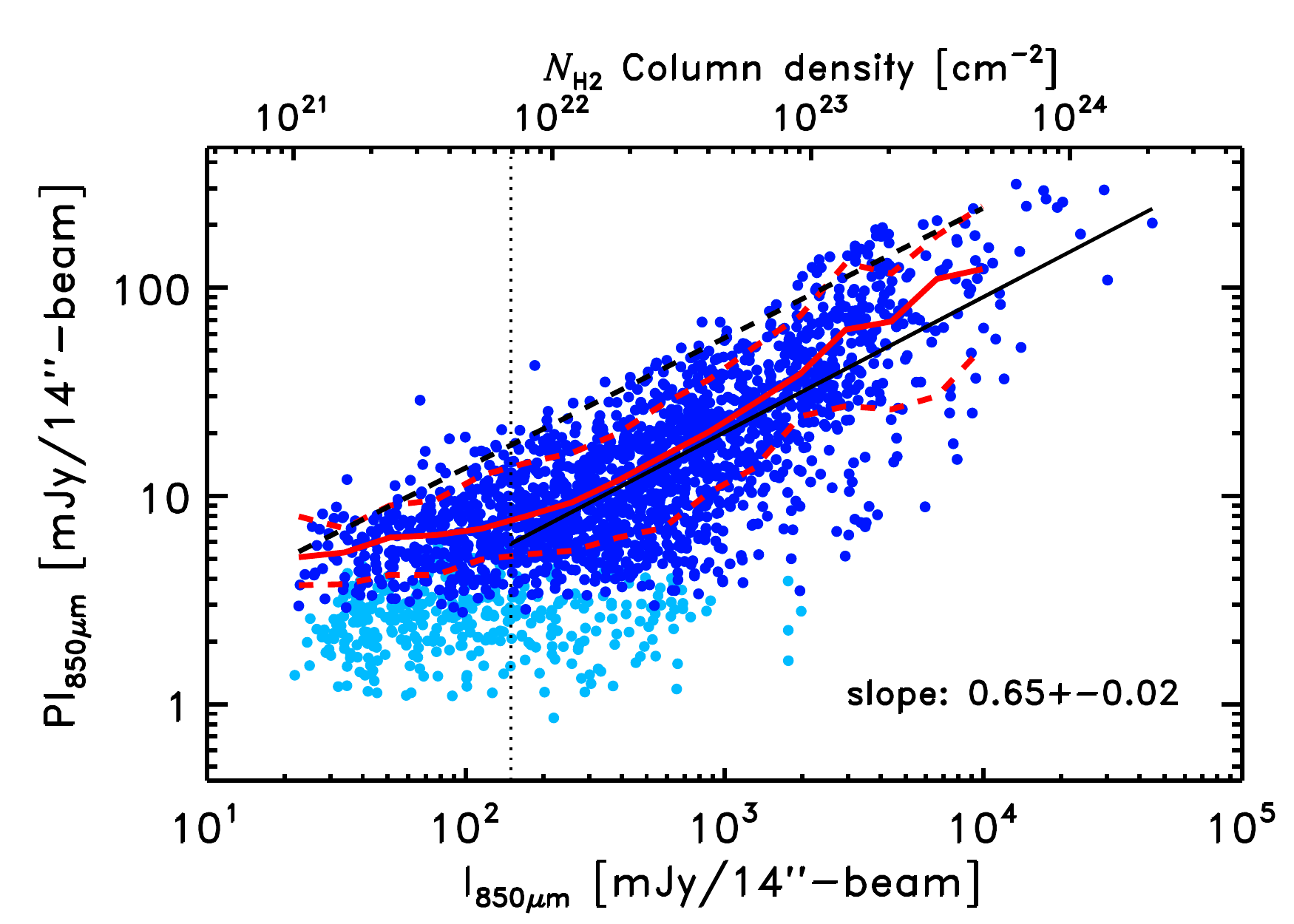}}
  \resizebox{8.5cm}{!}{\includegraphics[angle=0]{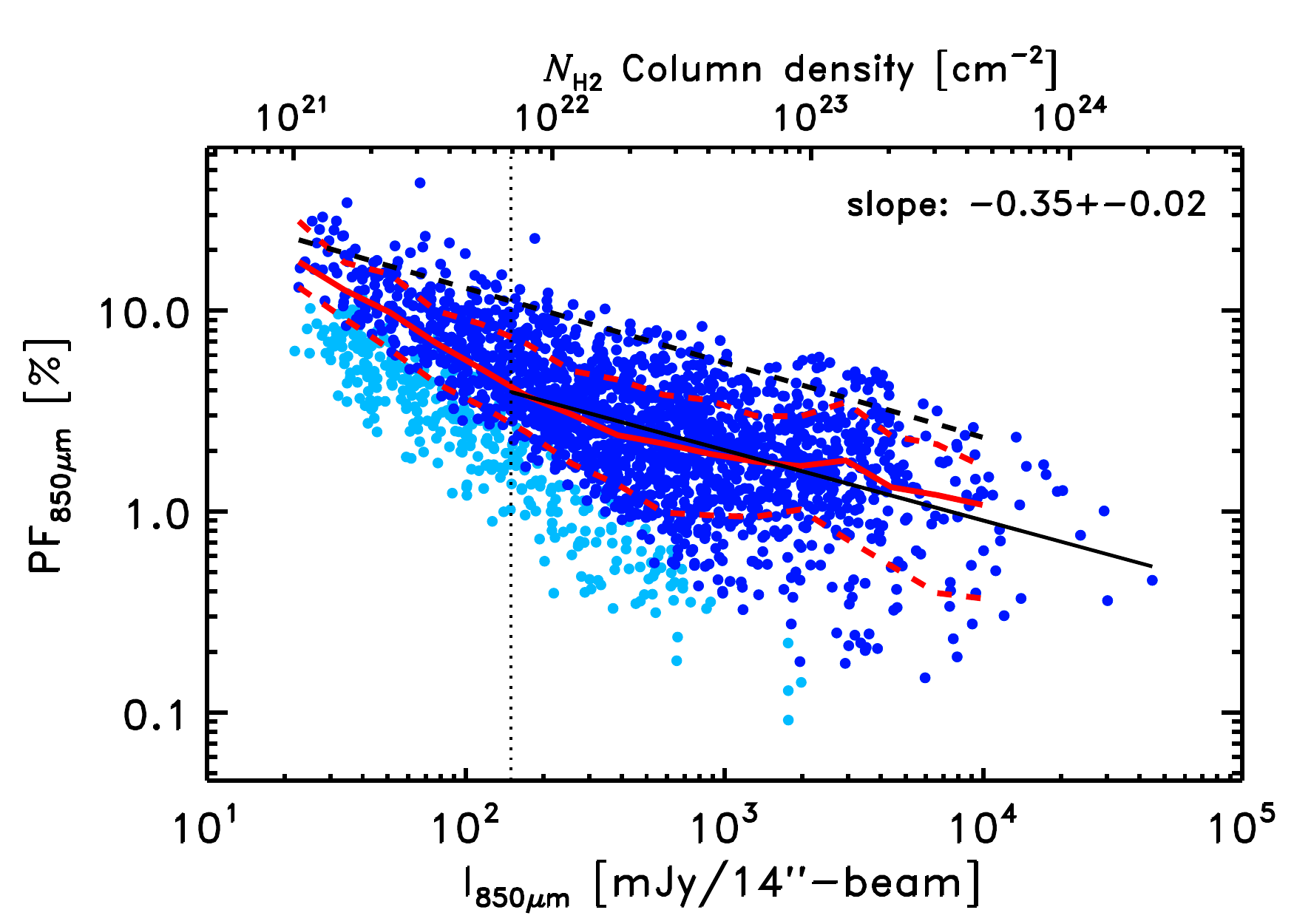}
}\vspace{-0.1cm}
  \caption{ {\it Top:} $PI$ versus  $I$  for {\it SNR}$(I)>25$. The light and dark blue symbols  are data points with $1<${\it SNR}$(PI)<3$ and {\it SNR}$(PI)>3$, respectively. %
    The dots correspond to  $12\arcsec$ pixels.
  The red solid curve shows the median  log($PI$) per bin of log($I$). The lower and the upper dashed red lines show the  $16\%$ and $84\%$ percentiles, respectively. 
  The black solid line is a linear fit to the distribution for {\it SNR}$(PI)>3$ and $I>150\,$mJy/beam (indicated by the dotted vertical line).  
  The value of the slope is given on the plot. 
  The black dashed line is a linear fit to the upper $95\%$ percentiles (see Sect.\,\ref{scatter}). 
The upper $x-$axis gives an estimate of the $\nhh$ column density (see Sect.\,\ref{Herschel}).     
  {\it Bottom:} $PF$ versus  $I$. The dots, curves, and colours are the same as  the upper plot for $PF$ versus $I$. 
}    
  \label{scatPlot}
    \end{figure}

 \subsection{Histogram of the polarization fraction}\label{histoPA}
 
Figure\,\ref{histoPlot}-{\it bottom} shows the distribution of the polarization fraction, $PF$, derived from the BISTRO data for 
  {\it SNR}$(I)>25$ and {\it SNR}$(PI)>3$. The mean, median, and  standard deviation are  $\sim4\%$, $\sim2.9\%$, and $\sim3.9\%$, respectively.  
    Typical statistical uncertainties\footnote{These statistical uncertainties are derived from sinusoidal function fitting to the sinusoidal time series data obtained with the POL2-DAISY observation mode of POL-2. The absolute $PF$ values discussed in this paper are corrected from the instrumental polarization (IP) offset, with a systematic uncertainty on the IP model estimated to be  about $0.3\%$ (c.f.,\,{\it https://www.eaobservatory.org/jcmt/2019/05/investigating-the-effects-of-uncertainties-in-the-pol2-ip-model/})\,as\,derived from  the BISTRO DR3 August 2019 IP model ({\it https://www.eaobservatory.org/jcmt/2019/08/new-ip-models-for-pol2-data/}).} 
    of the $PF$ measurements are $1.1\,\%$, $0.7\,\%$, and $0.2\,\%$  for samples with 3\,$<\,${\it SNR(PI)}$\,<$\,5,   5\,$<\,${\it SNR(PI)}$\,<$\,10, and {\it SNR(PI)}$>$10, respectively (c.f., Fig.\,\ref{Err_scatter}-middle).
   The  scatter of $PF$ %
   observed here is larger than the measurement uncertainties and may be used to study the physical origin of the $PF$ distribution.  
  The mean polarization fraction observed by $Planck$ towards the same region mapped by BISTRO observations is equal to $\sim1.6\%$ at the $5\arcm$ resolution. 
The lower polarization fraction at the coarser $Planck$ resolution is probably due to depolarization within the beam of unresolved B-field structure.

    \subsection{Scatter plots and correlations between the  parameters}\label{scatter}
    
    Figure\,\ref{scatPlot} shows scatter plots between $I$, $PI$, and $PF$.   
        We note that the {\rev two} plots shown in Fig.\,\ref{scatPlot} are different projections of three quantities ($I$, $PI$, and $PF$), which are, by definition, not independent (with $PF=PI/I$, see Eq.\,\ref{PF}). We thus expect clear relations among the distributions. We discuss this below. 
            
    We estimate  the mean and median values of log($PI$)  and log($PF$)  as well as their  $16\%$ and $84\%$ percentiles (corresponding to the standard deviation of a normal distribution) per  bin of log($I$)   
    for data points with  {\it SNR}$(I)>25$ and {\it SNR}$(PI)>3$.  
    We also show, the distribution of the data with $1<SNR(PI)<3$ in polarized intensity (that are not correctly debiased here) as a reminder of possible shortcomings in our interpretation 
     due to the removal of low {\it SNR}$(PI)$ and small $PF$ data points when only high-SNR ({\it SNR}$(PI)>3$) data points are selected. 
     For $I<150$\,mJy/beam (indicated as a dotted vertical line in Fig.\,\ref{scatPlot}) the mean/median of the distributions of  $PI$ versus $I$ and $PF$ versus $I$ are  significantly affected by the selection effect of high-SNR data points.

Since the polarized emission depends to first order (for an ordered B-field structure along the LOS and uniform grain alignment efficiency) on the column density of dust present along the LOS emitting polarized emission, $PI$ increases with increasing $I$. We can see this positive trend over  $\sim3$ orders of magnitude in both $PI$ and $I$, as $PI\propto I^{\,0.65\pm0.02}$, where the slope is derived from a linear fit (in  log-log scale) to the data points for $I\ge150$\,mJy/beam.     We can also  notice, for a given value of $I$, a large scatter in $PI$. This scatter by more than one order of magnitude in $PI$ is seen for all the $I$ values from a few 10\,mJy/beam to $10^4$\,mJy/beam, i.e.,  spanning $\sim3$ orders of magnitude in $I$. 
With such a scatter a linear fit to the upper envelope of the distribution may indicate the maximum observed $PI$  for a given $I$. In addition, this upper envelope is more robust with respect to high-SNR data selection effects, while it may be limited by the angular resolution of the observations. 
A fit to the upper envelope yields a slope of $0.62\pm0.03$ similar to the slope derived for the full distribution for {\it SNR}$(PI)>3$ and $I\ge150$\,mJy/beam.

Figure\,\ref{scatPlot}-{\it Bottom} shows the distribution of  $PF$ as a function of $I$. 
There is  an overall decreasing trend of $PF$ with increasing $I$ over $\sim2$ orders of magnitude in $PF$ (from $\sim0.2\%$ to  $\sim20\%$) with $PF\propto I^{\,-0.35\pm0.02}$. The slope  is derived from a linear fit (in  log-log scale) to the data points for {\it SNR}$(PI)>3$ and $I\ge150$\,mJy/beam. 
This best fit slope is expected from the result of Fig.\,\ref{scatPlot}-{\it Top} since $PF=PI/I$ and $PI\propto I^{\,0.65\pm0.02}$.
A steeper slope ($PF\propto I^{\,-0.46\pm0.01}$) is obtained if the full distribution (for {\it SNR}$(PI)>3$) is fitted.  
A fit to the upper envelope (not affected by the selection criteria of high-SNR) %
yields a slope of $-0.37\pm0.03$ similar to the slope derived for the full distribution for {\it SNR}$(PI)>3$ and $I\ge150$\,mJy/beam.

 The scatter  plot of $PF$ versus $I$ indicates a large scatter in $PF$  (of $\sim 2$ orders of magnitude) for a given value of $I$. 
 Such a scatter was shown earlier by the statistical analysis of the $Planck$ polarization data on large scales \citep{planck2015-XIX,planck2016-XLIV} as well as at smaller scales in other regions analysed within the BISTRO survey \citep{Kwon2018,Soam2018,Wang2019}.  
 We find here, however,  slopes significantly shallower than those shown before, e.g.,  \citet{Soam2018} and \citet{Wang2019} reported slopes of $-0.9$ and $-1$, respectively \citep[see also similar results from the analysis of SMA and ALMA data, e.g.,][]{Chen2012,Koch2018}.

 \begin{figure}[!h]
   \centering
     \resizebox{8.cm}{!}{ \hspace{-.5cm}
\includegraphics[angle=0]{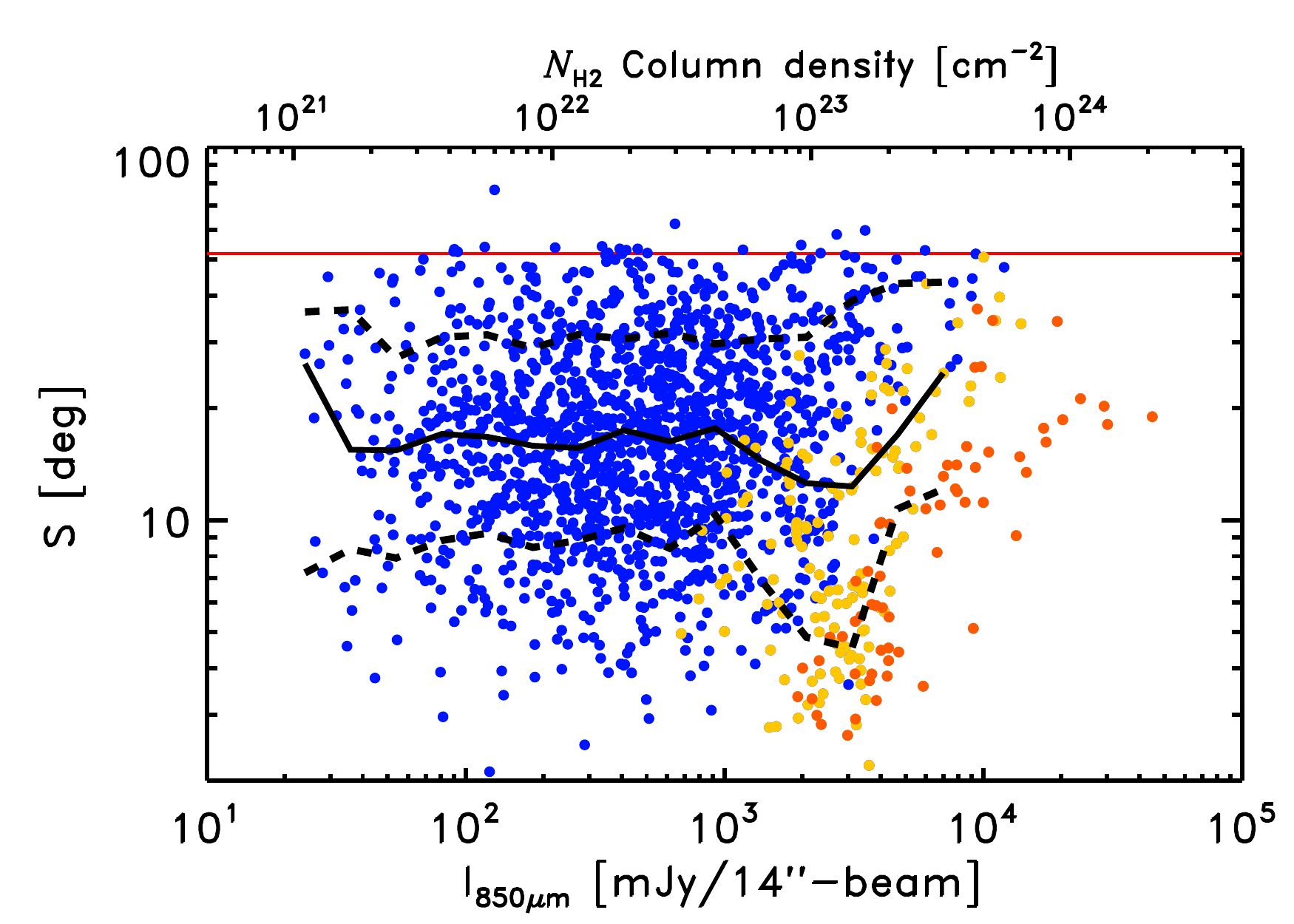}
}\vspace{-.7cm}
  \resizebox{8.cm}{!}{ \hspace{-.4cm}
\includegraphics[angle=0]{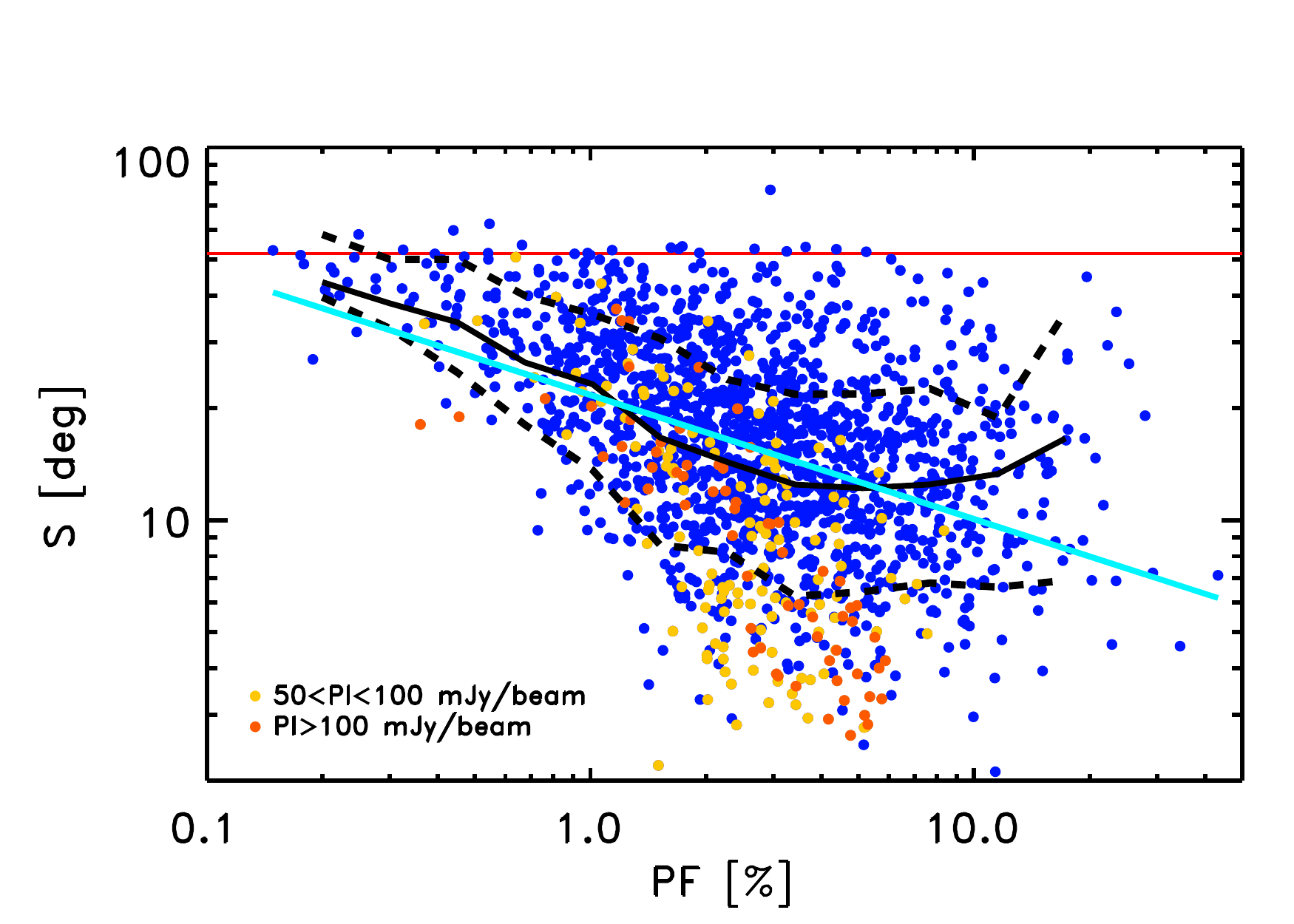}
}\vspace{-.2cm}
  \caption{ 
  Scatter plot of $S$ (dispersion function of  $\chi_{B_{\rm POS}}$) 
    as a function of $I$ ({\it top}) and $PF$ ({\it bottom})
    for data points with  {\it SNR}$(I)>25$ and {\it SNR}$(PI)>3$ (c.f., Fig.\,\ref{SrmsPA}).
The data points with $50<PI<100\,$mJy/beam and $PI>100\,$mJy/beam are shown in yellow and red, 
  respectively. 
  The black solid curves show the median  log($S$) per bin of log($I$) and log($PF$) in the top and bottom panels, respectively. The lower and the upper dashed black lines are the  $16\%$ and $84\%$ percentiles of the distributions, respectively. 
      The red horizontal line at  $S=52^\circ$ shows the value that $S$  would take for a B-field dominated by turbulence.       
In the bottom panel,  the cyan solid line is a linear fit to the distribution.  
}          
  \label{scatPlotS}
 \end{figure}

  Figure\,\ref{scatPlotS} shows the distribution of $S$ (dispersion function of  $\chi_{B_{\rm POS}}$, see Eq.\,\ref{eqS} and Fig.\,\ref{SrmsPA}) 
  as a function of $I$ or $PF$ for the whole field.  
  While there is no particular trend seen between $S$ and $I$,  an anti-correlation between $S$ and $PF$ can be noticed (with a slope of $-0.34$).
  We note, however, that this latter anti-correlation may be the result of the selection of data points with {\it SNR}$(PI)>3$ removing those measurements with $PF\lesssim1\%$.
 A large spread of the values of $S$ for a given value of $PF$ can be seen, especially for  $PF>1\%$. 
Such statistical distribution of   $S$ as a function of  $PF$  is also  seen with $Planck$ towards the whole sky  \citep[e.g.,][]{planck2015-XIX} as well as on much smaller scales with ALMA towards high-mass star forming regions \citep[e.g.,][]{Koch2018}.

We discuss possible origins of 
 these  scatters  of the observed polarization properties in Sects.\,\ref{disc:comp} and \ref{disc:feedback} below. %

\section{Properties of the different crests identified towards the NGC 6334 filament network}\label{PropCrests}

\begin{figure}[!h]
   \centering
     \resizebox{9.cm}{!}{ 
\includegraphics[angle=0]{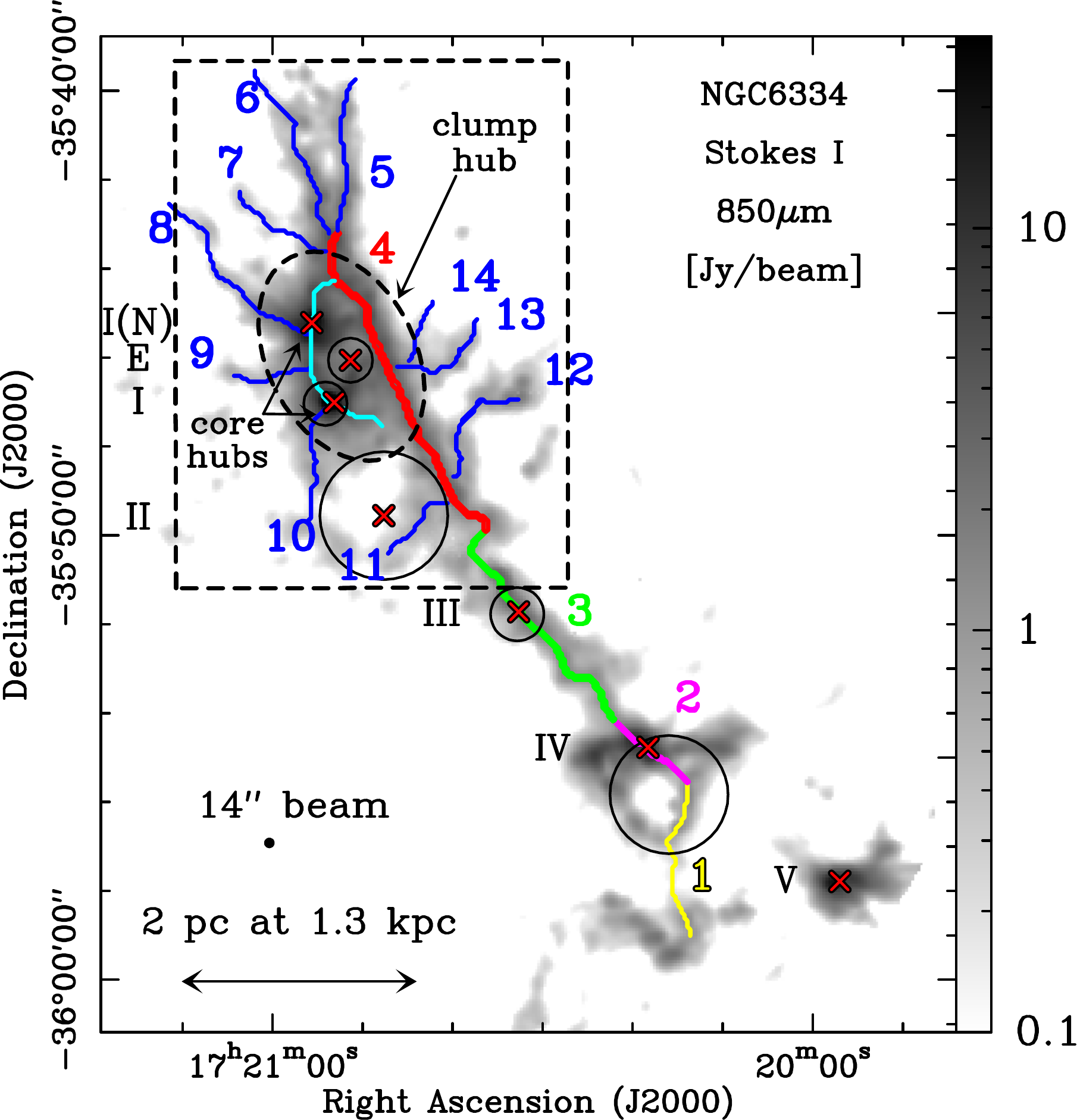}}
 \vspace{-.4cm} \caption{%
  The crests of the identified filamentary structures are over-plotted on the Stokes $I$ map. The connected sections from 1 to 4 trace the ridge crest  (cf., Figs.\,\ref{polCrest} and \ref{AngleCrest35}). The blue lines from 5 to 14 trace the crests of the sub-filaments.   The cyan line traces the crest of a filament connecting the two {\it core-hubs}, I and I(N).
  The black dashed rectangle indicates the hub-filament structure (see Fig.\,\ref{SkelFig2}). The black dashed ellipse  shows the {\it clump-hub}. 
 The crosses and circles are the same as in Fig.\,\ref{IQUmaps}-top-left. 
}          
  \label{SkelFig}
    \end{figure}

The molecular cloud structure of NGC 6334, on large scales,  may be described as a "ridge-dominated" structure (cf. Sect.\,\ref{intro})   with a main  $\sim10$\,pc-long ridge threaded by a network of shorter filamentary structures, the "sub-filaments", connected to the ridge from the side.
In the North of the field, two dense and compact star-forming hubs, I and I(N), can be seen 
 located at the junction of multiple sub-filaments  as presented by \citet[][]{Sandell2000} and observed  recently at high-resolution  with ALMA \citep{Sadaghiani2020}. 
These two sources are in turn observed towards a larger ellipsoidal $\sim1\,{\rm pc}\times2$\,pc-scale hub-like structure (see Fig.\,\ref{SkelFig}) encompassing also a section of the ridge, and the inner-parts of the sub-filaments. %
 To distinguish between these hub structures observed at different scales, hereafter we refer to the  $\sim0.1$\,pc-scale I and I(N)  sources as {\it core-hubs} and the $\sim1$\,pc scale structure as a {\it clump-hub} (cf., Fig.\,\ref{SkelFig}). 

We use the \disperse\footnote{See \citet{Sousbie2011} and \citet{Sousbie2011b} for a description of the method and \citet{Arzoumanian2011} for the first application of this algorithm in the context of tracing the crests of interstellar filaments.} algorithm to trace { \rev the elongated dusty crests seen}
on the 850\,$\mu$m Stokes $I$ map at a resolution of 14\arcsec\ (projected onto a grid with 4\arcsec\ pixel size). 
{ \rev The crests traced by \disperse\ may correspond to "filament-crests" or to "shell-crests"  associated with shell-like structures around \hii\ regions (such as those around sources II and IV). We mention briefly possible differences between "filament-crests" and "shell-crests" below \citep[see also][who present a study of the properties of shell-crests and filament-crests in the RCW 120 \hii\ region]{Zavagno2020}.}

In the following, we  study the physical properties of the 14 crests identified towards the  NGC 6334 filament network  (Fig.\,\ref{SkelFig}). %
We discuss the observed values along the ridge  (crests 1 to 4)\footnote{{ \rev For the purpose of this study, which is aimed at describing the variation of the properties along the NGC 6334 elongated dusty structure, we do not include other crests traced by \disperse\ around source IV. }} %
and the variation of the properties at both ends of the 10 sub-filaments (crests 5 to 14) from their outer parts to their inner parts connected to the  {\it clump-hub}. 
{ \rev We  include in this analysis crests 10 and 11, which have one of their ends connected to  the  {\it clump-hub} or the ridge, even though these two crests may be better described as shell-like structures associated with the \hii\ region of source II. We discuss this further below.} 
Table\,\ref{tab:TparamFromPol} summarizes the mean properties and their respective dispersions for the 14 filament crests. Table\,\ref{tab:paramCritic}  provides estimated quantities that describe the stability of these filaments. 

\begin{figure}[!h]
   \centering
     \resizebox{9.5cm}{!}{ \hspace{-.6cm}
\includegraphics[angle=0]{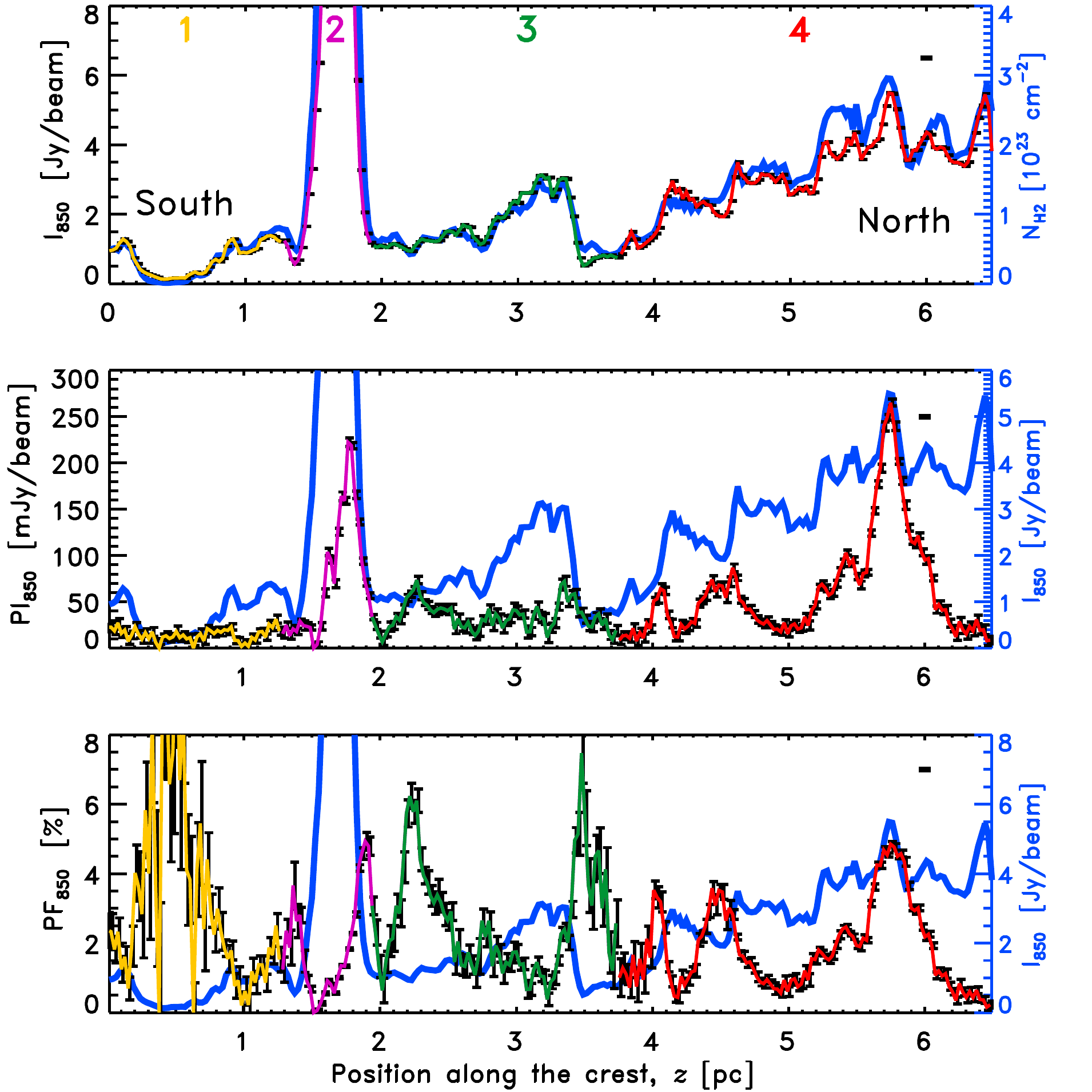}
}\vspace{-.1cm}
  \caption{%
  Observed $I$ (top), $PI$ (middle), and $PF$ (bottom) along the ridge crest  for the sections from 1 to 4 (indicated with different colours, see Fig.\,\ref{SkelFig}) from South to North (left to right on the horizontal axes).
   The short vertical black bars  show the $\pm\sigma$ statistical uncertainties. %
       The horizontal black segments on the top right hand  of the panels indicate the $14\arcsec$ (0.09\,pc) beam size of the data. 
   The blue curve on the top panel is the column density derived  from SPIRE+\Artemis\ data 
    after subtraction of a constant  "Galactic emission" value of  $3\times10^{22}\,\NHUNIT$ (cf. Sect.\,\ref{Herschel}). 
    The blue curves on the middle and bottom panels show  $I$ along the ridge crest (same as the coloured curve on the top panel).    
        }          
  \label{polCrest}
    \end{figure}

        \begin{figure}[!h]
   \centering
     \resizebox{9.cm}{!}{ \hspace{0.1cm}
\includegraphics[angle=0]{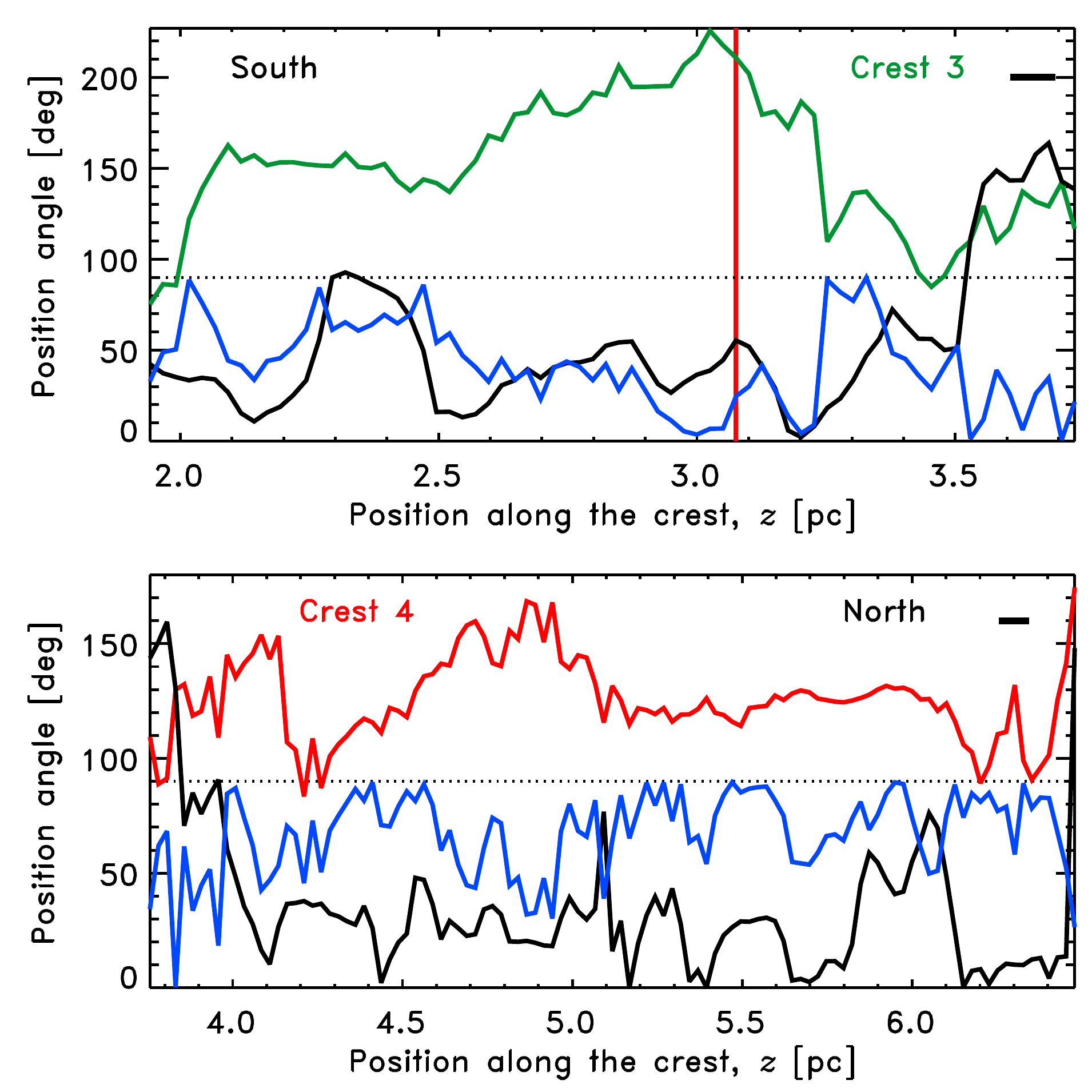}
}\vspace{-.1cm}
  \caption{%
  Variation  of the position angles of 
$\chi_{B_{\rm POS}}$ (green in the top and red in the bottom), $\theta_{\rm fil}$ (black), and  $\phi_{\rm diff}=|\theta_{\rm fil}-\chi_{B_{\rm pos}}|$ (blue) along  crests 3 (top) and 4 (bottom), from South to North. 
Only pixels along the crests are considered. 
The dotted horizontal lines show an angle of 90$^\circ$. The vertical red line in the top panel is the position of source III (cf., Fig.\,\ref{SkelFig}). The horizontal black segments on the top right hand  of the panels indicate the $14\arcsec$ (0.09\,pc) beam size of the data. 
       }          
  \label{AngleCrest35}
    \end{figure}

\subsection{Physical parameters along the ridge crest}\label{alongridge}

\subsubsection{Polarization properties}\label{alongridgePol}

Figure\,\ref{polCrest} shows the variation of $I$, $PI$, and $PF$ along the ridge crest from the South to the North.   
For comparison, we also plot the column density estimated from SPIRE+\Artemis\ data at 350\,$\mu$m  \citep{Andre2016} after subtracting the "Galactic emission" corresponding to the extended emission  filtered out in the BISTRO data (cf. Sect.\,\ref{Herschel}). 
We can see that both data trace essentially the same dust continuum emission.  

We notice two clear peaks in $PI$ at $z\sim1.8$\,pc and $z\sim5.8$\,pc. The former is associated to a peak in $I$ (with a maximum value of 12.6\,Jy/beam) and a trough in $PF$. This position corresponds to the active star-formation site IV (see Fig.\,\ref{SkelFig}).
The latter $PI$ peak, however, is not associated with a strong peak in $I$ %
nor with a decrease in $PF$ as usually expected  (c.f., Sect.\,\ref{scatter}). 
{\rev Indeed, the distribution of $PF$ along the ridge crest has a complex structure, with several relatively compact zones (roughly $0.2-0.4$\,pc) with moderately high values ($PF> 3\%$ and $\nhh\sim10^{23}\,\NHUNIT$), and these localised regions of $PF$ are neither correlated nor anti-correlated with $I$.}
{\rev Between $z\sim2.2$\,pc and $z\sim3.2$\,pc, $I$ increases by a factor $\sim3$, while $PI$ and $PF$ decrease by a factor of $\sim2$ and $\sim6$, respectively.
On the other hand, between
 $z\sim5$\,pc and $z\sim5.8$\,pc, there is an overall increase in the three quantities $I$, $PI$, and $PF$ by factors of $\sim2$, $\sim13$, and $\sim7$, respectively.}
{\rev We discuss physical possibilities for these variations in Sect.\,\ref{disc:feedback}.
}

On average, the position angle of the ridge is $\theta_{\rm fil}\sim30^\circ$ and the prevailing $B_{\rm pos}$ field orientation  {\rev (dominated by $\chi_{B_{\rm POS}}$ of the crest 4)} is close to perpendicular to the elongated ridge (cf., Table\,\ref{tab:TparamFromPol}), as  already suggested by $Planck$ observations (see Fig.\,\ref{BiPlanckMaps}). We  notice, however, some sections along the ridge crest where the %
 {\rev relative orientation between the B-field and the ridge changes.} 
These variations are shown in Fig.\,\ref{AngleCrest35} where we plot $\chi_{B_{\rm POS}}$, $\theta_{\rm fil}$ (the filament orientation), and  $\phi_{\rm diff}=|\theta_{\rm fil}-\chi_{B_{\rm pos}}|$ along  crests 3 and 4. The orientation of the crest, $\theta_{\rm fil}$, is  measured on the Stokes $I$ map using the Hessian matrix \citep[c.f., Eq. 3 in][]{Arzoumanian2019}.
{\rev  Along the $\sim3$\,pc crest 4,  $\phi_{\rm diff}$ is mostly uniform and close to $90^\circ$,  except for some localised regions (between  $z\sim3.8-4.0$\,pc  and $z\sim4.6-4.9$\,pc) where $\phi_{\rm diff}$ fluctuates between $30^\circ$ and $60^\circ$, i.e., neither perpendicular nor parallel orientation.} 
 {\rev  Along the $\sim2$\,pc  crest 3,  $\phi_{\rm diff}$ oscillates between mostly perpendicular to mostly random to mostly parallel } from South to North (see Sect.\,\ref{disc:hub} for a discussion).

\subsubsection{Magnetic field strength and stability parameters}\label{BCritCalcul}

Dust polarized thermal emission observations do not provide a direct measurement of the B-field strength, which is critical to study the stability and dynamical evolution of filaments and the role of the B-field in the star formation process. Indirect methods, such as the Davis-Chandrasekhar-Fermi (DCF) method \citep[][]{Davis1951,Chandrasekhar1953} calibrated on magnetohydrodynamic (MHD) simulations  \citep[e.g.,][]{Ostriker2001,Heitsch2001,Falceta-Goncalves2008}, can provide an estimate of the POS B-field strength. %

We here use the DCF method to estimate the $B_{\rm POS}$ field strength, combining estimates of the volume density $n_{\rm H_2}$(in cm$^{-3}$), the non-thermal  velocity dispersion  $\sigma_{\rm NT}$ (in km/s), and the POS B-field angle dispersion $\sigma_{\chi_{B_{\rm pos}}}$ (in degrees), using the following relation \citep[cf., e.g.,][]{Crutcher2004}:
\begin{equation}
B_{\rm POS}=9.3\,\sqrt{n_{\rm H_2}}\,\frac{\sigma_{\rm NT}\,\sqrt{8\ln2}}{\sigma_{\chi_{B_{\rm pos}}}}\label{Eq_Bstrength}
 \end{equation}
 
 \noindent
 in units of $\mu$G. 
We take  $n_{\rm H_2}=N_{\rm H_2}/W_{\rm fil}$  from the $N_{\rm H_2}$ values  (derived as explained in Sect.\,\ref{Herschel} for $T=20\pm5$\,K) and a filament width of $W_{\rm fil}=0.11\pm0.04$\,pc \citep{Andre2016}. 
The  $\sigma_{\chi_{B_{\rm POS}}}$ values correspond to the median absolute deviation, $mad$, of the $\chi_{B_{\rm pos}}$  measurements  observed at each pixel position along a given crest, i.e., 
$\sigma_{\chi_{B_{\rm POS}}}= $\,median$[|\chi_{B_{\rm POS}}-$\,median$(\chi_{B_{\rm pos}})|]$. These values are given in Table\,\ref{tab:TparamFromPol} (column 8). Here we compute the $mad$ value, which is a more robust measure of the dispersion than the standard deviation, {\rev which is more affected by outliers in non normal distributions.}  We also note that the $\sigma_{\chi_{B_{\rm POS}}}$ values calculated here are {\rev upper} limits since possible large-scale variations of the B-field structure (which have not been removed) may increase the measured dispersions. We refer to a future work for detailed modelling of the B-field structure along and across the filaments. This work will be done in  combination with the velocity structure derived from molecular line observations of similar resolution as the BISTRO data. 

There are no available spectroscopic observations tracing the velocity structure of the dense gas towards the 10\,pc-long NGC 6334 filamentary structure at comparable resolution of the BISTRO data\footnote{The N$_2$H$^+(1-0)$ spectra observed along the NGC 6334 ridge as part of the MALT90 survey \citep{Jackson2013}  with the MOPRA 22-m telescope at the angular resolution of 40\arcsec\  show $\sigma_{v}\gtrsim1$\,km\,s$^{-1}$.
 This latter velocity dispersion  is much larger than the values observed towards other dense filaments 
at better angular resolutions (see examples in the text)}. 
{\rev To estimate  $\sigma_{v}$,  we use the  $\sigma_{v}\sim\nhh^{0.35}$  relation suggested by \citet{Arzoumanian2013}
from the analysis of N$_2$H$^+(1-0)$ observations (with the IRAM-30m telescope) towards a sample of supercritical filaments with line masses between 20\,\sunpc\ and 500\,\sunpc,  corresponding to a similar line mass range for  the filaments studied in this paper. }
The derived $\sigma_{v}$\footnote{Since the $\sigma\sim\nhh^{0.35}$ relation in \citet{Arzoumanian2013} is calibrated for a gas temperature of $T=10$\,K, we rescale it here to the NGC 6334 temperature of $T=20$\,K.}
 values are given in column 5 of Table\,\ref{tab:paramCritic} {\rev  and range between  $\sim0.4 - 0.6$\,km\,s$^{-1}$ compatible with observational results towards other filaments of similar line masses. For example, observations 
towards the SDC13  filament system with the IRAM-30m at $\sim30\arcsec$  \citep[using N$_2$H$^+$,][]{Peretto2014}, towards dense filaments in a sample of Galactic giant molecular clouds with the GBT as part of the KEYSTONE project at $\sim30\arcsec$  \citep[using NH$_3$,][]{Keown2019}, {\rev  and towards the infrared dark cloud G$14.225-0.506$ with ALMA
\citep[using N$_2$H$^+$,][]{VivienChen2019}, }
{\rev  all suggest values of $\sigma_{v}\sim0.4 - 0.8$\,km\,s$^{-1}$. }
 In addition, the $\sigma_{v}=0.62$\,km\,s$^{-1}$ derived here for the crest 4 is similar to that derived from N$_2$H$^+$ ALMA data at $3\arcsec$  of the same region  \citep[][]{Shimajiri2019}. }
We calculate the non-thermal component $\sigma_{\rm NT}=\sqrt{\sigma_{v}^2-c_{\rm s}^2}$ for a temperature $T=20$\,K  for  NGC 6334.  
The  uncertainty on  $\sigma_{v}$   is inferred from the dispersion of $\nhh$ along each crest.  
The uncertainty on $\sigma_{\rm NT}$ is derived from the propagation of those on $\sigma_{v}$  and $c_{\rm s}$ (estimated for a temperature uncertainty of 5\,K). 
 
The values of $B_{\rm POS}$ along the ridge crest span a range from $\sim100\,\mu$G to $\sim800\,\mu$G with an average  value of $\sim270\pm100\,\mu$G. The uncertainty on $B_{\rm POS}$  is estimated from error propagation (see the uncertainties of the different parameters in Tables\,\ref{tab:TparamFromPol} and \,\ref{tab:paramCritic}).  

To estimate the support of the B-field against  gravitational collapse, we calculate the magnetic critical mass per unit length (assuming the 3D B-field structure to be close to the POS) using the relation  derived by \citet{Tomisaka2014}:
\begin{equation}
M_{\rm line,crit}^{B}=4.48\,(W_{\rm fil}/0.1\,{\rm pc})\,(B_{\rm POS}/10\,\mu{\rm G}).\label{Eq_MlinecritB}
\end{equation}
We compare this latter value with the mass per unit length ($M_{\rm line}$) estimating the magnetic virial parameter  $\alpha_{\rm vir}^{B}= M_{\rm line,crit}^B/M_{\rm line}$, where 
$M_{\rm line}= \mu_{\rm H_2}m_{\rm H}  \nhh W_{\rm fil}$, with $W_{\rm fil}=0.11$\,pc is the filament width ($\mu_{\rm H_2}$ and $m_{\rm H}$ are the same as in Sect.\,\ref{Herschel}).
Thus $\alpha_{\rm vir}^{B}\propto B_{\rm POS}/\nhh \propto (\sigma_{\rm NT}/\sigma_{\chi_{B_{\rm pos}}})/\sqrt{\nhh W_{\rm fil}}$.

We also derive the kinetic (thermal + turbulent) virial parameter $\alpha_{\rm vir}^{v}= M_{\rm line,vir}^v/M_{\rm line}$ where $M_{\rm line,vir}^v=2\sigma_{v}^2/G$ \citep[c.f., e.g.,][]{Fiege2000} and the 
effective virial parameter  that takes into account the thermal, turbulent, and magnetic support, $\alpha_{\rm vir}^{v,B}= M_{\rm line,vir}^{v,B}/M_{\rm line}$, where $M_{\rm line,vir}^{v,B}=M_{\rm line,vir}^{v}+M_{\rm line,crit}^{B}$ \citep[c.f.,][]{Tomisaka2014}.

The results (see Table\,\ref{tab:paramCritic}) indicate  that while the different sections of {\rev the ridge are highly thermally supercritical with %
 $\alpha_{\rm cyl}^T=M_{\rm line,crit}^T/M_{\rm line}\ll1$ }
 (with  $M_{\rm line, crit}^{T=20\,\rm K}\sim27\,$M$_\odot$/pc), magnetically supercritical with  $\alpha_{\rm vir}^{B}<1$ and gravitationally unstable with respect to the kinetic support $\alpha_{\rm vir}^{v}<1$, the combination of the magnetic and kinetic (thermal and turbulent) energies may provide effective support to the ridge as a whole against gravitational collapse  with $\alpha_{\rm vir}^{v,B}\sim1.1\pm0.6$. 
We notice, however, some differences: for crest 3, %
$\alpha_{\rm vir}^{B}\sim0.5\pm0.3$ suggesting that the gravitational energy dominates over  the magnetic energy, which may be compatible with the change of  ${\chi_{B_{\rm pos}}}$ along this section of the ridge (cf., Fig.\,\ref{AngleCrest35}), while for crest 4,  $\alpha_{\rm vir}^{B}=1.1\pm0.7$ suggesting  a more important internal magnetic support against gravitational collapse  (see discussion Sect.\,\ref{disc:hub}).

\begin{figure}[!h]
   \centering \hspace{-.2cm}
  \resizebox{9.cm}{!}{
\includegraphics[angle=0]{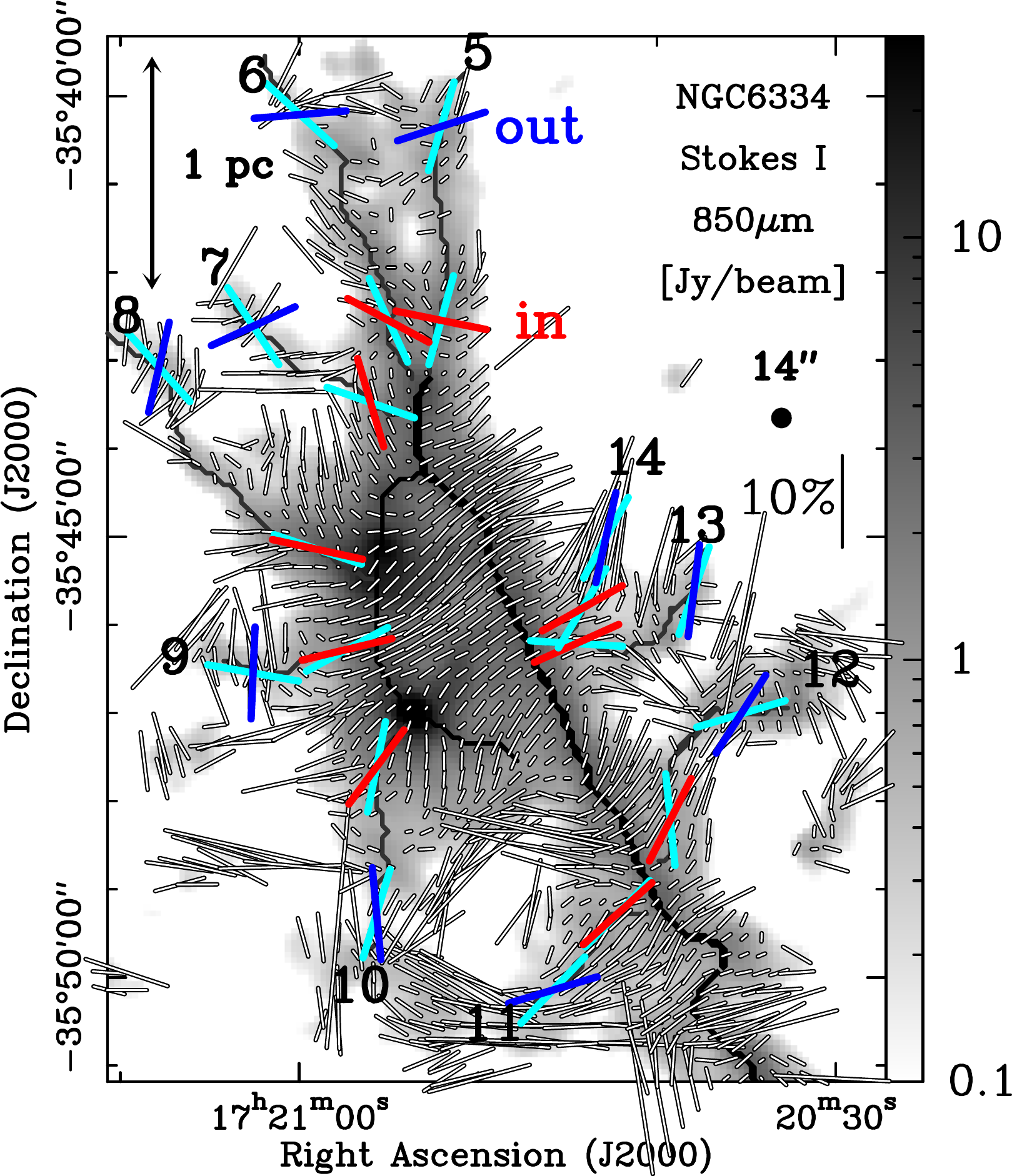}
}
  \caption{%
   Close-up of the northern region of NGC 6334 within the black dashed rectangle indicated in  Fig.\,\ref{SkelFig}. 
  The crests are those shown in Fig.\,\ref{SkelFig} and over-plotted on the Stokes $I$ map. 
  The white lines correspond to $\chi_{B_{\rm POS}}$ and  their lengths are proportional to $PF$.  
    On the right hand side,  a vertical line shows the  $10\%$ polarization fraction scale. The black filled circle corresponds to the $14\arcsec$ beam.   
 The blue and red lines indicate the mean values for $\chi_{B_{\rm POS}}$   for the $out$ and  $in$ parts of the sub-filaments, respectively. 
The cyan lines show the  orientation  ($\theta_{\rm  fil}$) of the $out$ and  $in$ parts of the sub-filaments (see Sect.\,\ref{alongsubfil}, Fig.\,\ref{InOut_histo}, and Table\,\ref{tab:TparamFromPol}). The length of the blue, red, and cyan lines is equal to 60\arcsec, corresponding to the extent of the circular area used to estimate the  parameters. 
}          
  \label{SkelFig2}
    \end{figure}

     \subsection{Physical parameters along the sub-filaments}\label{alongsubfil}

Figure\,\ref{SkelFig} shows  the crests tracing the network of the sub-filaments connected to the clump-hub encompassing  a section of the ridge (crest 4) and the two core-hubs I and I(N) that also lie along a filament traced with the cyan line in Fig.\,\ref{SkelFig}.

 We estimate the physical properties of the sub-filaments as explained in Sect.\,\ref{alongridge} for the ridge. 
To investigate possible variations of these properties along the sub-filaments from their outer-parts to their inner-parts connected to the clump-hub we give, in Tables\,\ref{tab:TparamFromPol} and\,\ref{tab:paramCritic},  two values {\rev of the  properties for each sub-filament towards  their inner and outer $\sim60\arcsec$  section of the crests, respectively. }
  The inner-parts ($in$) %
  {\rev correspond to} the $33\%$ (one-third) section   of the crests connected to the clump-hub. The outer-parts ($out$) are the $33\%$ sections of the other end of the crest away from the clump-hub (cf., Fig.\,\ref{SkelFig2}). In practice, the Stokes $I$  and the filament orientation $\theta_{\rm fil}$ values are measured along the inner and outer $\sim60\arcsec$  of the crests   ($\sim 4$ independent 14\arcsec\ beams). The $PI$, $PF$, and $\chi_{B_{\rm POS}}$ values are estimated in a circular area with a diameter of 60\arcsec\ towards the $in$ and $out$ parts, which yield between 10 and 14 independent measurements for a 14\arcsec\ beam (except for 6 $out$ and 14 $out$ where the values are measured towards 7  independent data points).

\begin{figure}[!h]
   \centering
  \resizebox{9.cm}{!}{ \hspace{-.6cm}
\includegraphics[angle=0]{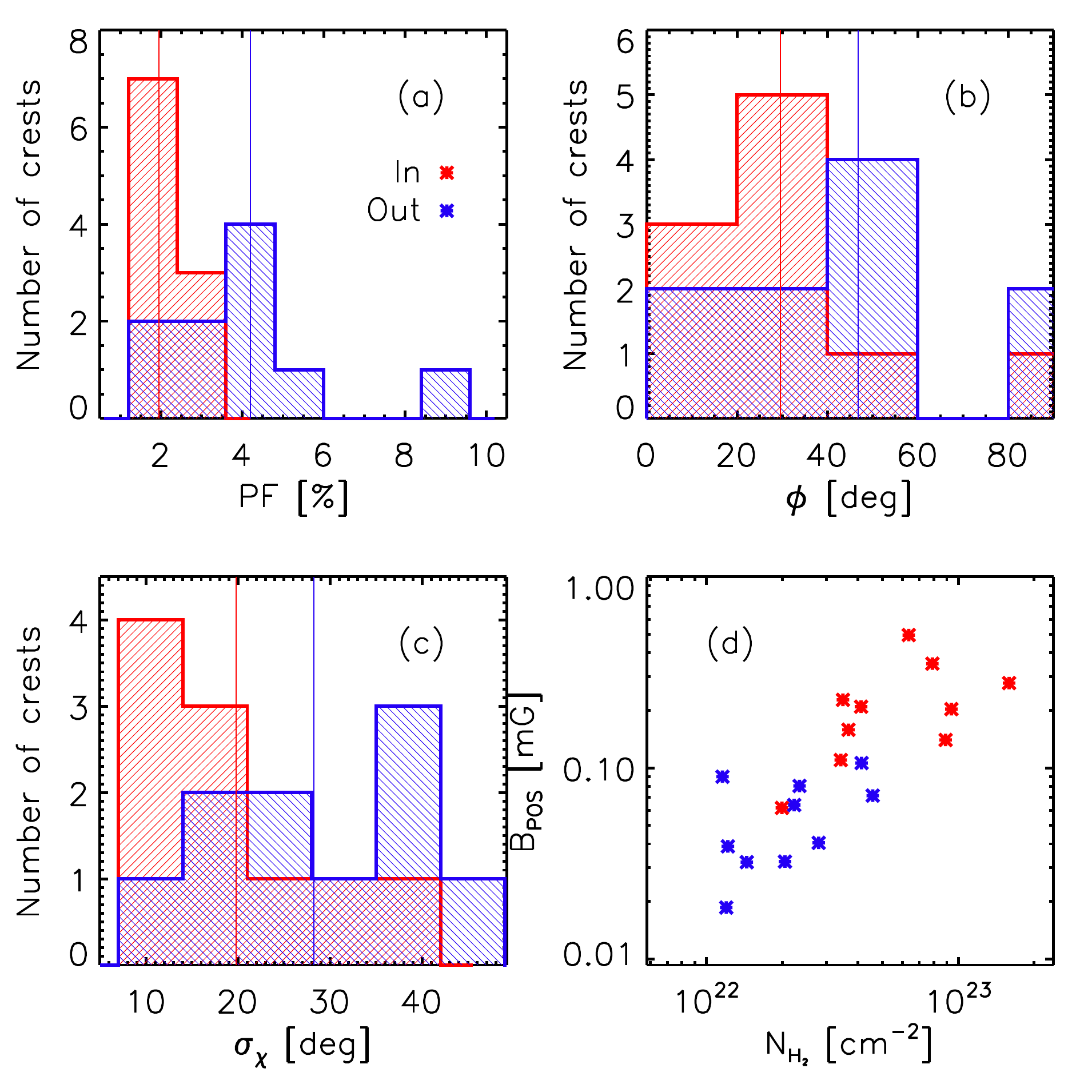}}
\vspace{-.5cm}
  \caption{%
  Distributions of properties measured towards the $in$ (red) and $out$ (blue) parts of the 10 sub-filaments: a) polarization fraction, b) $\phi_{\rm diff}=|\theta_{\rm fil}-\chi_{B_{\rm pos}}|$, difference between the filament orientation $\theta_{\rm fil}$ and the POS B-field angle $\chi_{B_{\rm pos}}$, c) dispersion of $\chi_{B_{\rm pos}}$, d) POS B-field strength as a function of the column density.  
  The red and blue vertical lines indicate the median of the distributions of the properties measured towards the $in$ and $out$  parts of the sub-filaments, respectively (see text for details and Tables\,\ref{tab:TparamFromPol} and\,\ref{tab:paramCritic}  for the values). 
       }          
  \label{InOut_histo}
    \end{figure}

\begin{figure}[!h]
   \centering
  \resizebox{9.cm}{!}{ \hspace{-.6cm}
\includegraphics[angle=0]{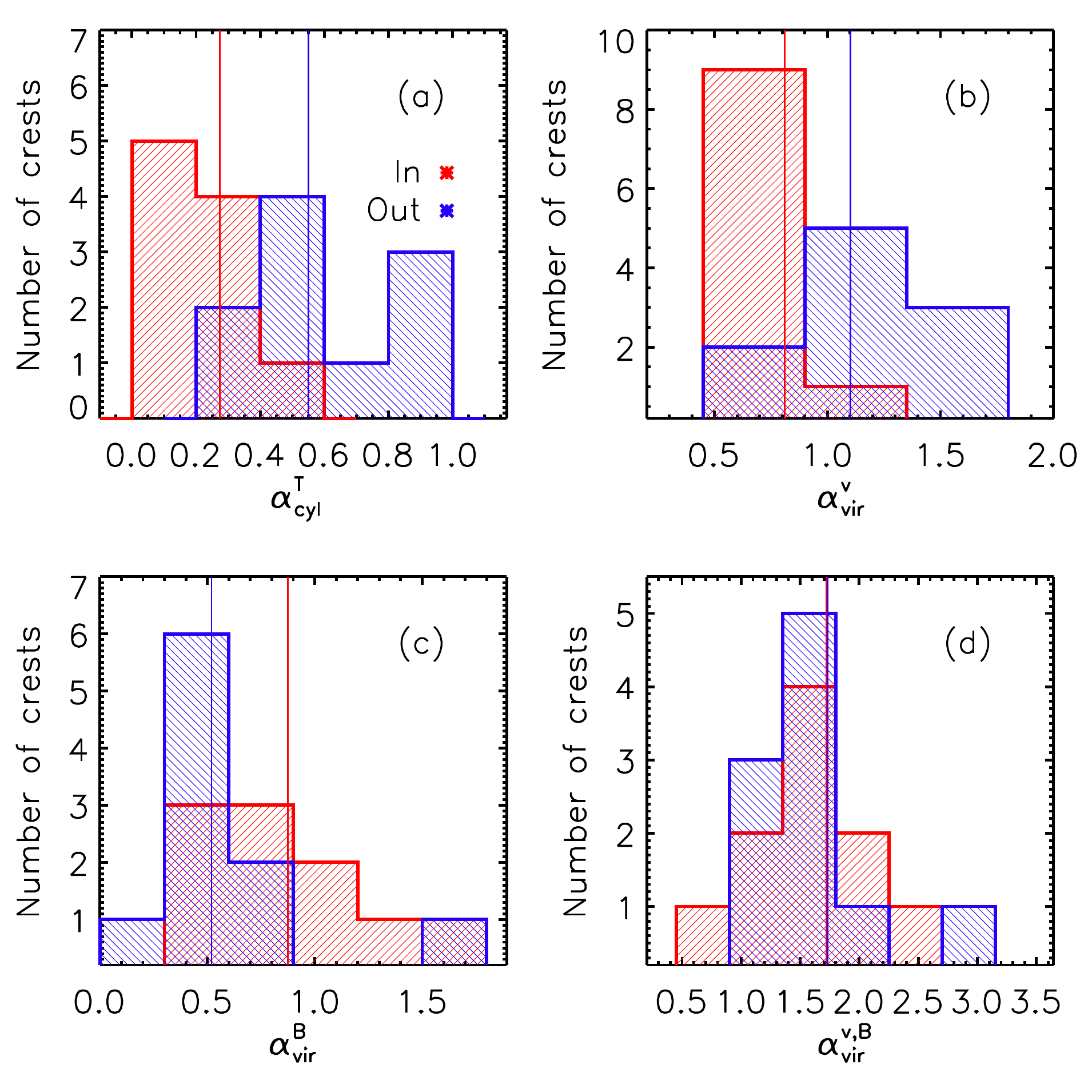}}
\vspace{-.8cm}
  \caption{%
 Similar to Fig.\,\ref{InOut_histo} for the parameters describing the stability of the $in$ (red) and $out$ (blue) sub-filament parts: a) Thermal critical stability parameter, b) kinetic (thermal + turbulent) virial  parameter, c) magnetic virial parameter, d) effective (thermal + turbulent + magnetic) virial parameter. 
       }          
  \label{InOut_histoCrit}
    \end{figure}

Figures\,\ref{InOut_histo} and\,\ref{InOut_histoCrit} show the distribution of the properties measured towards the $in$ and  $out$ parts of the sub-filaments. These histograms illustrate the variation  of the properties along the sub-filaments as they merge with the clump-hub. %
The total intensity $I$ and accordingly  \nhh\ and  $M_{\rm line}$ increase on average by  factors of $\sim2.8$ from the outer parts to the inner parts. %
 The increase of \nhh\ (calculated here for $T=20$\,K)  is larger than what would be expected from increases of temperature from  $\sim15$\,K to $\sim25$\,K from the outer to the inner parts. 
 The polarized intensity $PI$ increases on average by a factor of $\sim1.5$ and   the polarization fraction $PF$ decreases on average by a factor of $\sim2$ (corresponding to a decrease of $\sim2.2\%$ on average) %
  from the outer-parts  to the inner-parts. 
There is, however, a larger dispersion of the $PF$ distribution in the outer part, with $\sigma_{PF}^{out}\sim2\%$, while in the inner part $\sigma_{PF}^{in}\sim0.6\%$. 
The dispersion of the POS B-field angle is smaller by a factor $\sim1.4$ on average in the inner-parts compared to the outer-parts.  In the outer-parts, however, the distribution of $\sigma_{\chi_{B_{\rm POS}}}$  spans a wider range of values.
 In the inner-part of the sub-filaments, %
 the POS B-field seems to be mostly parallel to their crests, except for crest 5  
  (see Fig.\,\ref{SkelFig2}). 
  In the outer-parts, {\rev the histogram of $\phi_{\rm diff}^{out}$ (Fig.\,\ref{InOut_histo}b) does not show a clear trend for a preferred orientation.}
     {\rev Towards the crests 7 and 9   the POS B-field is mostly perpendicular to the sub-filaments with relative angles $\phi_{\rm diff}\sim80^\circ$. Towards the crests 5, 6,  and 8   the POS B-field has no preferred relative orientation, 
 with $\phi_{\rm diff}$ between $46^\circ$ and $55^\circ$.}  
 {\rev  Towards the crests 10  to 14   the POS B-field  is mostly parallel to the crests,   
 with  $\phi_{\rm diff}$ between $12^\circ$ and $40^\circ$.}  
 Crests 10 and 11 are observed towards the edge of an \hii\ region (i.e., Source II in Fig.\,\ref{SkelFig}) and may be affected by the interaction with the expanding bubble (see discussion in Sect.\,\ref{disc:feedback}). Crest 14 is very short and the outer and inner parts, as we defined them here, are not really distinguishable. 
These variations of the relative orientation between the filament and the B-field orientations may result from the coupled   dynamical evolution of the filaments and the B-fields. However,  they  may also be due to projection effects: A configuration that is perpendicular in projection, i.e., derived from the observations, has a high probability to be perpendicular in 3D, while a configuration that is parallel in projection may correspond to a perpendicular configuration in 3D 
\citep[see Fig.\,C.5. in][and \citealt{Doi2020}]{planck2016-XXXV}.
 We discuss  possible origins of the variation of the B-field orientation  along the sub-filaments in Sect.\,\ref{disc:hub}.

 {\rev The sub-filaments are thermally supercritical with $\alpha_{\rm cyl}^T<1$, and show an decrease of $\alpha_{\rm cyl}^T$ (by a factor of $\sim2$ on average) from $out$ to $in$.} The kinetic virial parameter $\alpha_{\rm vir}^v$ is on average smaller in the inner parts compared to the outer parts (by a factor of 1.4 on average), while the magnetic virial parameter 
 $\alpha_{\rm vir}^B$  increases (by a factor of $\sim1.7$ on average) from $out$ to $in$. In the outer parts,  $\alpha_{\rm vir}^B\sim0.5\pm0.3<1$ suggesting that  magnetic tension alone is not enough to balance gravity while the inner parts of the sub-filaments seem to be in a magnetic critical balance with $\alpha_{\rm vir}^B\sim1$. 
When both the magnetic and kinetic (thermal and turbulent) supports are combined, the estimated $\alpha_{\rm vir}^{v,B}\gtrsim1$ values suggest that the inner-parts and the outer-parts are in an overall balance between the effective pressure forces and gravity {\rev (see Sect.\,\ref{disc:hub} for a discussion on these results).} %

 \begin{table*}[!h] 
    \hspace{-1.cm}  
\centering
 \caption{Polarization properties derived from the BISTRO maps along 14 crests identified towards the NGC 6334 field. The crests 1 to 4 trace the ridge. The crests 5 to 14 correspond to the sub-filaments, for which $in$ and $out$ values are given (see Figs.\,\ref{SkelFig} and \ref{SkelFig2}). }  
\begin{tabular}{|c|c|c|cc|cc|cc|c|c|}   
\hline\hline   
Crest & $l_{\rm fil}$ &   $I_{\rm fil}\pm\sigma_{I_{\rm fil}}$   & $PI_{\rm fil}\pm\sigma_{PI_{\rm fil}}$ & $\delta PI_{\rm fil}$ & $PF_{\rm fil}\pm\sigma_{PF_{\rm fil}}$&$\delta PF_{\rm fil}$ & $\chi_{B_{\rm pos}}\pm\sigma_{\chi_{B_{\rm pos}}}$ & $\delta \psi_{\rm fil}$ & $\theta_{\rm fil}\pm\sigma_{\theta_{\rm fil}}$ &  $\phi_{\rm diff}\pm\sigma_{\phi_{\rm diff}}$ \\  %
$\#$& [pc] & [Jy/beam] &\multicolumn{2}{c|}{[mJy/beam]}& \multicolumn{2}{c|}{[$\%$]}& \multicolumn{2}{c|}{[deg]}& [deg]& [deg]\\
(1)&(2)& (3) & (4) &(5) &(6)&(7)&(8)&(9)&(10)&(11)\\
\hline
1&1.1&0.7$\,\pm\,$0.5&18.0$\,\pm\,$3.2&3.9&2.34$\,\pm\,$0.96&0.50&92$\,\pm\,$21&6&8$\,\pm\,$17&83$\,\pm\,$27\\
2&0.7&3.3$\,\pm\,$2.6&69.6$\,\pm\,$47.3&3.9&1.83$\,\pm\,$0.98&0.09&72$\,\pm\,$14&1&55$\,\pm\,$14&17$\,\pm\,$20\\
3&2.0&1.3$\,\pm\,$0.4&33.7$\,\pm\,$9.1&4.5&2.31$\,\pm\,$0.91&0.30&153$\,\pm\,$26&3&34$\,\pm\,$16&61$\,\pm\,$31\\
4&3.0&3.2$\,\pm\,$0.7&50.9$\,\pm\,$25.8&3.9&1.66$\,\pm\,$0.78&0.12&125$\,\pm\,$10&2&26$\,\pm\,$12&80$\,\pm\,$16\\
\hline\hline
1 to 4 &6.9&2.3$\,\pm\,$1.1&33.1$\,\pm\,$14.5&4.0&2.01$\,\pm\,$0.92&0.20&121$\,\pm\,$22&3&28$\,\pm\,$16&87$\,\pm\,$28\\
                            \hline
                               \hline      
5\,$in$ &\multirow{2}{*}{1.6}&1.9$\,\pm\,$0.2&16.3$\,\pm\,$7.0&5.1&1.40$\,\pm\,$0.49&0.38&78$\,\pm\,$37&8&164$\,\pm\,$8&85$\,\pm\,$38\\
5\,$out$ &&1.0$\,\pm\,$0.1&11.4$\,\pm\,$4.3&6.3&1.96$\,\pm\,$0.99&1.18&108$\,\pm\,$40&7&163$\,\pm\,$11&55$\,\pm\,$41\\
\hline
6\,$in$ &\multirow{2}{*}{1.7}&2.0$\,\pm\,$0.2&15.1$\,\pm\,$6.1&5.2&1.32$\,\pm\,$0.54&0.38&61$\,\pm\,$27&10&24$\,\pm\,$6&37$\,\pm\,$27\\
6\,$out$ &&0.4$\,\pm\,$0.2&14.2$\,\pm\,$4.4&7.5&4.06$\,\pm\,$2.30&2.17&94$\,\pm\,$40&8&47$\,\pm\,$13&46$\,\pm\,$42\\
\hline
7\,$in$ &\multirow{2}{*}{1.0}&0.8$\,\pm\,$0.6&18.3$\,\pm\,$8.3&4.6&1.96$\,\pm\,$0.68&0.42&16$\,\pm\,$9&7&70$\,\pm\,$11&54$\,\pm\,$14\\
7\,$out$ &&0.3$\,\pm\,$0.1&11.3$\,\pm\,$5.2&5.4&4.47$\,\pm\,$2.29&2.31&114$\,\pm\,$28&6&33$\,\pm\,$20&81$\,\pm\,$34\\
\hline
8\,$in$ &\multirow{2}{*}{1.8}&3.4$\,\pm\,$1.5&28.6$\,\pm\,$9.7&4.4&1.20$\,\pm\,$0.49&0.19&77$\,\pm\,$31&4&71$\,\pm\,$9&6$\,\pm\,$33\\
8\,$out$ &&0.3$\,\pm\,$0.1&15.3$\,\pm\,$5.5&6.2&5.22$\,\pm\,$2.28&2.41&167$\,\pm\,$19&4&42$\,\pm\,$23&54$\,\pm\,$30\\
\hline
9\,$in$ &\multirow{2}{*}{0.8}&1.4$\,\pm\,$0.6&49.3$\,\pm\,$37.3&3.8&2.58$\,\pm\,$0.43&0.20&103$\,\pm\,$7&2&118$\,\pm\,$15&14$\,\pm\,$16\\
9\,$out$ &&0.5$\,\pm\,$0.1&11.4$\,\pm\,$3.9&4.2&3.51$\,\pm\,$1.77&1.36&177$\,\pm\,$22&1&79$\,\pm\,$20&82$\,\pm\,$30\\
\hline
10\,$in$ &\multirow{2}{*}{1.2}&1.7$\,\pm\,$0.8&19.2$\,\pm\,$8.5&4.2&1.64$\,\pm\,$0.37&0.28&143$\,\pm\,$13&6&168$\,\pm\,$21&25$\,\pm\,$25\\
10\,$out$ &&0.3$\,\pm\,$0.1&9.5$\,\pm\,$3.3&4.7&4.60$\,\pm\,$2.48&2.56&5$\,\pm\,$39&5&162$\,\pm\,$18&22$\,\pm\,$43\\
\hline
11\,$in$ &\multirow{2}{*}{0.8}&0.7$\,\pm\,$0.1&18.1$\,\pm\,$5.4&4.3&2.35$\,\pm\,$0.87&0.59&132$\,\pm\,$19&6&135$\,\pm\,$17&3$\,\pm\,$26\\
11\,$out$ &&0.5$\,\pm\,$0.1&14.5$\,\pm\,$5.4&4.5&4.19$\,\pm\,$1.40&1.32&106$\,\pm\,$18&6&136$\,\pm\,$6&29$\,\pm\,$19\\
\hline
12\,$in$ &\multirow{2}{*}{1.4}&0.9$\,\pm\,$0.1&18.6$\,\pm\,$10.1&4.4&2.47$\,\pm\,$0.88&0.57&153$\,\pm\,$12&6&5$\,\pm\,$30&32$\,\pm\,$33\\
12\,$out$ &&0.9$\,\pm\,$0.2&12.1$\,\pm\,$5.0&4.8&1.96$\,\pm\,$0.80&0.70&147$\,\pm\,$24&6&106$\,\pm\,$28&40$\,\pm\,$37\\
\hline
13\,$in$ &\multirow{2}{*}{1.1}&0.8$\,\pm\,$0.2&17.6$\,\pm\,$5.6&3.8&1.64$\,\pm\,$0.60&0.47&114$\,\pm\,$14&6&86$\,\pm\,$20&27$\,\pm\,$24\\
13\,$out$ &&0.6$\,\pm\,$0.1&9.2$\,\pm\,$3.3&4.3&2.43$\,\pm\,$1.41&1.11&172$\,\pm\,$44&5&160$\,\pm\,$14&12$\,\pm\,$46\\
\hline
14\,$in$ &\multirow{2}{*}{0.6}&0.4$\,\pm\,$0.1&15.3$\,\pm\,$4.9&3.7&2.58$\,\pm\,$0.98&0.73&119$\,\pm\,$20&7&148$\,\pm\,$9&29$\,\pm\,$22\\
14\,$out$ &&0.3$\,\pm\,$0.1&14.0$\,\pm\,$3.5&3.8&8.50$\,\pm\,$3.24&2.51&167$\,\pm\,$7&6&151$\,\pm\,$6&16$\,\pm\,$9\\
\hline
    \hline
       median\,$in$ &\multirow{2}{*}{1.2}&1.4$\,\pm\,$0.2&18.3$\,\pm\,$8.3&4.4&1.96$\,\pm\,$0.60&0.42&114$\,\pm\,$19&6&118$\,\pm\,$15&29$\,\pm\,$26\\
median\,$out$ &&0.5$\,\pm\,$0.1&12.1$\,\pm\,$4.4&4.8&4.19$\,\pm\,$2.28&2.17&147$\,\pm\,$28&6&136$\,\pm\,$18&46$\,\pm\,$37\\
   \hline  \hline                        
                  \end{tabular}
\begin{list}{}{}
 \item[]{{\bf Notes:} The values given in this table are derived from BISTRO observations at a spatial resolution of $14\arcsec$.  The row "1 to 4" corresponds to the full length of the ridge combining the crests from 1 to 4. The last two rows "median\,$in$" and "median\,$out$" give the median values measured towards the sub-filaments "$in$" and "$out$" parts, respectively (see Fig.\,\ref{SkelFig2}). Columns 3, 4, 6, 8, 10, and 11 give the median and their median absolute deviation per crests. These values are calculated along the ridge crests (for crests 1 to 4) and towards one-third of the "$in$" and "$out$" parts  of the sub-filaments (c.f., Sect.\,\ref{alongsubfil}). 
 Columns 5, 7, and 9 give the statistical uncertainties. 
 Columns: (1) Crest number. (2) Crest length. (3) Stokes $I$. (4) Polarized intensity. (6) Polarization fraction. (8) POS B-field orientation, $\chi_{B_{\rm pos}}$.  To avoid artefacts introduced by the discontinuity in the distribution of angles around $0^\circ$ and $180^\circ$, which is the B-field orientation ambiguity inherent in polarimetric observations, we estimate two values of $\sigma_{\chi_{B_{\rm pos}}}$, one in the range  $[0^\circ,180^\circ]$
 and another one in the range $[-90^\circ,90^\circ]$. We then select the smallest value of $\sigma_{\chi_{B_{\rm pos}}}$ and the corresponding median $\chi_{B_{\rm pos}}$ value (mapped back to the $[0^\circ,180^\circ]$ range when needed). 
 (10) Orientation of the filament crest, $\theta_{\rm fil}$, measured on the Stokes $I$ map. %
  (11) Difference between the orientation of the POS B-field angle and the filament crest $\phi_{\rm diff}=|\theta_{\rm fil}-\chi_{B_{\rm pos}}|$. 
   }
 \end{list}      
 \label{tab:TparamFromPol}    
  \end{table*}

\begin{sidewaystable*}
\centering
 \caption{%
 Properties along the 14 crests identified towards the NGC 6334 field. The crests 1 to 4 trace the ridge. The crests 5 to 14 correspond to the sub-filaments, for which $in$ and $out$ values are given (see Figs.\,\ref{SkelFig2}, \,\ref{InOut_histo},  and \ref{InOut_histoCrit}). See Sect.\,\ref{BCritCalcul} for the calculation of the parameters and Sect.\,\ref{BCritCalcul} for the the definition  of the  $in$ and $out$ parts of the sub-filaments.
 } 
  \label{tab:paramCritic}   
\begin{tabular}{|c||ccc|cc|cc|c||cccc|}   
\hline\hline   
Crest &   $N_{\rm H_2}$& $ n_{\rm H_2}$ & $M_{\rm line}$ & $\sigma_{v}$ & $M_{\rm line,vir}^{v}$    & $B_{\rm pos}$ & $M_{\rm line,crit}^{B}$ & $M_{\rm line,vir}^{v,B}$ & $\alpha_{\rm cyl}^{T}$  &  $\alpha_{\rm vir}^{v}$&  $\alpha_{\rm vir}^B$&$\alpha_{\rm vir}^{v,B}$ \\ 
$\#$ & [$10^{23}\NHUNIT$] & [$10^{5}$cm$^{-3}$] &[\sunpc]&[km/s]&[\sunpc]& [$\mu$G] &[\sunpc]&[\sunpc] &&&&\\
  (1)&(2)& (3) & (4) &(5) &(6)&(7)&(8)&(9)&(10)&(11)&(12) &(13)\\
      \hline  \hline
   1&0.35$\,\pm\,$0.21&1.0$\,\pm\,$0.7&84$\,\pm\,$60&0.40$\,\pm\,$0.07&73$\,\pm\,$26&101$\,\pm\,$54&49$\,\pm\,$32&123$\,\pm\,$41&0.33$\,\pm\,$0.25&0.9$\,\pm\,$0.7&0.6$\,\pm\,$0.6&1.5$\,\pm\,$0.9\\
2&1.53$\,\pm\,$1.21&4.5$\,\pm\,$3.9&372$\,\pm\,$325&0.63$\,\pm\,$0.16&183$\,\pm\,$95&585$\,\pm\,$320&288$\,\pm\,$189&471$\,\pm\,$212&0.07$\,\pm\,$0.07&0.5$\,\pm\,$0.5&0.8$\,\pm\,$0.8&1.3$\,\pm\,$1.0\\
3&0.62$\,\pm\,$0.18&1.8$\,\pm\,$0.9&151$\,\pm\,$70&0.47$\,\pm\,$0.05&103$\,\pm\,$19&144$\,\pm\,$43&71$\,\pm\,$33&175$\,\pm\,$39&0.18$\,\pm\,$0.10&0.7$\,\pm\,$0.3&0.5$\,\pm\,$0.3&1.2$\,\pm\,$0.5\\
4&1.46$\,\pm\,$0.33&4.3$\,\pm\,$1.8&356$\,\pm\,$152&0.62$\,\pm\,$0.05&178$\,\pm\,$27&818$\,\pm\,$275&403$\,\pm\,$199&581$\,\pm\,$201&0.08$\,\pm\,$0.04&0.5$\,\pm\,$0.2&1.1$\,\pm\,$0.7&1.6$\,\pm\,$0.8\\      
         \hline    \hline
   1 to 4 &1.07$\,\pm\,$0.52&3.1$\,\pm\,$1.9&259$\,\pm\,$158&0.56$\,\pm\,$0.09&145$\,\pm\,$45&270$\,\pm\,$107&133$\,\pm\,$71&278$\,\pm\,$85&0.11$\,\pm\,$0.07&0.6$\,\pm\,$0.4&0.5$\,\pm\,$0.4&1.1$\,\pm\,$0.6\\               
                 \hline    \hline 
           5\,$in$&0.89$\,\pm\,$0.10&2.6$\,\pm\,$1.0&216$\,\pm\,$82&0.53$\,\pm\,$0.02&129$\,\pm\,$11&139$\,\pm\,$42&68$\,\pm\,$32&198$\,\pm\,$34&0.13$\,\pm\,$0.06&0.6$\,\pm\,$0.2&0.3$\,\pm\,$0.2&0.9$\,\pm\,$0.3\\
5\,$out$&0.46$\,\pm\,$0.03&1.3$\,\pm\,$0.5&111$\,\pm\,$41&0.43$\,\pm\,$0.02&86$\,\pm\,$7&71$\,\pm\,$19&35$\,\pm\,$16&121$\,\pm\,$17&0.25$\,\pm\,$0.11&0.8$\,\pm\,$0.3&0.3$\,\pm\,$0.2&1.1$\,\pm\,$0.4\\
\hline
6\,$in$&0.94$\,\pm\,$0.10&2.8$\,\pm\,$1.0&228$\,\pm\,$86&0.54$\,\pm\,$0.02&134$\,\pm\,$11&203$\,\pm\,$84&100$\,\pm\,$55&234$\,\pm\,$56&0.12$\,\pm\,$0.05&0.6$\,\pm\,$0.2&0.4$\,\pm\,$0.3&1.0$\,\pm\,$0.4\\
6\,$out$&0.21$\,\pm\,$0.11&0.6$\,\pm\,$0.4&49$\,\pm\,$32&0.34$\,\pm\,$0.05&54$\,\pm\,$16&32$\,\pm\,$15&15$\,\pm\,$9&70$\,\pm\,$19&0.55$\,\pm\,$0.38&1.1$\,\pm\,$0.8&0.3$\,\pm\,$0.3&1.4$\,\pm\,$0.8\\
\hline
7\,$in$&0.35$\,\pm\,$0.28&1.0$\,\pm\,$0.9&84$\,\pm\,$74&0.40$\,\pm\,$0.09&73$\,\pm\,$34&227$\,\pm\,$212&111$\,\pm\,$112&185$\,\pm\,$117&0.32$\,\pm\,$0.30&0.9$\,\pm\,$0.9&1.3$\,\pm\,$1.8&2.2$\,\pm\,$2.0\\
7\,$out$&0.14$\,\pm\,$0.03&0.4$\,\pm\,$0.2&35$\,\pm\,$14&0.32$\,\pm\,$0.03&46$\,\pm\,$8&32$\,\pm\,$10&15$\,\pm\,$7&61$\,\pm\,$11&0.78$\,\pm\,$0.38&1.3$\,\pm\,$0.6&0.4$\,\pm\,$0.3&1.8$\,\pm\,$0.7\\
\hline
8\,$in$&1.59$\,\pm\,$0.71&4.7$\,\pm\,$2.7&387$\,\pm\,$223&0.64$\,\pm\,$0.09&188$\,\pm\,$55&277$\,\pm\,$99&136$\,\pm\,$70&325$\,\pm\,$89&0.07$\,\pm\,$0.04&0.5$\,\pm\,$0.3&0.4$\,\pm\,$0.3&0.8$\,\pm\,$0.4\\
8\,$out$&0.12$\,\pm\,$0.03&0.4$\,\pm\,$0.2&29$\,\pm\,$13&0.30$\,\pm\,$0.03&42$\,\pm\,$8&38$\,\pm\,$13&19$\,\pm\,$9&61$\,\pm\,$12&0.93$\,\pm\,$0.47&1.4$\,\pm\,$0.7&0.6$\,\pm\,$0.4&2.1$\,\pm\,$0.8\\
\hline
9\,$in$&0.64$\,\pm\,$0.27&1.9$\,\pm\,$1.0&154$\,\pm\,$86&0.48$\,\pm\,$0.06&105$\,\pm\,$28&495$\,\pm\,$228&244$\,\pm\,$143&349$\,\pm\,$146&0.18$\,\pm\,$0.11&0.7$\,\pm\,$0.4&1.6$\,\pm\,$1.3&2.3$\,\pm\,$1.3\\
9\,$out$&0.22$\,\pm\,$0.03&0.7$\,\pm\,$0.3&54$\,\pm\,$20&0.35$\,\pm\,$0.02&57$\,\pm\,$7&63$\,\pm\,$13&31$\,\pm\,$13&89$\,\pm\,$15&0.51$\,\pm\,$0.23&1.1$\,\pm\,$0.4&0.6$\,\pm\,$0.3&1.6$\,\pm\,$0.5\\
\hline
10\,$in$&0.79$\,\pm\,$0.39&2.3$\,\pm\,$1.4&192$\,\pm\,$117&0.51$\,\pm\,$0.08&120$\,\pm\,$37&351$\,\pm\,$205&172$\,\pm\,$119&293$\,\pm\,$125&0.14$\,\pm\,$0.09&0.6$\,\pm\,$0.4&0.9$\,\pm\,$0.8&1.5$\,\pm\,$0.9\\
10\,$out$&0.12$\,\pm\,$0.04&0.4$\,\pm\,$0.2&29$\,\pm\,$14&0.30$\,\pm\,$0.03&42$\,\pm\,$9&18$\,\pm\,$6&9$\,\pm\,$4&51$\,\pm\,$10&0.94$\,\pm\,$0.51&1.4$\,\pm\,$0.8&0.3$\,\pm\,$0.2&1.8$\,\pm\,$0.8\\
\hline
11\,$in$&0.34$\,\pm\,$0.05&1.0$\,\pm\,$0.4&83$\,\pm\,$32&0.40$\,\pm\,$0.02&72$\,\pm\,$8&109$\,\pm\,$42&54$\,\pm\,$28&126$\,\pm\,$30&0.33$\,\pm\,$0.15&0.9$\,\pm\,$0.4&0.7$\,\pm\,$0.4&1.5$\,\pm\,$0.6\\
11\,$out$&0.23$\,\pm\,$0.06&0.7$\,\pm\,$0.3&57$\,\pm\,$25&0.36$\,\pm\,$0.03&59$\,\pm\,$11&80$\,\pm\,$35&39$\,\pm\,$22&98$\,\pm\,$25&0.48$\,\pm\,$0.25&1.0$\,\pm\,$0.5&0.7$\,\pm\,$0.5&1.7$\,\pm\,$0.7\\
\hline
12\,$in$&0.41$\,\pm\,$0.06&1.2$\,\pm\,$0.5&100$\,\pm\,$39&0.42$\,\pm\,$0.03&81$\,\pm\,$9&208$\,\pm\,$120&102$\,\pm\,$70&184$\,\pm\,$70&0.27$\,\pm\,$0.13&0.8$\,\pm\,$0.3&1.0$\,\pm\,$0.8&1.8$\,\pm\,$0.9\\
12\,$out$&0.41$\,\pm\,$0.10&1.2$\,\pm\,$0.5&100$\,\pm\,$44&0.42$\,\pm\,$0.04&81$\,\pm\,$13&105$\,\pm\,$38&52$\,\pm\,$26&133$\,\pm\,$30&0.27$\,\pm\,$0.14&0.8$\,\pm\,$0.4&0.5$\,\pm\,$0.4&1.3$\,\pm\,$0.5\\
\hline
13\,$in$&0.37$\,\pm\,$0.11&1.1$\,\pm\,$0.5&89$\,\pm\,$42&0.40$\,\pm\,$0.04&75$\,\pm\,$15&158$\,\pm\,$82&77$\,\pm\,$49&153$\,\pm\,$51&0.31$\,\pm\,$0.17&0.8$\,\pm\,$0.4&0.9$\,\pm\,$0.7&1.7$\,\pm\,$0.8\\
13\,$out$&0.28$\,\pm\,$0.05&0.8$\,\pm\,$0.3&67$\,\pm\,$27&0.37$\,\pm\,$0.03&65$\,\pm\,$9&40$\,\pm\,$10&19$\,\pm\,$8&84$\,\pm\,$12&0.40$\,\pm\,$0.19&1.0$\,\pm\,$0.4&0.3$\,\pm\,$0.2&1.2$\,\pm\,$0.4\\
\hline
14\,$in$&0.20$\,\pm\,$0.05&0.6$\,\pm\,$0.3&48$\,\pm\,$21&0.34$\,\pm\,$0.03&54$\,\pm\,$9&61$\,\pm\,$27&30$\,\pm\,$17&84$\,\pm\,$19&0.57$\,\pm\,$0.28&1.1$\,\pm\,$0.5&0.6$\,\pm\,$0.4&1.7$\,\pm\,$0.7\\
14\,$out$&0.12$\,\pm\,$0.04&0.3$\,\pm\,$0.2&28$\,\pm\,$13&0.30$\,\pm\,$0.03&41$\,\pm\,$9&89$\,\pm\,$80&44$\,\pm\,$42&85$\,\pm\,$43&0.97$\,\pm\,$0.53&1.5$\,\pm\,$0.8&1.6$\,\pm\,$1.7&3.0$\,\pm\,$1.9\\
\hline\hline
median\,$in$&0.64$\,\pm\,$0.11&1.9$\,\pm\,$1.0&154$\,\pm\,$82&0.48$\,\pm\,$0.04&105$\,\pm\,$15&208$\,\pm\,$99&102$\,\pm\,$70&198$\,\pm\,$70&0.27$\,\pm\,$0.13&0.8$\,\pm\,$0.4&0.9$\,\pm\,$0.7&1.7$\,\pm\,$0.8\\
median\,$out$&0.22$\,\pm\,$0.04&0.7$\,\pm\,$0.3&54$\,\pm\,$25&0.35$\,\pm\,$0.03&57$\,\pm\,$9&63$\,\pm\,$15&31$\,\pm\,$13&85$\,\pm\,$17&0.55$\,\pm\,$0.38&1.1$\,\pm\,$0.6&0.5$\,\pm\,$0.3&1.7$\,\pm\,$0.7\\
\hline\hline
                  \end{tabular}
\begin{list}{}{} 
 \item[]{{\bf Notes:} 
The values given in this table are derived from BISTRO observations at a spatial resolution of $14\arcsec$. 
These values are calculated along the ridge crests (crests 1 to 4) and towards one-third of the "$in$" and "$out$" parts  of the sub-filaments (c.f., Sect.\,\ref{PropCrests}). 
The row "1 to 4" corresponds to the full length of the ridge combining the crests 1 to 4, {\rev hence the crest 4, which is the longest, contributes the most to these values. } The last two rows "median\,$in$" and "median\,$out$" give the median values measured towards the sub-filaments "$in$" and "$out$" parts, respectively (see Fig.\,\ref{SkelFig2}). 
  Columns: (1) Crest number. (2) Column density  calculated from the total intensity  Stokes $I_{850}$ for 
   $T=20\,$K (see Sect.\,\ref{Herschel}). %
(3) Volume density  estimated from the $N_{\rm H_2}$ values for a filament width of $W_{\rm fil}=0.11\pm0.04$\,pc \citep[c.f.,][]{Andre2016}. 
  (4) Mass per unit length calculated using the relation: $M_{\rm line}=\Sigma \times W_{\rm fil}$, where $\Sigma= \mu_{\rm H_2}m_{\rm H}  \nhh$ is the gas surface density, $W_{\rm fil}=0.11$\,pc,  
$\mu_{\rm H_2}=2.8$  the mean molecular weight per hydrogen molecule, and $m_{\rm H}$ the mass of a hydrogen atom.
 (5) Total velocity dispersion estimated  using the relation  $\sigma_{v}\sim\nhh^{0.35}$ suggested by \citet{Arzoumanian2013} rescaled to $T=20\,$K. 
  (6) Virial mass per unit length  $M_{\rm line,vir}^v=2\sigma_{v}^2/G$, where $\sigma_{v}$   is the total (turbulent and thermal) velocity dispersion (Col 5) and $G$ is the gravitational constant. 
  (7)  Plane-of-the-sky B-field strength $B_{\rm POS}$ calculated using Eq.\,\ref{Eq_Bstrength}. 
 (8) Magnetic critical mass per unit length $M_{\rm line,crit}^{B}$ calculated using Eq.\,\ref{Eq_MlinecritB}. 
(9) Effective virial mass per unit length, taking into account the kinetic (thermal and turbulent) and B-field support:   
 $M_{\rm line,vir}^{v,B}=M_{\rm line,vir}^{v}+M_{\rm line,crit}^{B}$.  
 (10) {\rev Thermal stability parameter   $\alpha_{\rm cyl}^T= M_{\rm line,crit}^T/M_{\rm line}$ 
 \citep[defined as the inverse of the $f_{\rm cyl}$ parameter discussed in][]{Fischera2012, Arzoumanian2019},}  with $M_{\rm line,crit}^{T=20\pm5\,{\rm K}}= 2\,c_{\rm s}^2/{\rm G}\sim27\pm7\,\sunpc$.
  (11) Virial parameter  $\alpha_{\rm vir}^v= M_{\rm line,vir}^v/M_{\rm line}$. 
(12) Magnetic virial parameter  $\alpha_{\rm vir}^{B}= M_{\rm line,vir}^B/M_{\rm line}$. 
(13)   Effective virial parameter, taking into account the kinetic and magnetic support, $\alpha_{\rm vir}^{v,B}= M_{\rm line,vir}^{v,B}/M_{\rm line}$.
   }
 \end{list}     
\end{sidewaystable*}

\section{Power spectrum of intensity and B-field angle along the ridge crest}\label{powspecSection}

Here we present an analysis of the one-dimensional (1D) power spectrum of  $I$, $\chi_{B_{\rm pos}}$, and $\phi_{\rm diff}=|\theta_{\rm fil}-\chi_{B_{\rm pos}}|$ along the ridge crest. 
The power spectrum $P(s)$  of $Y(z)$ $-$ one of the properties observed along a crest $-$
 is proportional to the square of its Fourier transform and can be expressed in 1D as
\begin{equation}
P(s)=\frac{1}{L}|\tilde Y(s)|^2,
\end{equation}
where $z$ is the spatial position along the crest (see, e.g., Figs.\,\ref{polCrest} and 
      \ref{AngleCrest35}) and $s$ is   the angular frequency. %
$\tilde Y(s)=\int Y(z)\exp^{-2i\pi s z}dz$ is the Fourier transform of $Y(z)$ and $L=\int dz$ is the total length of the studied crest.

\begin{figure}[!h]
   \centering
  \resizebox{9.cm}{!}{ 
\includegraphics[angle=0]{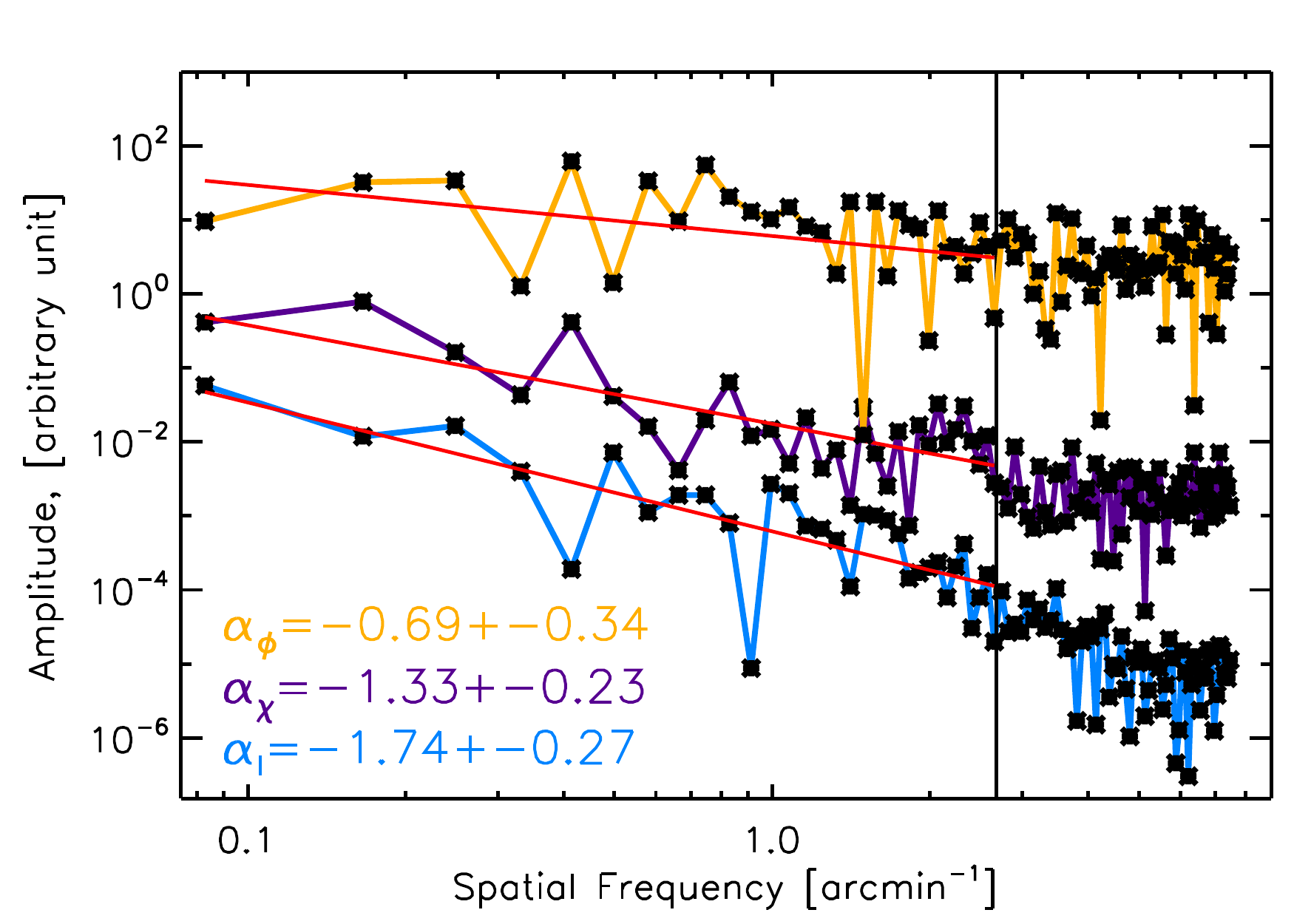}}
\vspace{-.5cm}
  \caption{%
 Power spectra of $I$, $\chi_{B_{\rm pos}}$, and $\phi_{\rm diff}$ in blue, purple, and yellow, respectively,  observed along the crest ridge combining crests 3 and 4 (Fig.\,\ref{SkelFig}). 
  The amplitudes   are scaled to conveniently show the three  power spectra on the same plot. The red solid lines indicate the best fit linear function  (in log-log) to the observed power spectra. The derived best fit slopes are given in the bottom left of the plot,  $\alpha_I$ in blue, $\alpha_\chi$ in purple, and  $\alpha_\phi$ in yellow, for the power spectra of $I$, $\chi_{B_{\rm pos}}$, and $\phi_{\rm diff}$, respectively. 
The vertical black solid line  indicates the $14\arcsec$ HPBW of the data at 850\,$\mu$m 
($s\sim2.7$\,arcmin$^{-1}$),  
which is also the highest frequency data point used to fit the power-law.
       }          
  \label{PS_crest3_5}
    \end{figure}

Figure\,\ref{PS_crest3_5}  shows the power spectra of  $I$, $\chi_{B_{\rm pos}}$, and $\phi_{\rm diff}$ along crests 3 and 4 combined. The amplitudes of the power spectra are scaled so they can be plotted  on the same panel. 

No characteristic scales can been seen in the power spectra of $I$, $\chi_{B_{\rm pos}}$, or $\phi_{\rm diff}$.  The observed power spectra of  $I$, $\chi_{B_{\rm pos}}$, and $\phi_{\rm diff}$  are well represented by power laws down to the angular resolution of the data ($14\arcsec$ or $s\sim2.7$\,arcmin$^{-1}$). We fit the observed power spectra with a power law function (where $s=2.7$\,arcmin$^{-1}$  is the highest frequency data point used for the fit). 
The best fit slope for $I$ is $\alpha_I=-1.74\pm0.27$. This value is compatible with  the statistical results found by \citet{Roy2015} from the (1D) analysis of the column density along the crest  of a sample of 80 subcritical/transcritical filaments observed by $Herschel$ in the Gould Belt. 
The slopes of the power spectra of $\chi_{B_{\rm pos}}$ and $\phi_{\rm diff}$  are $\alpha_\chi=-1.33\pm0.23$ and $\phi_{\rm diff}=-0.69\pm0.34$, $\sim20\%$ and $\sim60\%$ shallower than that of $I$, respectively. 
{\rev While the $\chi_{B_{\rm pos}}$ distribution, which is defined in the range $[0^\circ,180^\circ]$, may in some cases be affected by discontinuities around $0^\circ$ and $180^\circ$, 
that result from the B-field orientation ambiguity inherent in polarimetric observations,
the power spectrum of $\chi_{B_{\rm pos}}$ is not affected by such localised discontinuities, which 
 would only  contribute to the  amplitude of the power spectrum at the smallest scales, i.e., at the largest spatial frequencies. %
Moreover, a clean power-law is also found for the power spectrum of $\theta_{\rm diff}$ defined in the range $[0^\circ,90^\circ]$, which  is not affected by discontinuities at the edges of its interval of definition, suggesting that the power spectra of both $\chi_{B_{\rm pos}}$ and $\theta_{\rm diff}$ are physically meaningful. } 

We checked whether  the well defined power law shape of the observed power spectra  could be obtained from uncorrelated emission of the properties along the filament crest. To do this test, we randomly reordered the observed values along the crest and derived the power spectra for 50 realisations of  test-data. As can be seen on Fig.\,\ref{PS_test}, the power spectra for these test-data are not well represented by a power law and when fitted with a power law function the derived slopes are compatible with a flat spectrum. We hence conclude that the observed  power spectrum slopes,
$\alpha_I$, $\alpha_\chi$, and  $\alpha_\phi$, are a result of correlated fluctuations along the filament crest and they can be used to infer physical properties of the filaments. %
{\rev We discuss physical interpretations of this proposed power spectrum analysis in Sect.\,\ref{powerSpec}.}

\section{Discussion}\label{disc}

\subsection{Comparing the observational results of NGC 6334 with other regions observed as part of  BISTRO } 
\label{disc:comp}

 NGC 6334, at $1.3$\,kpc,  is the first high-mass star-forming region at a distance  larger than that of the Gould Belt clouds ($\sim500$\,pc) that has been analyzed as part of the BISTRO survey. 
These observations reveal the B-field structure at a resolution of $\sim0.1$\,pc towards a $\sim10\,$pc-long ridge and a 
parsec-scale clump-hub structure. 
The results of the analysis of our data show  similarities and differences with the regions already analysed as part of the BISTRO survey. 

\begin{figure}[!h]
   \centering
  \resizebox{8.7cm}{!}{ \hspace{-.7cm}
\includegraphics[angle=0]{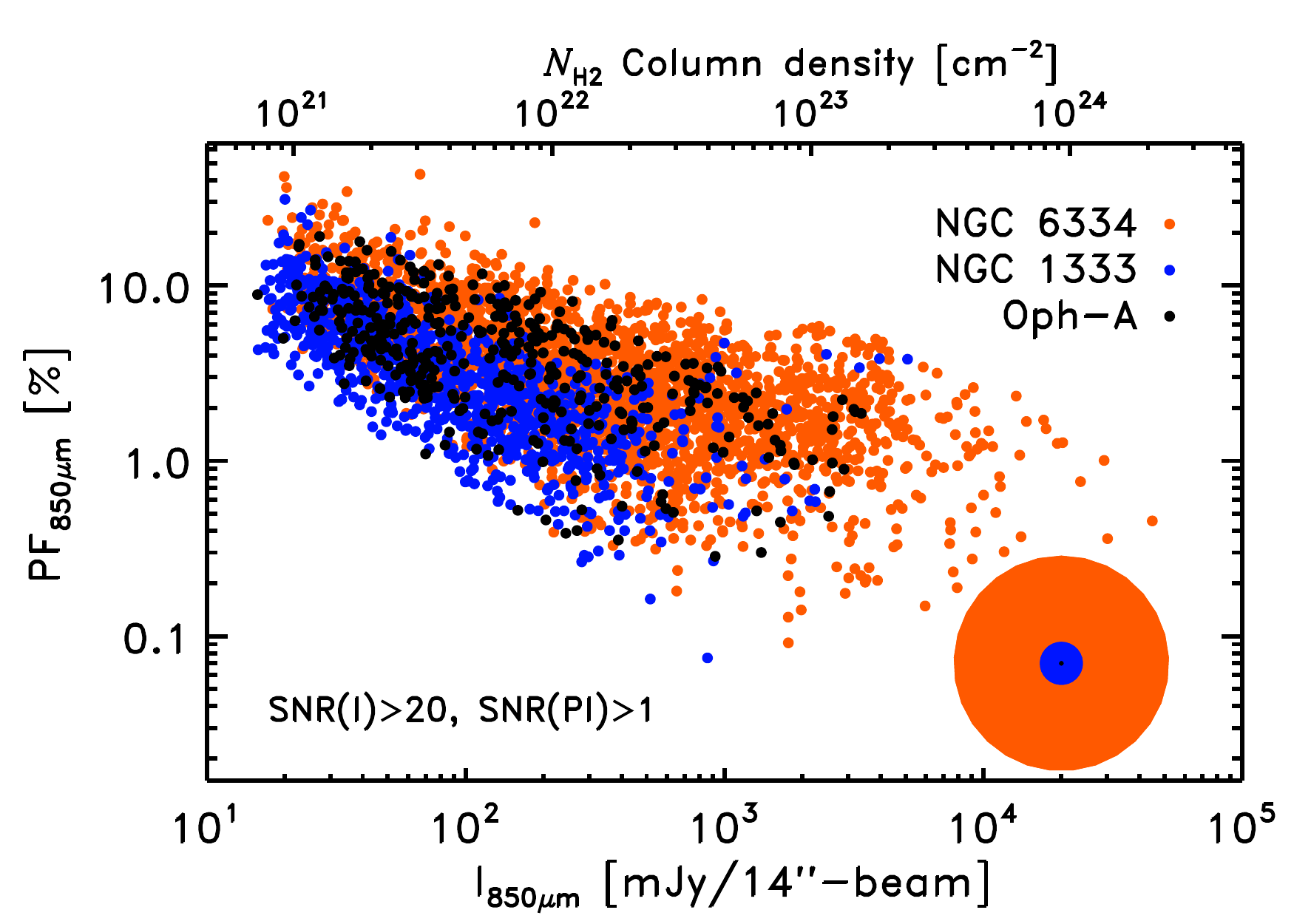}}
\vspace{-.2cm}
  \caption{Comparison plot of $PF$ as a function of  $I$ derived from BISTRO observations in three different regions: NGC 6334 in orange (this work, similar to the middle panel of Fig.\,\ref{scatPlot}), NGC 1333 in blue \citep{Doi2020},  and Oph-A in black \citep[presented in][and also Furuya et al., in prep.]{Kwon2018}.
  The concentric filled circles on the bottom right of the plot indicate the relative sizes of the regions corresponding to about 10\,pc, 2\,pc, and 0.2\,pc in total extent, for NGC 6334, NGC 1333, and Oph-A, respectively. The   angular resolution of the BISTRO data are  0.09\,pc, 0.02\,pc, and 0.01\,pc in physical scales for NGC 6334, NGC 1333,  and Oph-A, at the distance of 1.3\,kpc, 299\,pc, and 150\,pc,  respectively.
       }          
  \label{compPFI}
    \end{figure}

 Figure\,\ref{compPFI} presents a comparison  of the $PF$ values against $I$ observed towards three regions: NGC 6334 (this work), NGC 1333 \citep{Doi2020}, and Oph-A \citep[][]{Kwon2018,Pattle2019}. The values of $PF$  span  very similar ranges from $\sim20\%$ down to $\sim0.2\%$. In the three regions, we observe an overall decrease of $PF$ for increasing $I$ (with similar upper-envelope slopes). %
  The range in $I$ and column density 
 observed towards these three regions 
 are also very similar from  $\sim10^{21}\NHUNIT$ to $\sim10^{23}\NHUNIT$ (the lower value being due to the selection of data; in this case $SNR(I)>20$ for all three regions). In NGC 6334, values up to  $\gtrsim10^{24}\NHUNIT$ are measured (towards the two core-hubs) as a hint of the extreme conditions of this region forming  high-mass stars and stellar clusters compared to the other Gould Belt clouds.  
  All three regions show a large scatter in $PF$ for a given value of $I$ as well as a large scatter in $I$ for a given value of $PF$. 
   
  These similarities are all the more intriguing considering that  the three regions are at different distances from us. The closest is Oph-A at a distance of just 150\,pc, then  NGC 1333 at 299\,pc, and the furthest is NGC 6334 at 1.3\,kpc. The physical sizes of the observed fields span 2 orders of magnitude from $\sim0.2\,$pc, to $\sim2\,$pc, and $\sim10\,$pc (in total extent), for the nearest to the furthest regions (see Fig.\,\ref{compPFI}). 
     The observed similarities despite the range of scales, could suggest that the physical properties of the dust grains emitting the linearly polarized light are comparable in all these  regions  and that the integration  along the magnetized LOS (towards different scales) affects the observed emission in a similar way. 
  This {\rev simple} comparison could also suggest that the self-similar description of the physical processes in a possibly fractal magnetized ISM has  an imprint in the dust polarized thermal emission.  
{\rev We refer the readers to future publications presenting the detailed comparison of  different regions observed as part of the BISTRO survey. %
}

In all three regions, the prevailing POS B-field angle observed at 14\arcsec\ is compatible with the orientation derived at 5\arcmin\ from $Planck$ data and starlight polarization. The high-resolution observations with BISTRO, however, reveal the complex small-scale structures of the B-field. This B-field spans a wide range of POS angle with no characteristic orientation when integrated over the whole region. Detailed inspection of the maps, however, shows the locally ordered B-field geometries in connection with the underlying intensity structure, i.e., the line mass distribution.
In NGC 1333 for example, the POS B-field is observed to be ordered along each filament but its relative orientation with respect to the filament crests varies from one filament to the other.  This variation of the relative orientation between $\chi_{B_{\rm pos}}$ and $\theta_{\rm fil}$ is arguably due to projections effects \citep[cf., the analysis presented in][]{Doi2020}.

In NGC 6334, $\chi_{B_{\rm pos}}$  varies significantly along the different filament crests where we can see the imprint of the formation, evolution, and interaction of the sub-filaments, ridge, and hubs. With these observations, we can also trace  the effects of feedback  from past and on-going high-mass star formation, {\rev such as in the surroundings of the sources from I to V indicated on Fig.\,\ref{IQUmaps}-top-left.} We  qualitatively discuss these two aspects   below and leave more detailed analysis to future works (e.g., Tahani et al., in prep.).

\subsection{The  B-field structure of a high-mass star forming hub-filament system}\label{disc:hub}

The B-field structure is on average perpendicular to the {\rev northern section (crest 4) of the} main ridge of NGC 6334, which would suggest a  formation process  compatible with the  theoretical model  proposed by \citet[][]{Inoue2018}, where a filament is formed by the convergence of a flow of matter along the B-field lines hence forming a filament perpendicular to these lines \citep[see also][]{Inoue2013,Vaidya2013,Arzoumanian2018,Bonne2020}. In this scenario, the flow of matter would  be initially generated by the bending of the ambient B-field lines induced by an interstellar shock compression. 
Such a bent structure of the B-field lines around elongated molecular clouds is  compatible with  the  observational findings  of \citet{Tahani2018} and \citet{Tahani2019}.
Large scale CO molecular line observations 
have suggested that NGC 6334 may be located at the shock front of the collision of two clouds \citep{Fukui2018}.

The network of sub-filaments surrounding the ridge is connected to it in 
a roughly perpendicular orientation %
(see Fig.\,\ref{SkelFig2}).  The {\rev POS} B-field is mostly perpendicular {\rev (or has a random orientation with respect)}  to the sub-filaments in their outer parts. In the inner parts, where the sub-filaments merge with  the clump-hub, the relative orientation between the sub-filaments and the POS B-field angle becomes mostly parallel 
(see Figs.\,\ref{SkelFig2} and\,\ref{InOut_histo}). 
{\revbis This trend has also been found recently towards the G33.92+0.11 \citep{Wang2020} and Serpens South \citep{Pillai2020} hub-filament systems.} 
This finding suggests that the sub-filaments {\rev may be}  formed orthogonal to the ambient B-field and 
then merged with the ridge,  forming a clump-hub structure, due to, e.g., shock compression events, interstellar turbulence, or under the gravitational attraction by  the ridge itself. {\rev Indeed, the whole region, i.e., the clump-hub and the sub-filaments  (crests 5 to 10, 13, and 14), is virially bound with a total mass of $\sim7700\,$M$_\odot$ and  total virial masses of $\sim300-1000\,$M$_\odot$, assuming a spherical geometry with an effective radius of  $\sim1$\,pc and $\sigma_{v}\sim0.5 - 1.0$\,km\,s$^{-1}$.}
 
 In addition to a large-scale longitudinal velocity gradient observed in HCO$^+$ along the ridge,  
 some of the sub-filaments (as identified in Fig.\,\ref{SkelFig}) exhibit longitudinal velocity gradients perpendicular to the ridge \citep[see Fig.\,2 of][]{Zernickel2013}. ALMA high-resolution molecular line  observations of dense gas show also multiple velocity components along a section of the southern part of crest 4 \citep{Shimajiri2019}, that may be tracing dynamical interactions between the ridge and the surrounding sub-filaments. 
 These velocity gradients may be tracing matter flowing onto the ridge/clump-hub or the merging of the sub-filaments with the ridge.  The B-field structure at the tip of the sub-filaments evolves under the influence of the gravitational potential of the clump-hub, while the B-field in the outer part of the sub-filaments still keeps the initial orientation, which better reflects the physical process leading to their formation. 
 The variation of the B-field structure along the sub-filaments from their outer parts to their inner parts may be tracing and be coupled to these local velocity flows of matter in-falling onto the clump-hub, as also suggested by MHD simulations  \citep[][]{Gomez2018}.

Along  crest 3 of the ridge, the B-field orientation varies from being mostly perpendicular to parallel to the crest  (Fig.\,\ref{AngleCrest35}). 
Comparing with the velocity field derived from HCO$^+(3-2)$ molecular line data \citep{Zernickel2013}, this variation of the B-field structure may be induced by longitudinal motions along the crest, possibly by the gravitational potential of source III, a young cluster with high-mass stars. This variation of $\chi_{B_{\rm POS}}$ observed towards the source  III may also be due to the  effect of feedback from the expansion of the UC\hii\ region associated with this young high-mass star-forming site.

     \begin{figure*}[!h]
   \centering
    \resizebox{4.7cm}{!}{ 
\includegraphics[angle=0]{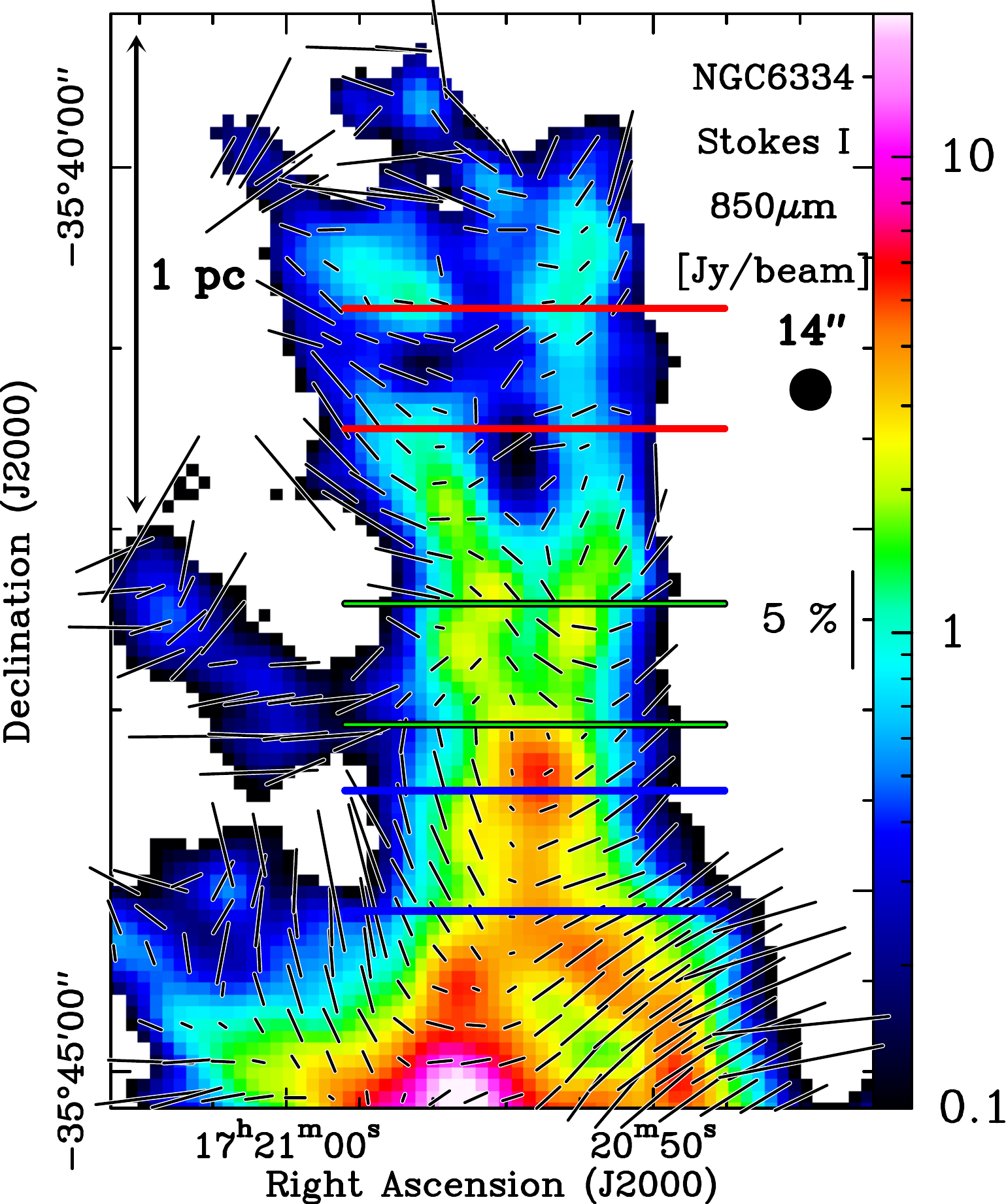}
}
  \resizebox{6.7cm}{!}{ 
\includegraphics[angle=0]{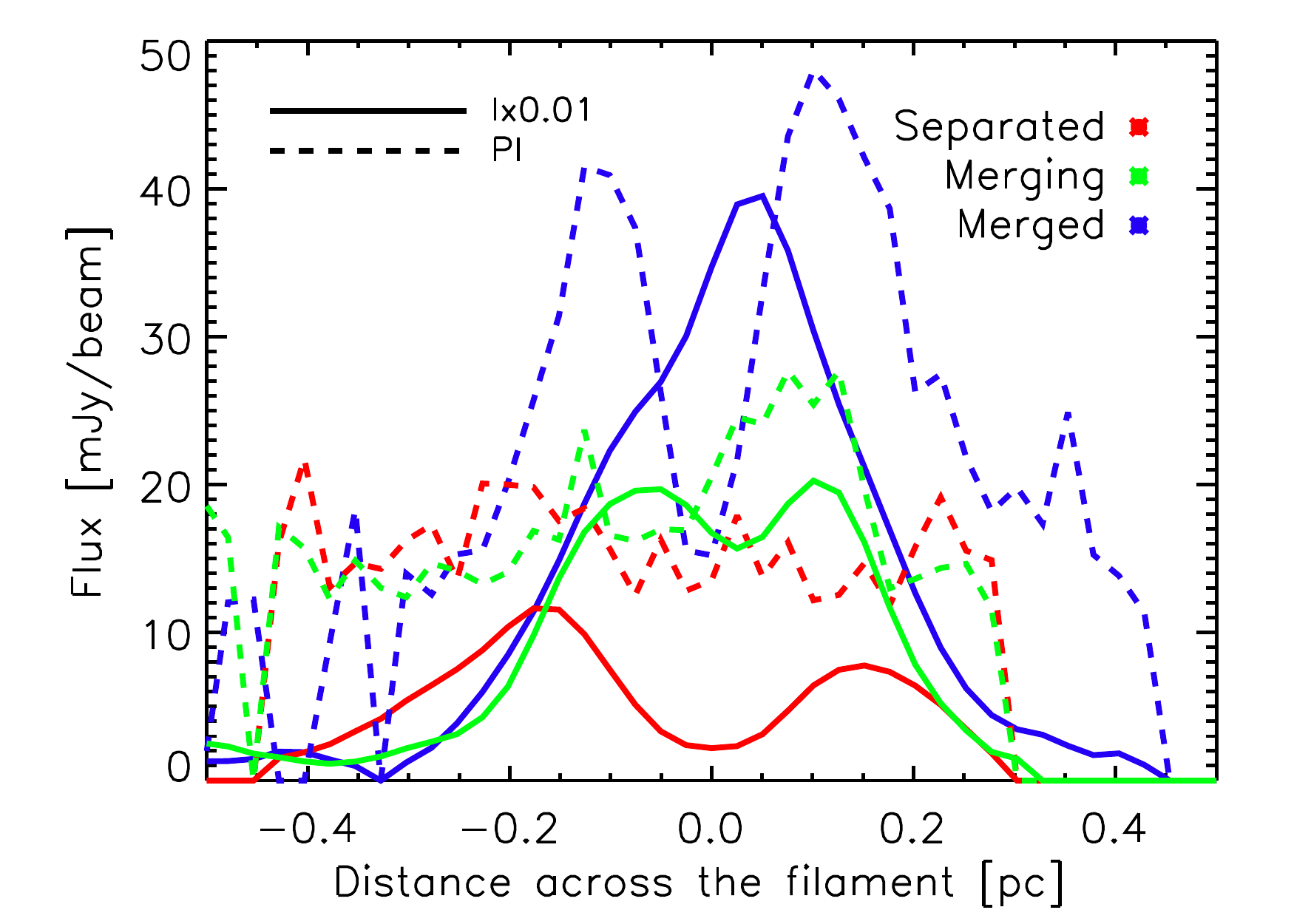}
}
  \resizebox{6.7cm}{!}{ 
\includegraphics[angle=0]{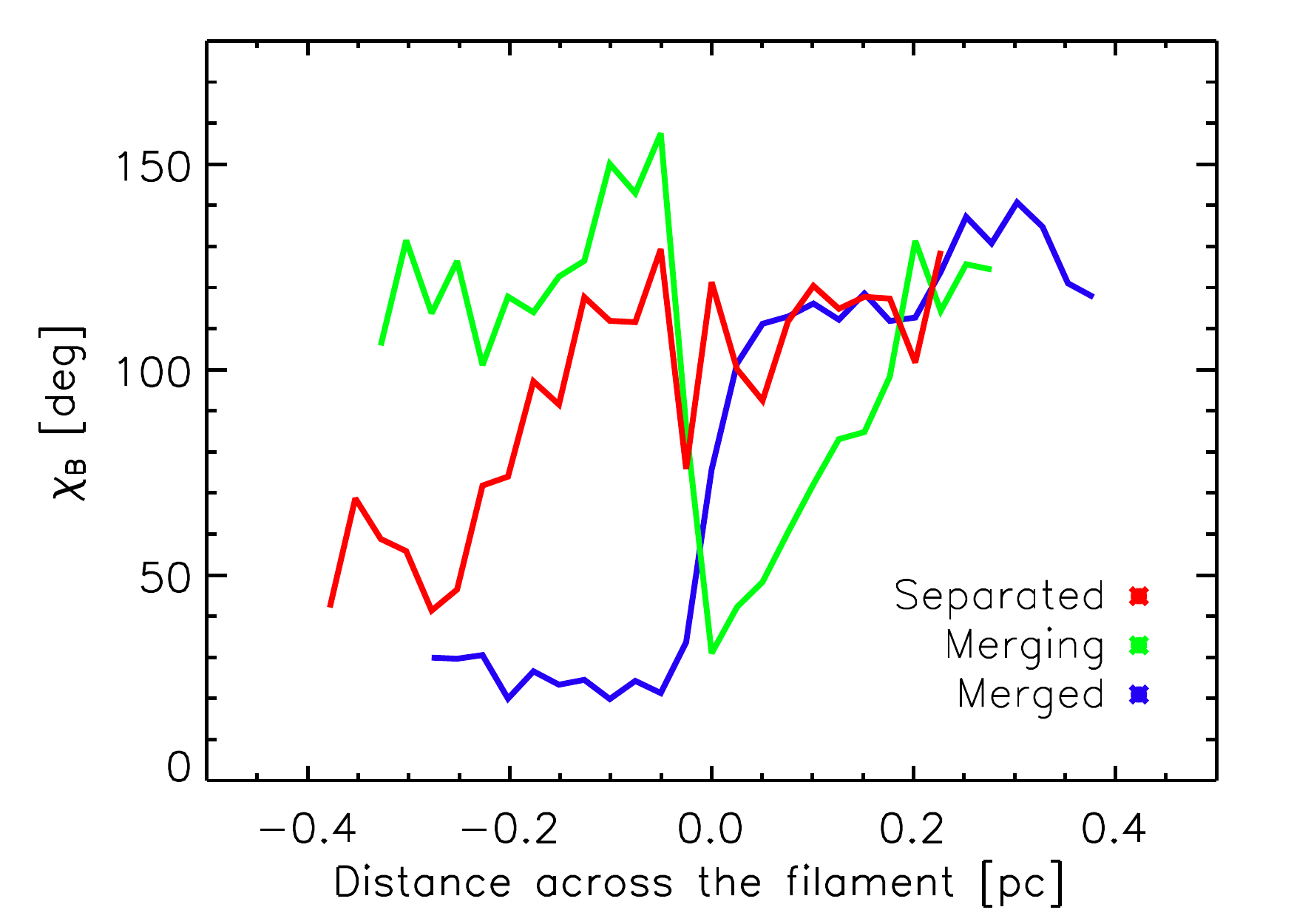}
}
  \caption{%
  {\it Left}: 
    Stokes $I$ map of the northern region of NGC 6334 towards the sub-filaments  5 and 6   (see Fig.\,\ref{SkelFig}).    
  The black lines correspond to $\chi_{B_{\rm POS}}$ and  their lengths are proportional to $PF$.  
    On the right hand side,  a vertical line indicates the  $5\%$ $PF$ scale. The black filled circle indicates the $14\arcsec$ beam.   
  The horizontal blue, green, and red lines show the (0.8\,pc) cuts used to derive the profiles shown on the middle and right panels, for the merged, merging, and separated profiles, respectively. 
  {\it Middle}: 
  Three transverse cuts, from  East to West,  where the sub-filaments are still separated (red), where they are merging (green), and where they are merged (blue) and not distinguishable any more as two peaks in  Stokes $I$. 
Each profile  has been averaged vertically between the two lines of the same colour indicated on the left hand side plot. 
  The solid and dashed line profiles show Stokes $I$ and $PI$, respectively. 
  We see a drop of $PI$ towards the peak in $I$ when the two filaments merge, that may be due to depolarization via beam-smearing of  the B-field where the two sub-filaments  merge. 
  {\it Right}:   Same as the middle panel for the variation of  $\chi_{\rm B_{\rm POS}}$  across the two sub-filaments. 
}          
  \label{FilMergeProfile}
    \end{figure*}
      
Figure\,\ref{FilMergeProfile} shows 
the variation of the polarization properties as two sub-filaments (crests 5 and 6) merge into a single filament (in projection at least) connected to the clump-hub. 
Transverse profiles of $I$ and $PI$ across the two separated crests 5 and 6  shows two peaks in $I$ and is flat in $PI$ (red curves in Fig.\,\ref{FilMergeProfile}). The two peaks in the $I$ profile show larger values and their separation gets smaller as the distance to the clump-hub shrinks. When the two crests are observed to merge (corresponding to the blue segments in Fig.\,\ref{FilMergeProfile}-Left), a single peak in $I$ is seen, with a value  $\sim4$ times larger compared to the peak intensity when the two crests were separated. This increase in $I$ towards the crests of the merged-filaments is larger than a simple addition of the intensity  of the two separated-filaments, suggesting 1) this merging is not  due to projection effects and integration along the LOS of two physically independent filaments, but 2) there is an additional efficient infall of matter onto the merged filaments (resulting in an increase of the density). We might be witnessing here the precursor of a core-hub.   The $PI$ profile across the merged filaments shows a dip (seen also in $PF$) at the position of the peak in $I$.      The POS B-field angle across these two filaments also changes and is observed to be perpendicular to the crest after they merge, while the B-field lines are bent on both sides of the crest in a horseshoe pattern (Figs.\,\ref{PI_PA2} and\,\ref{FilMergeProfile}).  
This correlation between the variation of the POS B-field structure and the dip in $PI$ and $PF$ may be tracing depolarization of the polarized emission due to the integration along the LOS or within the beam of unresolved B-field structures. Similar variations  of the  polarization properties can be seen along crest 8, which seems to be connected to the core-hub I(N)  (see Figs.\,\ref{PI_PA2} and\,\ref{SkelFig2}). These variations are also associated with an increase of the dispersion of $\chi_{\rm POS}$ (c.f., Fig.\,\ref{SrmsPA}) suggesting the change  of the geometry of the B-field  as the sub-filaments merged with the clump- and core-hubs.

As can be seen in Table\,\ref{tab:paramCritic} and Fig.\,\ref{InOut_histoCrit},
the ridge and most of the sub-filaments of NGC 6334 are {\rev thermally supercritical with $\alpha_{\rm cyl}^T<1$.} Towards the ridge, both turbulent $\alpha_{\rm vir}^v$ and magnetic $\alpha_{\rm vir}^B$ virial parameters $<1$,    suggesting that  when taken separately the thermal, turbulent, or magnetic energy alone cannot support the ridge against gravity. 
Towards the sub-filaments, $\alpha_{\rm vir}^v$ decreases 
and $\alpha_{\rm vir}^B$ increases
from their outer parts to their inner parts, suggesting stronger role of the gravity and magnetic field in the inner parts of the sub-filaments when they merge with the clump-hub.   
When all three main pressure forces are considered simultaneously,    $\alpha_{\rm vir}^{v,B}\gtrsim1$, implying that while gravity is important in the evolution of and the interactions between the NGC 6334 filaments, 
the magnetic field as well as  kinetic support play a key role in their overall energy balance.

\subsection{Impact of the star formation feedback  on the observed polarization properties}\label{disc:feedback}
 
  A main difficulty in the interpretation of polarization observations is that dust polarized emission  depends on 1) the dust grain properties (size, shape, composition), 2) the efficiency of grain alignment  with the local B-field, and 3) on the magnetic field structure (mean orientation with respect to the LOS and dispersion on the POS and along the LOS). In addition, the observed emission is integrated along the LOS and averaged within the observational beam.  Thus understanding the origin of the variation of the polarization properties, e.g., the relation between the polarization fraction and the intensity (column density), cannot be gained without taking into account these effects simultaneously. For example, large $PF$ values require well aligned grains and ordered B-field structures with optimal orientations with respect to the LOS, while low $PF$ values may be due to a loss of grain alignment, variation of the grain properties, or a change of the B-field configuration. 
  At the observed densities, the alignment efficiency of  dust grains with the local B-field lines is independent from the absolute B-field strength. In addition, alignment itself  is considered instantaneous given the dynamical timescales of the molecular cloud even when there are  interactions and shock compressions  between  different  structures/filaments \citep{Hildebrand1988,Voshchinnikov2012,Hoang2016}.   
  The observed $PF$ values, however, may be affected by the B-field strength due to the increase or decrease of depolarization from a turbulent or ordered B-field structure, respectively.

   As mentioned earlier, 
 NGC 6334  is actively forming stars and stellar clusters. The dense and cold molecular gas is affected by  feedback from the high-mass stars formed along and next to the ridge. This region is hence an excellent  laboratory to study the effect of  feedback on the polarized dust emission properties and the B-field structure. While we refer the reader  to a companion  paper (Tahani et al. in prep.) for a  detailed analysis,  we give below a qualitative overview on the impact of the expanding \hii\ regions %
on the surrounding B-field and  polarized emission.

\begin{figure}[!h]
   \centering
     \resizebox{9.5cm}{!}{   \hspace{-.3cm}
     \includegraphics[angle=0]{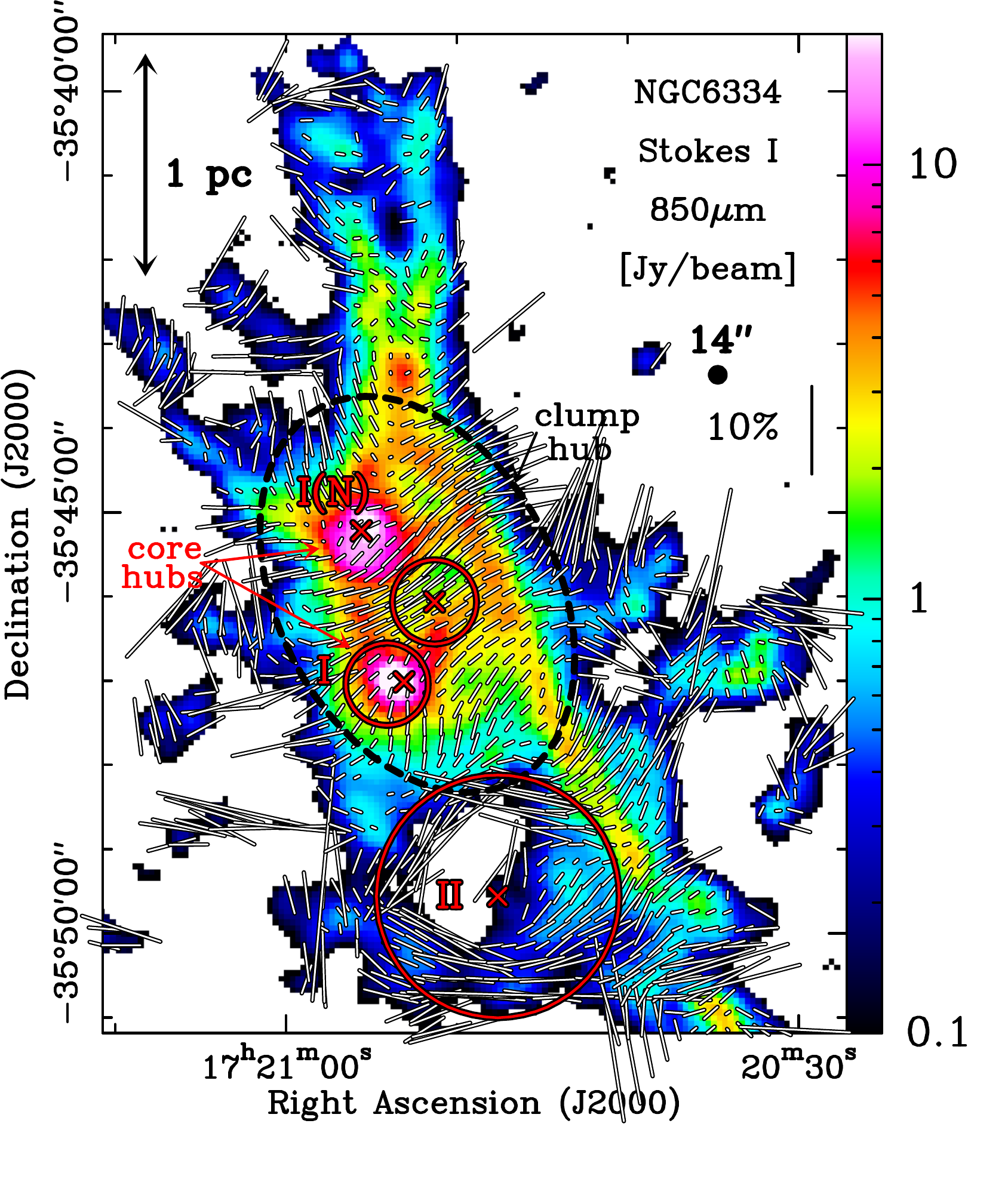}
}\vspace{-.5cm}
  \caption{ 
    Zoom-in of the northern region of NGC 6334 (same as Fig.\,\ref{SkelFig2}).  
  The white lines show $\chi_{B_{\rm POS}}$ and  their lengths are proportional to $PF$.  
    On the right hand side,  a vertical line indicates the  $10\%$ $PF$ scale. The black filled circle indicates the $14\arcsec$-beam.   
  The black dashed ellipse shows the clump-hub. The two core-hubs I and I(N) are also indicated. The crosses and circles are the same as in Fig.\,\ref{IQUmaps}-top-left.
}          
  \label{UpZoom}
    \end{figure}

  As an example of the impact of feedback, we note the 
 POS B-field structure bends  along the dense shell around the \hii\ region corresponding to source II, where the expansion of the \hii\ region has swept up most of the local dust and has created 
 a cavity of  submm continuum emission seen in our observations (c.f., Fig.\,\ref{UpZoom}). This expansion 
 likely influenced the pre-existing B-field structure, and the field was dragged along with the swept-up material. 
 The cloud was reshaped   
 into the now-observed shell with enhanced polarized emission due to 
  an ordered B-field structure curved along the compressed shell  \citep[see, e.g.,][for similar discussions]{Tang2009HII,planck2016-XXXIV}.
  {\rev A similar feedback effect may be seen towards the shell associated with the \hii\ region  to the South of source IV, albeit with a more complicated B-field structure and polarization pattern. Detailed 3D modelling is needed to understand the origin of these observed emission. This study will be partly addressed in Tahani et al. (in prep.).}\\

\begin{figure}[!h]
   \centering
  \resizebox{9.cm}{!}{ \hspace{-1.cm}
\includegraphics[angle=0]{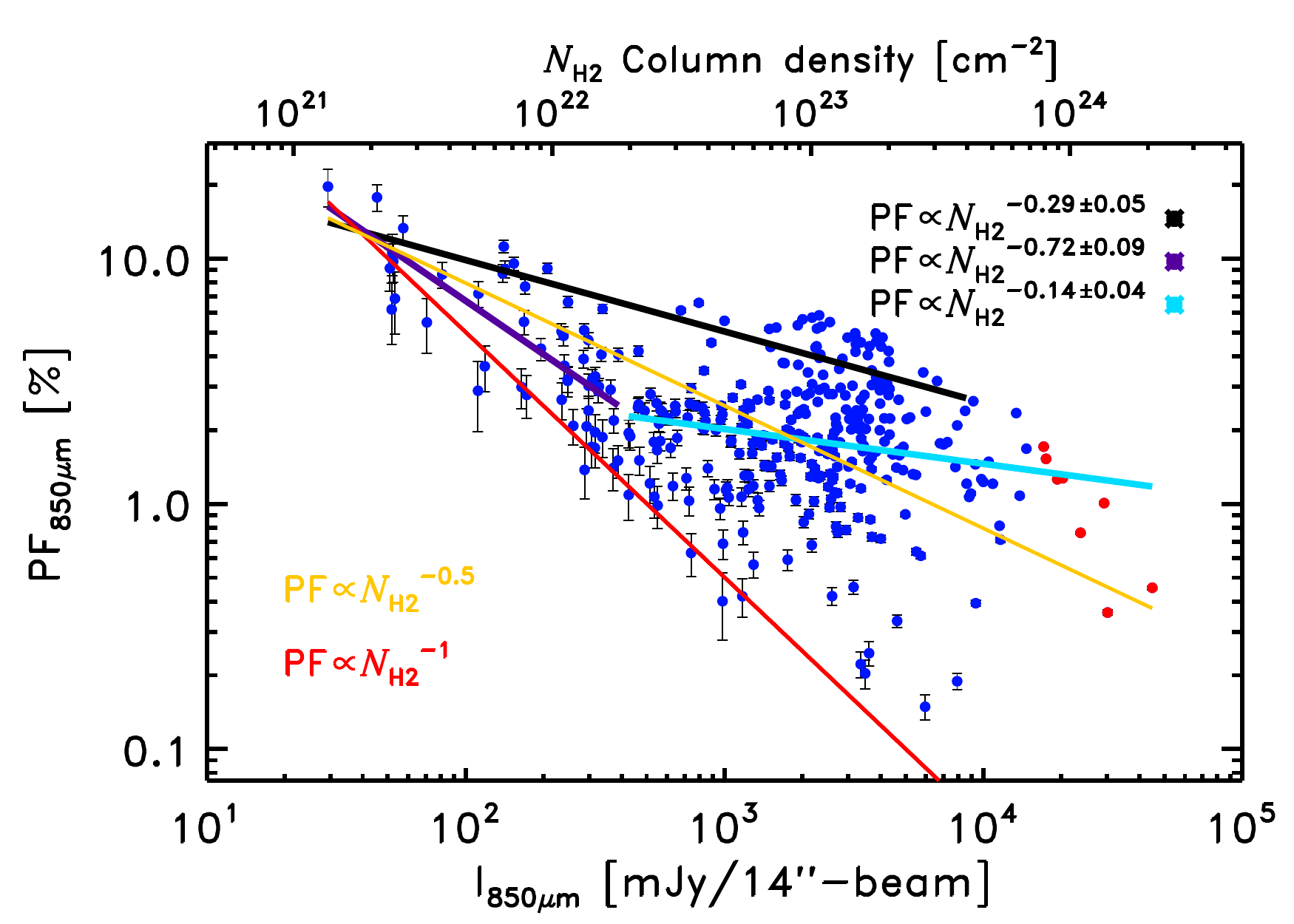}
}
  \resizebox{9.cm}{!}{ \hspace{-.2cm}
\includegraphics[angle=0]{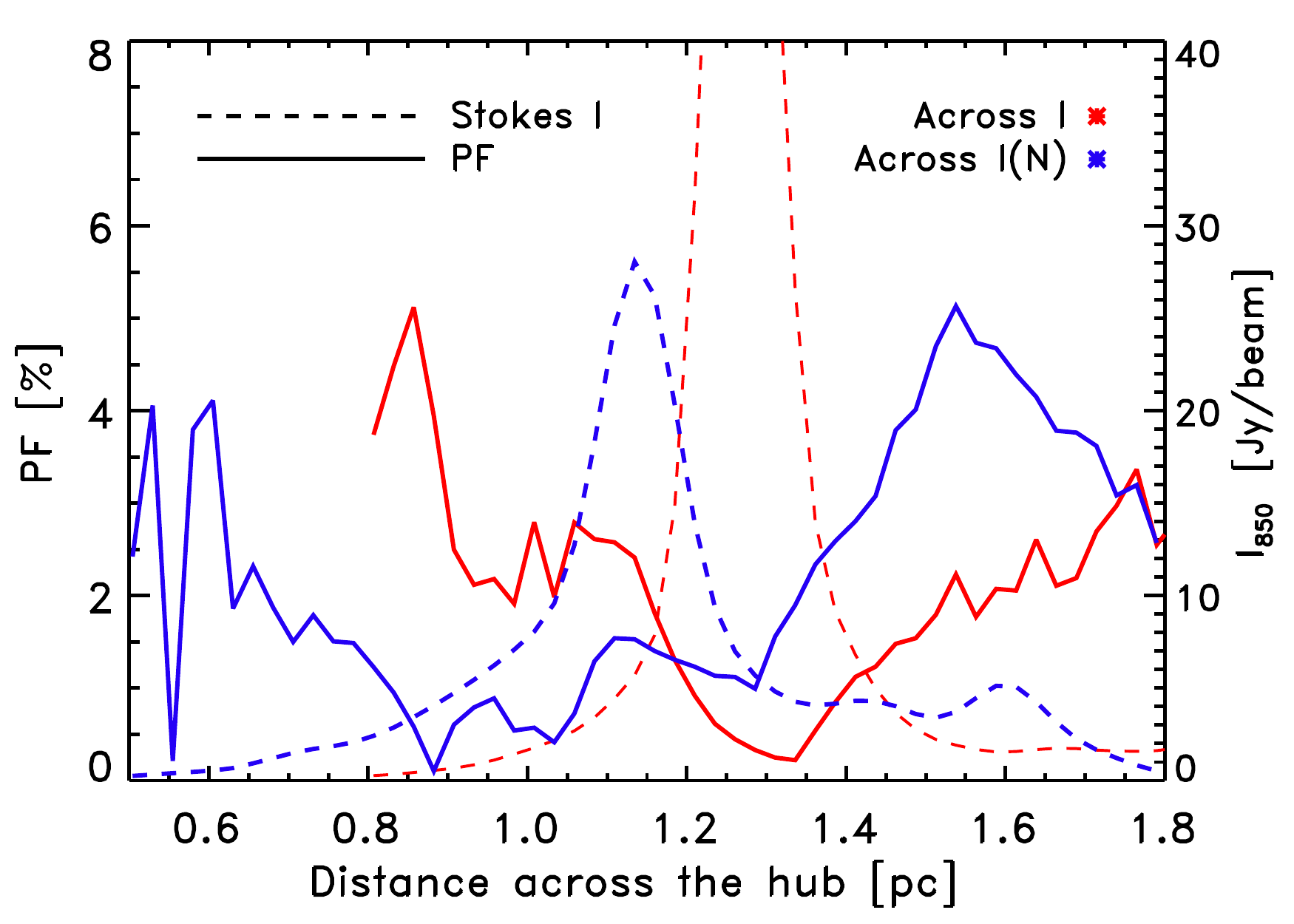}
}\vspace{-.2cm}
  \caption{%
  Properties observed towards the clump-hub (black dashed ellipse in Fig.\,\ref{UpZoom}).
  {\it Top}: $PF$ as a function of  $I$ for $SNR(I)>25$ and $SNR(PI)>3$. The vertical small black lines show the statistical uncertainties on $PF$. The black, purple, and cyan lines indicate the best linear fits to the upper $95\%$ percentiles of the distribution, the data points with $\nhh<2\times10^{22}\NHUNIT$, and the data points with $\nhh>2\times10^{22}\NHUNIT$, respectively. The best fit slopes are indicated on the plot. For comparision, the yellow and red lines indicate slopes of $-0.5$ and $-1$, respectively.
{\it Bottom}: Variation of  $PF$ (solid lines) and $I$ (dashed lines) for two horizontal cuts from East to West, crossing the star forming regions I %
and I(N) %
 in red and blue, respectively. The peak in $I$ of the blue dashed profile  at $\sim1.6$\,pc corresponds to the position of the ridge (crest 4).
}          
  \label{Filhub}
    \end{figure}

 The clump-hub shows another example of the impact of the feedback. 
 Figure\,\ref{Filhub}-top shows a scatter plot of $PF$ as a function of $I$ towards the clump-hub 
  (black dashed ellipse in Fig.\,\ref{UpZoom}). 
  All the data points towards the clump-hub have $SNR(I)>25$ and $SNR(PI)>3$. 
  We see an overall decreasing trend of $PF$ as a function of $I$ (c.f., Fig\,\ref{scatPlot}). A decrease in $PF$ with $I$ (or $\nhh$) has been reported in previous studies and ascribed to the loss of grain alignment efficiency  or the random component of the magnetic field \citep[e.g.,][]{Jones1992,Whittet2001,Whittet2008,Jones2015,planck2015-XIX}. 
  Notably, the observed $PF$ versus $I$ data points from  the clump-hub are not well described by a single mean power law. We can see that the linear fit $log\,PF$ versus $log\,\nhh$ for the data  points with $\nhh<2\times10^{22}\NHUNIT$  provides  a best fit slope of $-0.72\pm0.09$, while for the data points with $\nhh>2\times10^{22}\NHUNIT$ the fitted slope $-0.14\pm0.04$ is much shallower.  
A fit to the upper envelope of the distribution yields a slope of $-0.29\pm0.05$. All of these slopes are shallower than the $PF\propto\nhh^{-1}$ relation  expected from the progressive loss of grain alignment expected in the framework of radiative torques (RAT) grain alignment theory due to the attenuation of the interstellar radiation field with increasing column density
\citep[e.g.,][]{Cho2005,Lazarian2007,LazarianHoang2007,Hoang2014}.

\begin{figure*}[!h]
   \centering
     \resizebox{19.5cm}{!}{  \hspace{-2.cm}
\includegraphics[angle=0,width=2.5\textwidth]{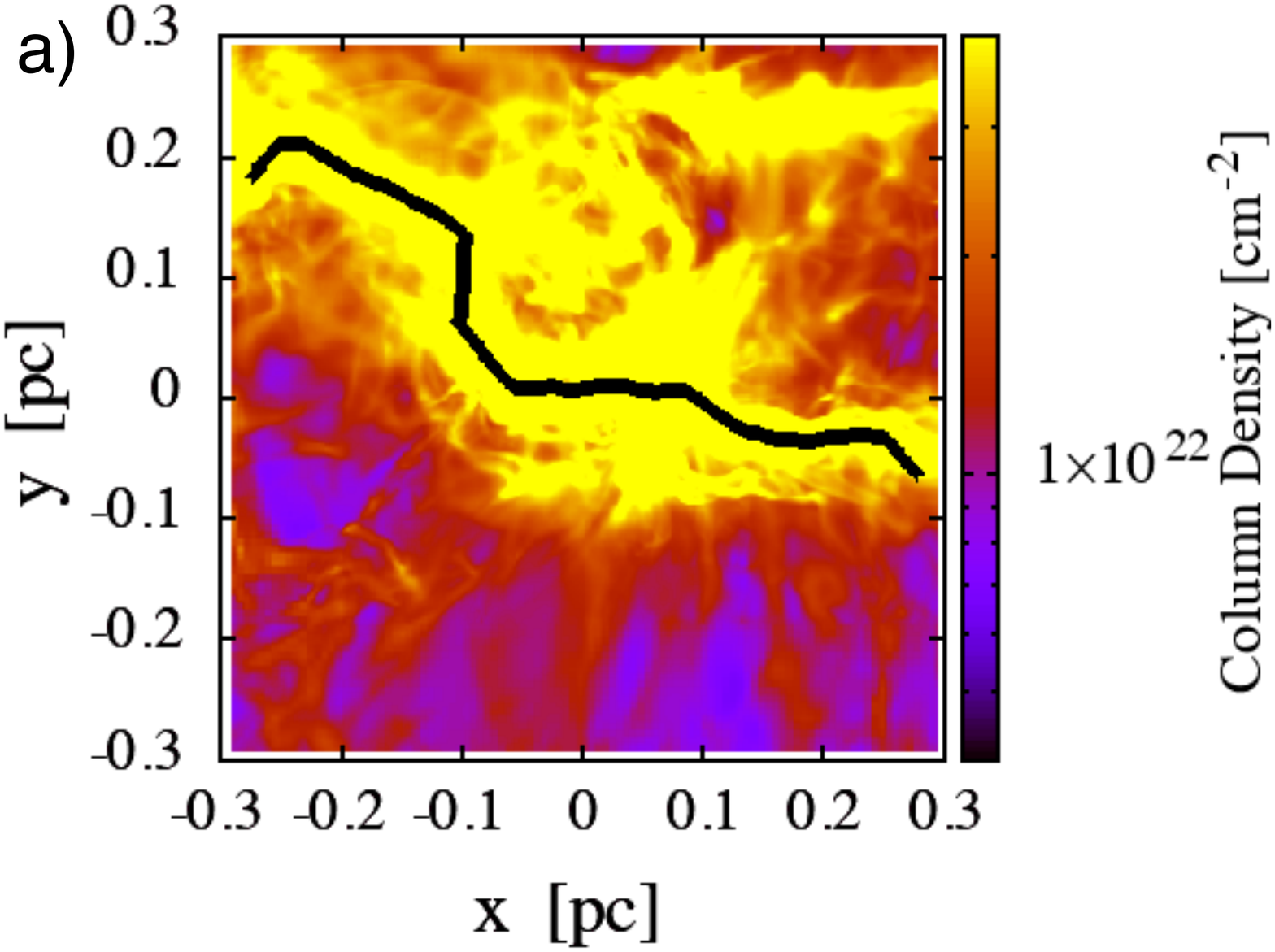} %
\includegraphics[angle=0,width=2.5\textwidth]{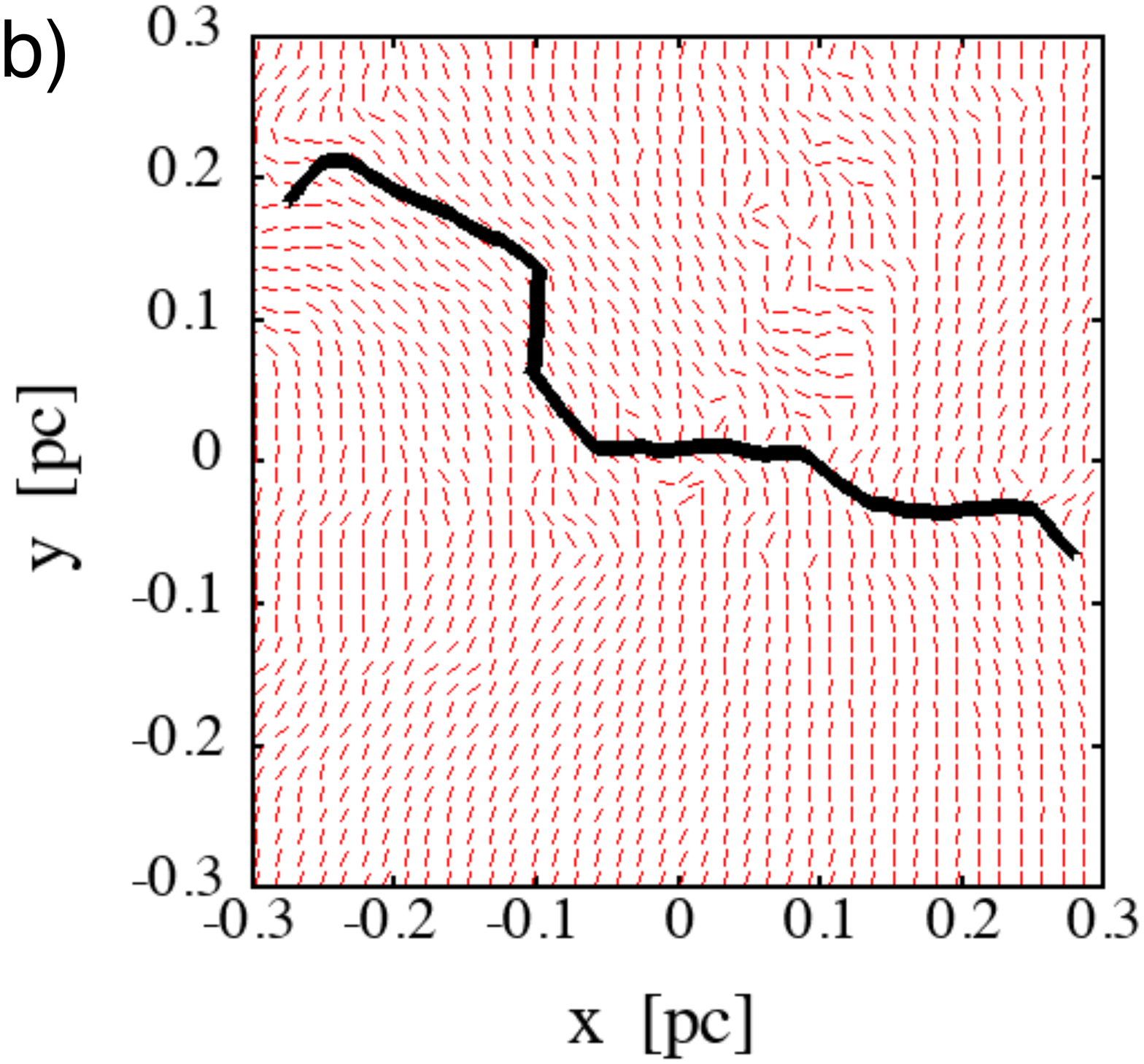}
\includegraphics[angle=0,width=2.5\textwidth]{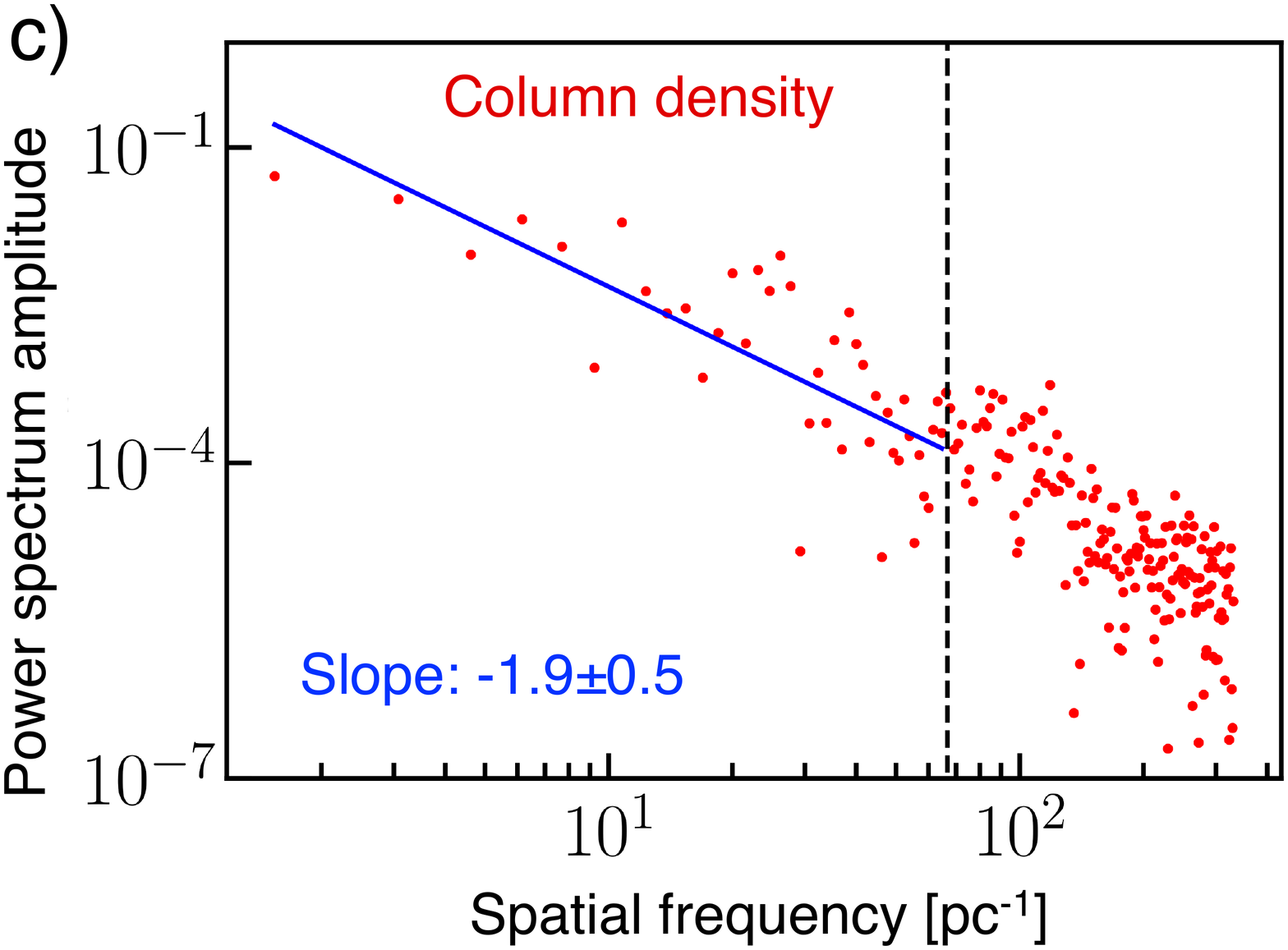} %
}
   \resizebox{19.5cm}{!}{\hspace{-2.cm}
\includegraphics[angle=0]{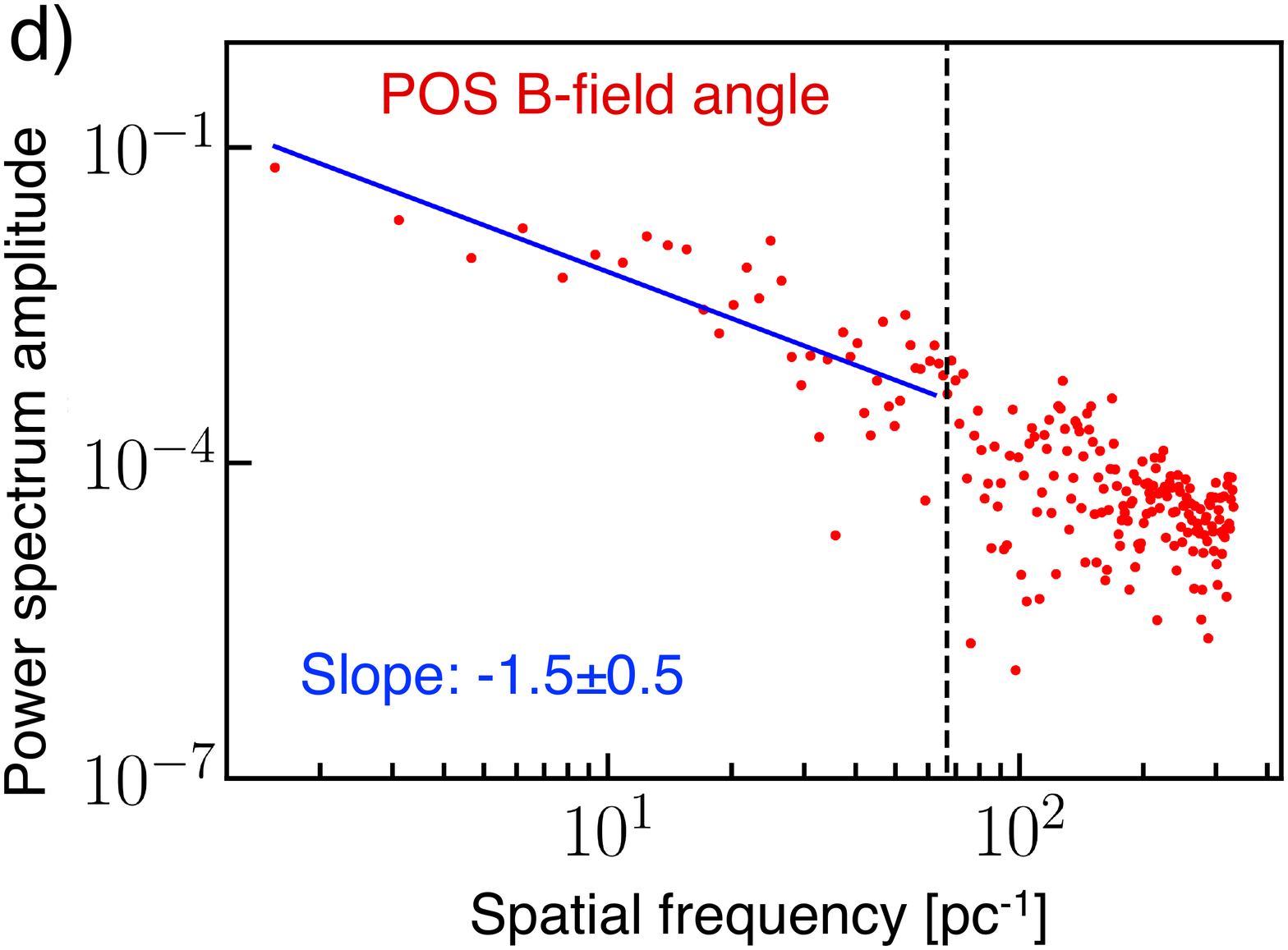}
\includegraphics[angle=0]{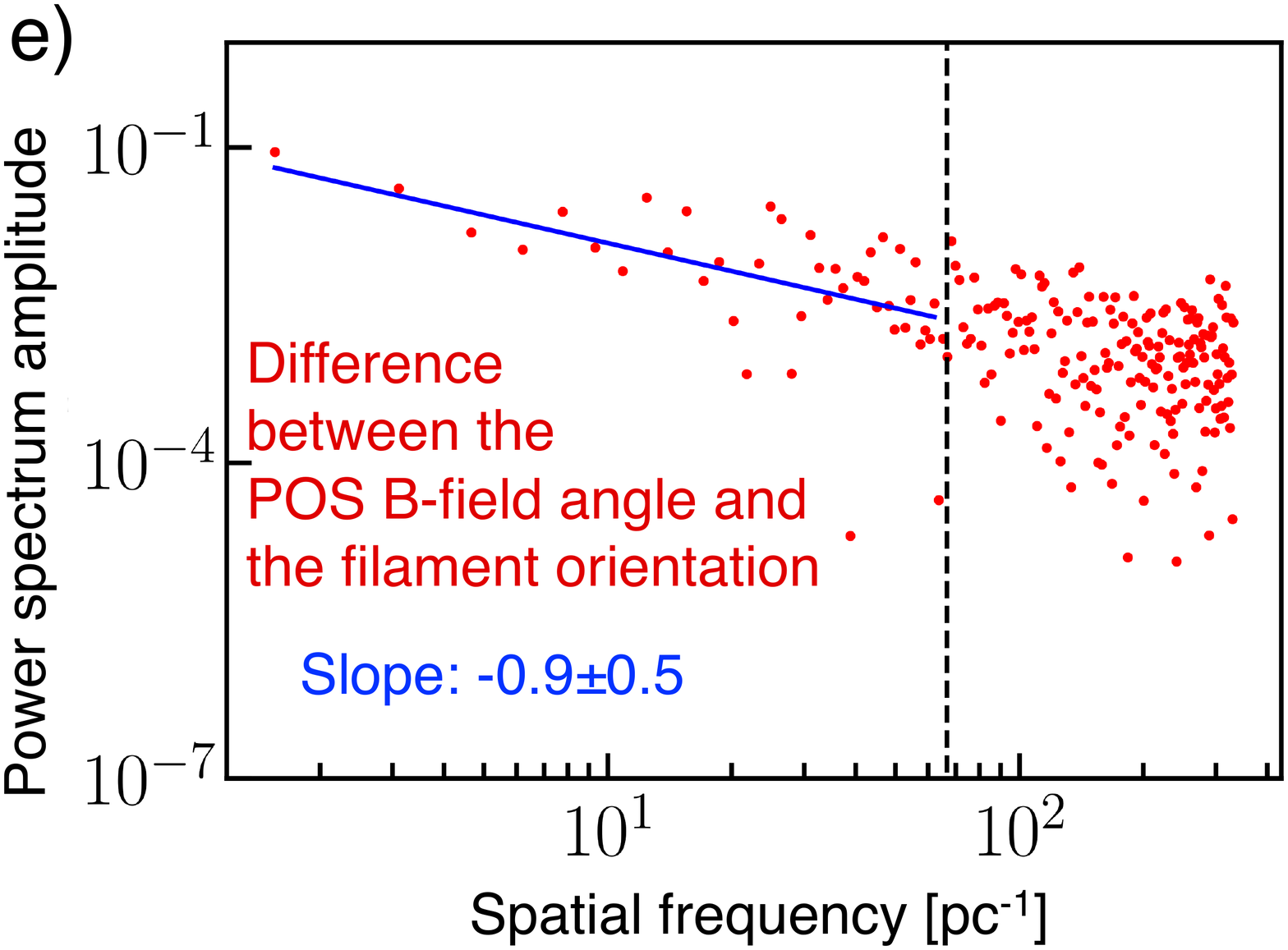}
\includegraphics[angle=0]{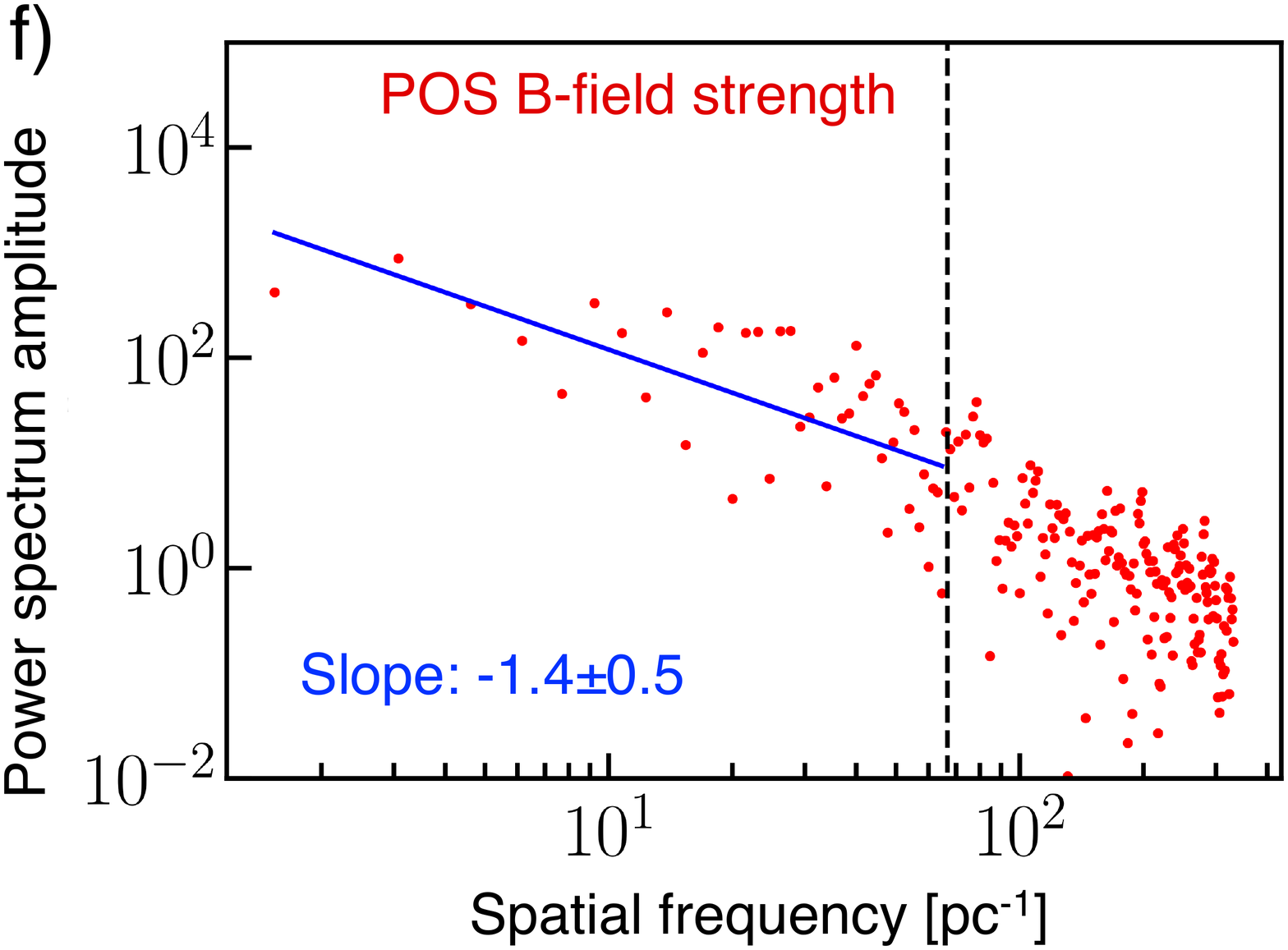}
}
\vspace{-.5cm}
  \caption{ Results  derived from the MHD numerical simulation presented by \citet{Inoue2018}.
  The selected snapshot corresponds to $t = 0.7$\,Myr after the filament formation by shock compression. The mass per unit length of the filament at this epoch is $\sim120\,\sunpc$.
a) Column density map of the filament. The black curve shows the crest of the filament  traced with {\tt FilFinder} \citep{Koch2015}. b) The small red lines indicate the POS B-field angle towards the same region shown in panel a). The black curve is the crest of the filament. From c) to f) Power spectra of the column density, the POS B-field angle, the difference of the angle between the POS B-field orientation and the filament orientation, and the B-field strength, respectively. The slopes derived from the power law fits (shown in blue) to each of the power spectra are indicated on the panels. The vertical black dashed line  indicates the highest frequency data point used to fit the power-law corresponding to 10 computational cells. 
}          
  \label{BstrengthPSsimu}
    \end{figure*}

For these high column densities $\nhh\gg10^{22}\NHUNIT$, the observed shallower slopes may be 
 tracing an increase of the grain alignment efficiency due to the enhanced anisotropic radiation from  newborn stars embedded in the clump-hub.  Indeed, according to the RAT theory, anisotropic radiation from stars embedded in high-column density clumps increases  the efficiency of grain alignment \citep[][]{LazarianHoang2007,LeeHoang2020,Hoang2020}, which may result in the observed shallower slopes \citep[see also][]{Pattle2019}. 
Moreover, \citet[][]{Hoang2019} showed that when the  intensity of the radiation from the embedded stars is large, the polarization fraction decreases due to the removal of large grains. We may be witnessing this effect towards the core-hubs (red data points in Fig.\,\ref{Filhub}-top), which show a steep decrease of $PF$ for $\nhh\gtrsim10^{24}\NHUNIT$. 
This effect may also be seen in 
the bottom panel of  Fig.\,\ref{Filhub}, which  traces  the variations of  $PF$ and $I$ across the  two core-hubs. 
Towards the older source I, a trough in $PF$ is associated with an increase in Stokes $I$. Towards the youngest source I(N), however, there is a spatial shift between the position of the dip in $PF$ and the peak in $I$. Lower $PF$ values are observed towards the edge of this latter source compared to larger $PF$ values towards its high column density centre. In the cut  towards the source I, the emission from the ridge (crest 4) can also be seen (at $\sim1.6$\,pc from the $x$-axis in Fig.\,\ref{Filhub}) which interestingly is not associated with a drop in $PF$, but with large $PF$ values.  The observed POS B-field  towards the crest 4 embedded in the clump-hub has an ordered structure, which may be also contributing to the observed large $PF$ values.
The slopes derived from the $PF$ versus $I$ plot in Fig.\,\ref{Filhub}-top are also shallower than the $-0.5$ slope expected from a purely turbulent B-field \citep[][]{Jones2015,Andersson2015}, consistent with the
 observed ordered B-field structure towards the clump-hub (c.f., e.g., Fig.\,\ref{UpZoom}).

In summary, the 
polarization  features observed towards  NGC 6334 may result from the impact of  star formation feedback. 
Such features include:  
1) the re-organization of the structure of the B-field on the scale of  \hii\ regions (within the compressed shells) where  the mean B-field orientation is reoriented closer to the POS,
2) an increase of the B-field strength,
ordering its structure and reducing depolarization due to  integration of  emission along the LOS and within the beam,
3) an increase of the efficiency of grain alignment with the local B-field due to an enhanced local radiation field from the newborn high-mass stars,  and 4) a drop of  polarization fraction towards the densest core-hubs due to the  removal of large grains from stronger stellar feedback or a more tangled and unresolved configuration of the B-field.

\subsection{Understanding the origin of the slope of the B-field power spectrum}\label{powerSpec}

A recent analysis of the column density variations along the crest of  a sample of filaments observed by $Herschel$ in three 
star-forming clouds  indicates that the  1D column density power  spectra of these filaments  are well defined by a power law function (down to $Herschel$ resolution of $\sim0.02$\,pc) with  a mean slope of $-1.6\pm0.3$ \citep{Roy2015}. 
This observed power law slope is   compatible with the theoretical value suggested by  \citet{Inutsuka2001} close to twenty years ago to explain the origin of the  \citet{Salpeter1955} power-law slope of the IMF at the high-mass end.
More recently, in an attempt to investigate the origin of the observed angular momentum of star-forming cores,  \citet{Misugi2019} proposed a new theoretical model based on the filament  velocity power spectrum.  This model
suggests that cylindrical filaments characterized  by a 1D Kolmogorov velocity power spectrum slope of $-5/3$  will fragment into cores showing a distribution of  angular momentum compatible with the observations.

The results presented by \citet[][]{Roy2015}  and  \citet[][]{Misugi2019} suggest that the column density (line mass) and velocity variations/fluctuations along a filament  may help to illuminate the origin of the CMF/IMF shape and the properties of the cores (e.g., angular momentum) formed from filament fragmentation. 
 Since the gas and the magnetic field are dynamically well coupled in the ISM (i.e., frozen-in conditions), the magnetic field power spectrum should be  related to the density and velocity power spectra and this link would depend on the physical processes resulting in the formation of the ISM structures. Thus, the observational determination of the power spectrum of the magnetic field configuration could provide valuable information on the origin of the formation of these structures and the  link between  the kinetic, magnetic, and gravitational energies at play in the ISM {\rev(see also \citealp{LiGX2018} for a theoretical approach linking the power spectra of the density and magnetic energy fields in the ISM)}.

  Here we present for the first time the  power spectrum of the B-field POS angles $\chi_{B_{\rm pos}}$  observed along the  NGC 6334 ridge (cf., Fig.\,\ref{PS_crest3_5}).  We show that the power spectrum of  $\chi_{B_{\rm pos}}$, similarly to the power spectrum of $I$,  can also be well represented with a power law function and no characteristic scales are detected  down to the resolution of the data. 
The slope of the power spectrum of $\chi_{B_{\rm pos}}$ $-1.33\pm0.23$ is measured to be $\sim20\%$ shallower than that of $I$ ($-1.74\pm0.27$, cf., Fig.\,\ref{PS_crest3_5}). We also present the power spectrum of  the difference of angle between $\chi_{B_{\rm pos}}$ and the ridge crest orientation; $\theta_{\rm diff}$ {\rev defined in the interval $[0^\circ,90^\circ]$}. This latter power spectrum has a slope of $-0.69\pm0.34$, shallower than the power spectrum slope of $I$ and $\chi_{B_{\rm pos}}$. 

We compare this observational result with  the power spectrum of the magnetic field angle along a supercritical filament formed in the  MHD numerical simulation of \citet{Inoue2018}.  Figure\,\ref{BstrengthPSsimu} presents the power spectra of the column density, the POS B-field angle, and the difference between the POS B-field angle and the filament orientation 
derived from the three-dimensional data cubes of this simulation. The slopes of the power spectra of the different quantities derived from the observations and the simulations are in agreement. 
This similarity suggests that the observed column density (line mass) and B-field fluctuations along the star forming NGC 6334 filament may be inherited from the filament formation process and have a coupled evolution due to, e.g., the accretion of matter and fragmentation  under the interplay of the turbulent, gravitational, and B-field energies as incorporated in this MHD simulation.
We also derive the power spectrum of the POS B-field intensity along the filament from the MHD simulation. This latter power spectrum has a power law slope similar to that of  the POS B-field angle power spectrum, both $\sim20\%$ shallower than that of $I$  (c.f., Fig.\,\ref{BstrengthPSsimu}).

Following these preliminary results,  we  suggest that understanding the theoretical link between the power spectra of column density, velocity, and magnetic field (both angle and strength) and its evolution in time may lead to  a new and independent method to infer the B-field strength from polarization observations, which do not provide a direct measurement of the B-field strength. 

Here we presented the analysis of the power spectra of different quantities towards a single observed filament  and a single simulated filament. 
 Further detailed analysis of the column density and B-field fluctuations along  statistical samples of both observed and simulated filaments are necessary to understand better the underlying physical processes captured by the observations.

\section{Summary and conclusions}\label{Summary}

In this paper,  we present the first BISTRO results of the JCMT/SCUBA-2/POL-2 observations 
at 850\,$\mu$m towards the NGC 6334 high-mass star-forming hub-filament network. 
Our  
main results may be summarized as follows:

 \begin{itemize}
\item BISTRO observations provide for the first time the submm polarization properties at high angular resolution (14\arcsec or 0.09\,pc at  1.3\,kpc distance) and high SNR ({\it SNR}$(I)\gg25$ and {\it SNR}$(PI)\gg3$) towards the $10\,{\rm pc}\times2\,$pc %
hub-filament network
hosting a 10\,pc-long ridge, sub-filaments, and core-hubs
 where stellar clusters and high-mass stars are forming (Figs.\,\ref{IQUmaps}, \ref{PAPF}, \ref{PI_PA1}, and  \ref{PI_PA2}). 
\\

\item The observations show a  positive correlation between the polarized intensity $PI$ and the total intensity  $I$ ($PI\propto I^{\,0.65\pm0.02}$), where  both $PI$ and $I$ span $\sim3$ orders of magnitude in polarized and total emission, respectively (Fig.\,\ref{scatPlot}-$left$). 
 This correlation, however, exhibits  a scatter by more than one order of magnitude in $PI$ for  $I$ values 
 spanning $\sim3$ orders of magnitude. %
The observations show an overall  negative correlation between the polarization fraction $PF$ and $I$  over $\sim3$ orders of magnitude in $PF$ 
with $PF\propto I^{\,-0.35\pm0.02}$ (Fig.\,\ref{scatPlot}-$middle$).  As for $PI$,  a large scatter in $PF$  (of $\sim 2$ orders of magnitude) is observed for a given value of $I$, with $PF$ values as high as $6\%$ for 
$\nhh\sim10^{23}\NHUNIT$ (see also, Fig.\,\ref{Filhub}-$top$). This indicates that the dust grains remain well-aligned with the B-field to the highest observed column densities.  \\

   \item 
 The observed slope between the $PF$ and $I$ values (see previous paragraph) is 
shallower than the  $-1$ slope expected from the progressive loss of grain alignment  due to the attenuation of the interstellar radiation field with increasing column density (i.e., from the RAT theory) and shallower than the $-0.5$ slope expected from a purely turbulent B-field (see Fig.\,\ref{Filhub}-$top$). 
In Sect.\,\ref{disc:feedback}, we suggest  that the large $PF$ values  observed towards  NGC 6334  may be tracing  ongoing  feedback  effects on the observed polarized dust emission that may result from the combination of 1) an enhanced radiation field from the newborn (high-mass) stars, 2) the re-organization of the structure of the B-field on the scale of  \hii\ regions (within      compressed shells),  
and 3) the increase of the B-field strength 
ordering its structure and reducing  depolarization due to the integration of  emission along the LOS and within the beam.\\

\item The ridge and the clump-hub structure are surrounded by sub-filaments (Fig.\,\ref{SkelFig}). In Sect.\,\ref{alongsubfil}, we investigate the variation of the polarization and physical properties along the sub-filaments  from their outer ($out$) to their inner ($in$) parts.
 {\rev In the outer parts, the POS B-field show mostly perpendicular or random orientation with respect to the  sub-filament crests.} In their inner parts, where the sub-filaments are merging with  the clump-hub, the B-field  is parallel to their crests  (cf., Fig.\,\ref{SkelFig2}). 
  This change of relative orientation along the sub-filaments may be  due to material flowing along their crests onto the ridge and  hubs. {\rev These flows of matter may drag the field lines, aligning them with the filament crests}  (c.f., Sect.\,\ref{disc:hub}). 
   The polarization fractions and  the dispersion of the POS B-field angle are smaller on average in the inner parts of the sub-filaments compared to their outer parts (cf., Fig.\,\ref{InOut_histo}).\\ 

\item We also observe a variation of the stability of these sub-filaments (cf., Fig.\,\ref{InOut_histoCrit}).
The kinetic virial parameter $\alpha_{\rm vir}^v$ is on average smaller in the inner parts compared to the outer parts, while the magnetic virial parameter 
 $\alpha_{\rm vir}^B$  increases (by a factor of $\sim1.8$ on average) from $out$ to $in$. In the outer parts,  $\alpha_{\rm vir}^B\sim0.5\pm0.3<1$ suggesting that the magnetic tension alone is not enough to balance gravity while the inner parts of the sub-filaments seem to be in a magnetic critical balance with $\alpha_{\rm vir}^B\sim1$. 
When both the magnetic and kinetic (thermal and turbulent) supports are combined, the estimated $\alpha_{\rm vir}^{v,B}\gtrsim1$ values suggest that the sub-filaments are in an overall balance between the effective pressure forces and gravity (Sect.\,\ref{alongsubfil}). This behaviour is also seen along the ridge (Sect.\,\ref{alongridge}). \\ 

\item We identify a signature in the polarized emission towards two sub-filaments (with two spatially separated crests) that seem to be merging (as seen in dust continuum) to form a filament with a single crest (Fig.\,\ref{FilMergeProfile}).  We see the gradual  increase of the total intensity due to the merging of the two sub-filaments and detect a drop in the polarized intensity  towards the crest of the merged filaments. We suggest that this drop in $PI$ (and $PF$), while $I$ increases, is due to the depolarization from the integration along the LOS and within the beam of the unresolved B-field structures resulting from the merging of   two different (partly independent) B-field structures along each of the initially separated filaments. 
This increase in $I$ towards the crests of the merged-filaments is larger than a simple addition of the intensity  of the two separated-filaments, suggesting 
 efficient infall of matter onto the merged-crest. This behaviour may be indicative of an  early stage of a core-hub formation.\\

\item We compare the properties observed towards NGC 6334 with two other regions observed as part of the BISTRO survey: Oph-A and NGC 1333 (Fig.\,\ref{compPFI}). While these  three regions have  physical sizes that span 2 orders of magnitude in spatial scales, from $\sim0.2\,$pc to  $\sim10\,$pc, they all show  a similar scatter of  $PF$ values, of more than 2 orders of magnitude in $PF$ (from $\sim0.2\%$ to $\sim20\%$). This scatter  is also observed for each $I$ value, where a large scatter in $I$ is also observed for a  given value of $PF$. 
These observed similarities 
  suggest that the self-similar description of the physical processes in a possibly fractal magnetized ISM also has an imprint in the dust polarized thermal emission (c.f., Sect.\,\ref{disc:comp}).  \\

\item  We present the first power spectrum of the POS B-field angles observed along the NGC 6334 ridge crest  and compare it to the power spectrum of the intensity emission (cf., Fig.\,\ref{PS_crest3_5} and Sect.\,\ref{powspecSection}).  The 
former  is well represented by a power law function with a slope $\sim20\%$ shallower than that of the latter. We compare this  result with  the power spectrum of filaments obtained from  an  MHD numerical simulation of massive filament formation  (Fig.\,\ref{BstrengthPSsimu})
and find consistent values.
These preliminary results
suggest that the structure of the column densities (line mass) and POS B-field angles along the  NGC 6334 filament may be each inherited from the filament formation process and have a coupled evolution (c.f., Sect.\,\ref{powerSpec}). 
Understanding the theoretical link between the power spectra of column density and magnetic field (POS angle and strength) may lead to a new and independent method to infer the B-field strength from polarization observations. 

 \end{itemize}

\noindent
{\rev
In this  paper we have presented a first look at these rich BISTRO observations towards NGC 6334.  The different topics addressed in Sect.\,\ref{disc}  will be investigated in dedicated and in-depth forthcoming studies (e.g., Tahani et al. in prep.).  To assess the reliability of the scenario presented above regarding  the change  of the properties along the sub-filaments and towards the clump-hub, large-scale molecular line mapping (at the appropriate resolutions) will be needed, 
to better understand the dynamics of these structures.
The combination of the B-field structure observed at the BISTRO scale with that  at higher angular resolution will also be very valuable to study the role of the B-field in the matter assembly and fragmentation processes that lead to the formation  of massive stars. 
Quantitative comparisons with dedicated MHD numerical simulations will also be important to  understand  the  physical processes at play in theses high-mass star-forming hub-filament systems. 
}

\begin{acknowledgements}
D.A. acknowledges support by FCT/MCTES through national funds (PIDDAC) by the grants UID/FIS/04434/2019 $\&$ UIDB/04434/2020.
This research is partially supported by Grants-in-Aid for Scientific Researches from the Japan Society for Promotion of Science
(KAKENHI 19H0193810).
C.L.H.H. acknowledges the support of the NAOJ Fellowship and JSPS KAKENHI grants 18K13586 and 20K14527.
P.M.K. acknowledges support from the Ministry of Science and Technology (MOST) grants MOST 108-2112-M-001-012
and MOST 109-2112-M-001-022 in Taiwan, and from an Academia Sinica Career Development Award.
C.W.L. is supported by the Basic Science Research Program through the National Research Foundation of Korea (NRF) funded by the Ministry of Education, Science and Technology (NRF-2019R1A2C1010851).
W.K. is supported by the New Faculty Startup Fund from Seoul National University and by the Basic Science Research Program through the National Research Foundation of Korea (NRF-2016R1C1B2013642).
This program is part of the JCMT BISTRO Large Program observed under project code M17BL011.
Team BISTRO-J is in part financially supported by 260 individuals. 
The James Clerk Maxwell Telescope is operated by the East Asian Observatory on behalf of The National Astronomical Observatory of Japan; Academia Sinica Institute of Astronomy and Astrophysics; the Korea Astronomy and Space Science Institute; Center for Astronomical Mega-Science (as well as the National Key R$\&$D Program of China with No. 2017YFA0402700). Additional funding support is provided by the Science and Technology Facilities Council of the United Kingdom and participating universities in the United Kingdom, Canada, and Ireland. 
Additional funds for the construction of SCUBA-2 were provided by the Canada Foundation for Innovation.
The authors wish to recognize and acknowledge the very significant cultural role and reverence that the summit of Maunakea has always had within the indigenous Hawaiian community. We are most fortunate to have the opportunity to conduct observations from this mountain.
\end{acknowledgements}

\bibliographystyle{aa}
\bibliography{aa} 


\begin{appendix}
\section{Details on the quality assessments of the BISTRO data of NGC 6334}\label{App1}
    
 \begin{figure*}[!h]
   \centering
     \resizebox{17.cm}{!}{
\includegraphics[angle=0]{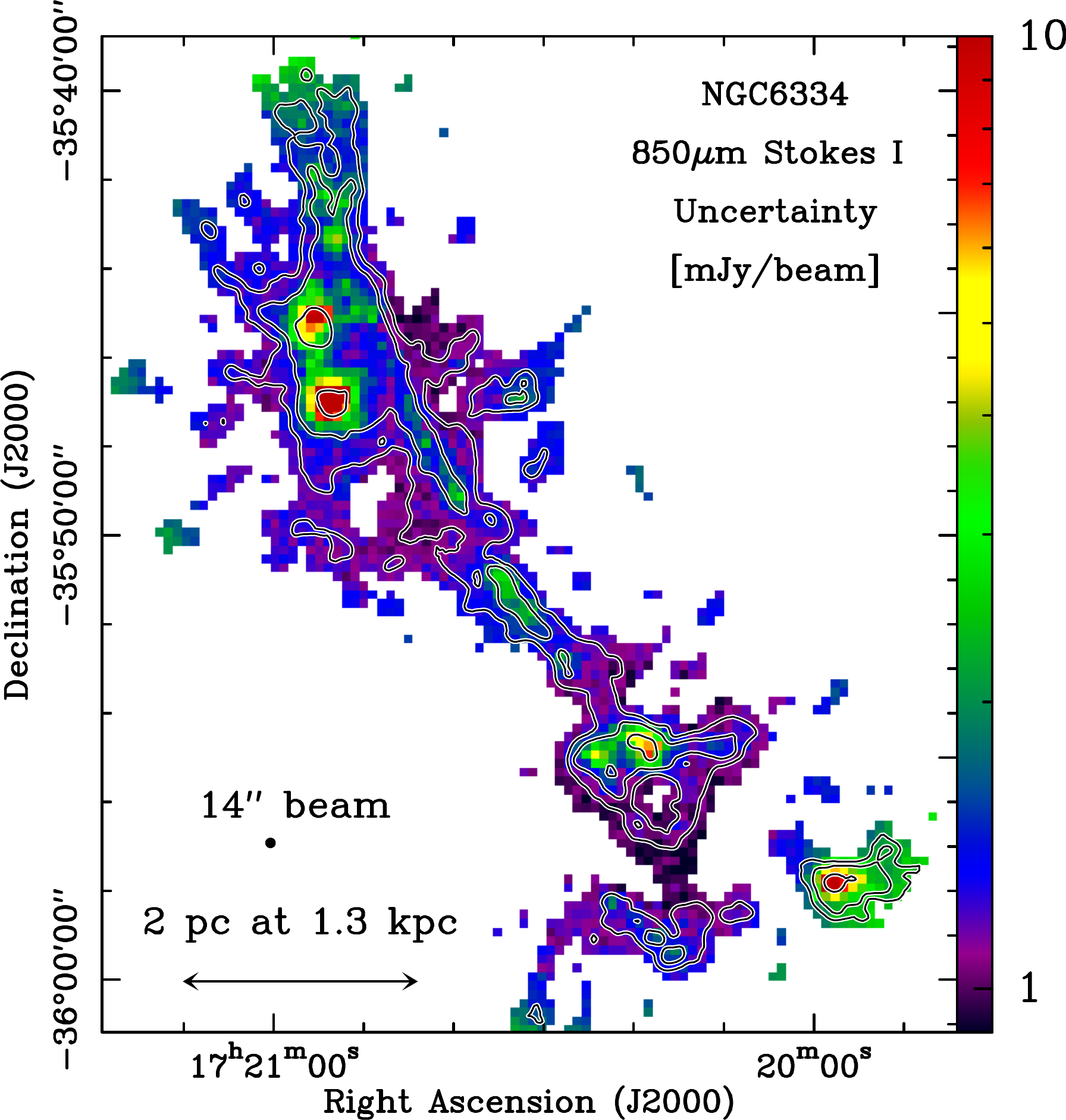}
\includegraphics[angle=0]{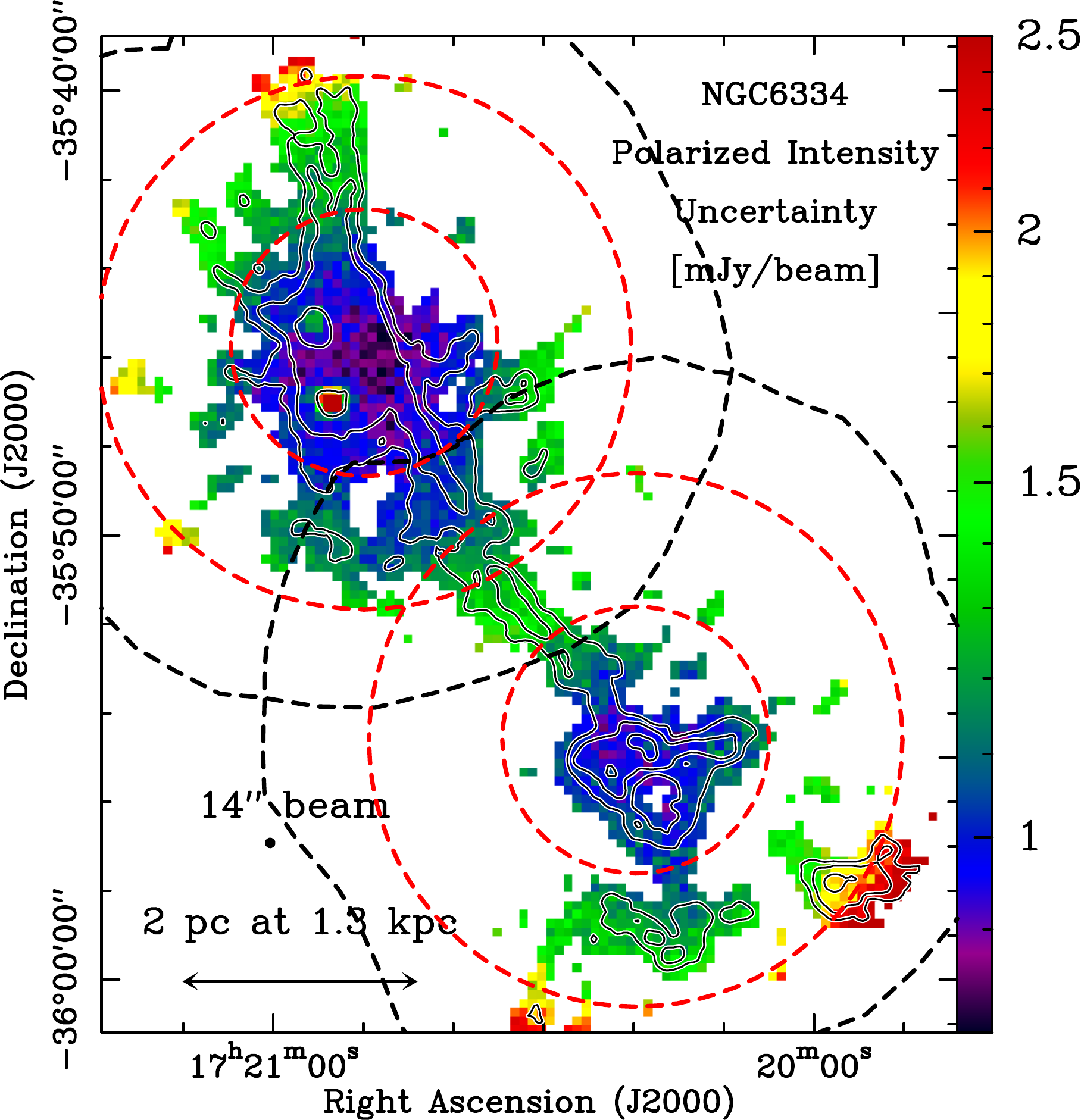}}
  \resizebox{17.cm}{!}{
\includegraphics[angle=0]{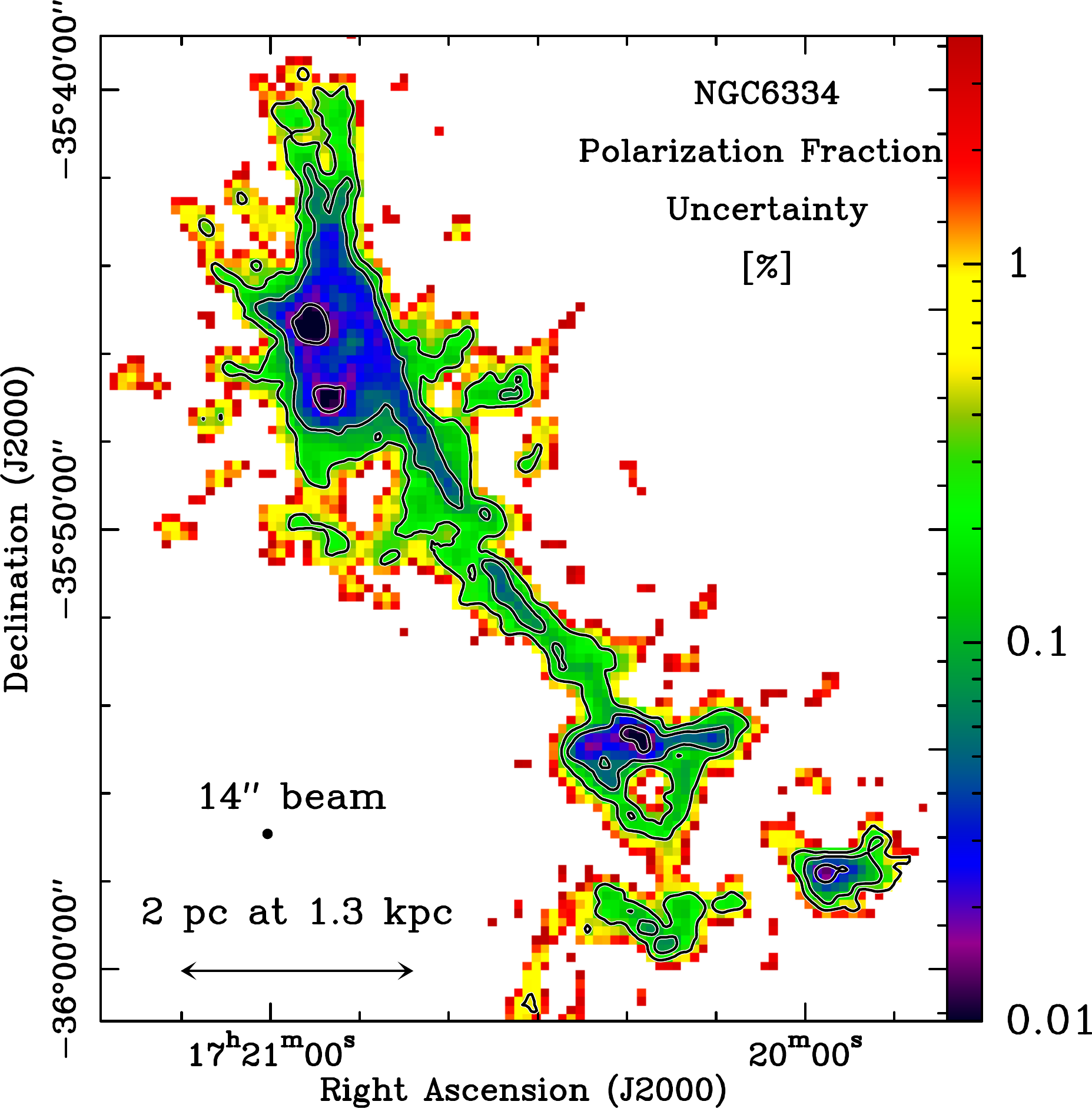}
\includegraphics[angle=0]{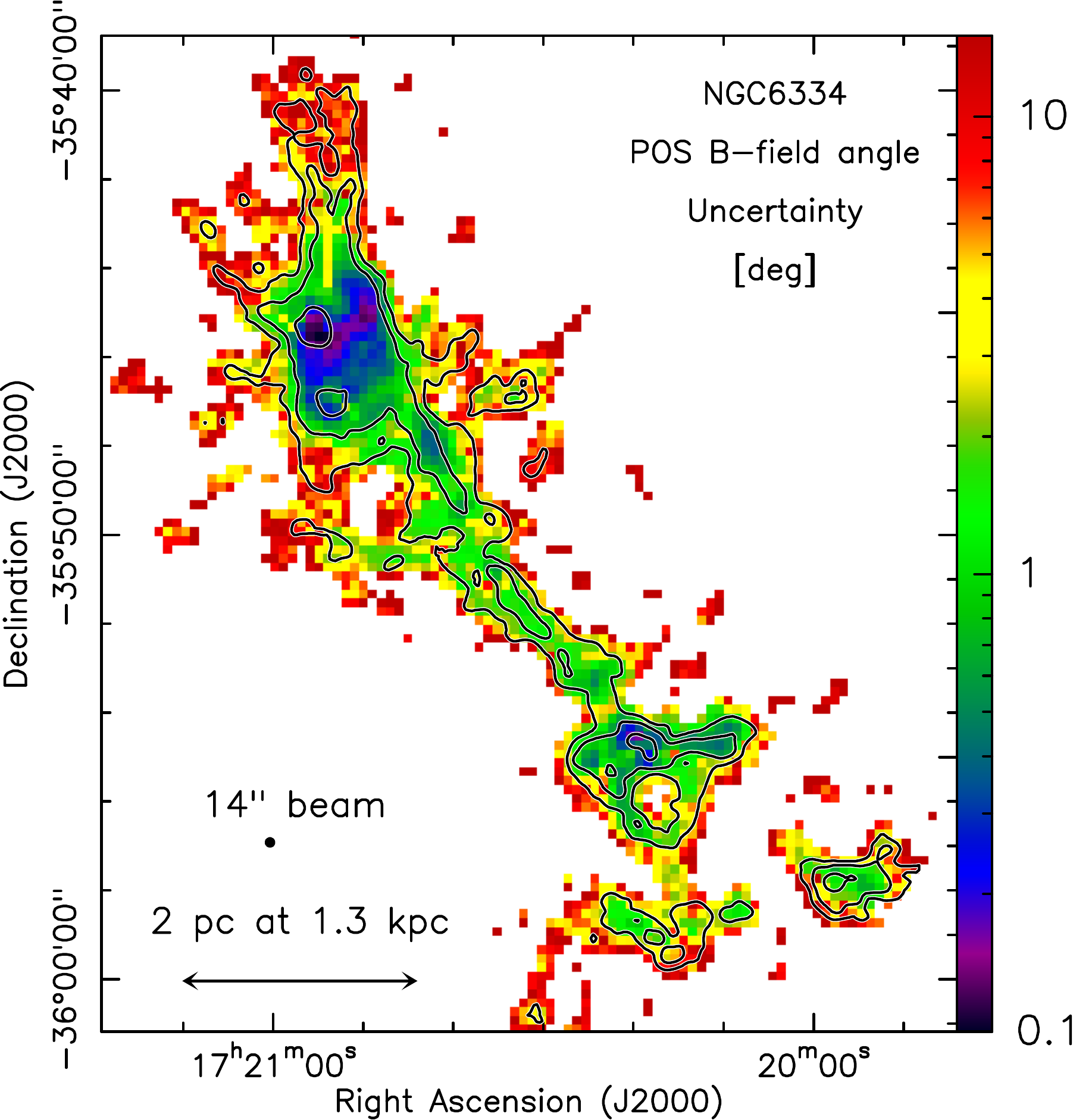}
}
\vspace{-0.1cm}
  \caption{Maps showing the distribution of the uncertainties derived for   Stokes $I$, $PI$, $PF$, and $\chi_{B_{\rm POS}}$ from left to right and top to bottom. The HPBW resolution of these maps is $14\arcsec$.
  The same emission area is plotted for all the four maps corresponding to {\it SNR}$(I)>25$. The  contours, which are the same for all the plots, correspond to 
 $I=0.4, 1.4,$ and $8\,$Jy\,beam$^{-1}$ or $\nhh\sim(2, 6,\, $and$\, 37) \times10^{22}\NHUNIT$ 
(as in Fig.\,\ref{IQUmaps}). 
  The black dashed circles over-plotted on the $PI$ map (top right) indicate the limits of the two observed DAISY sub-fields 
  centred 
    on (17:20:50.011, $-$35:45:34.33) and  (17:20:19.902, $-$35:54:30.45) in (R.A. J2000, Dec. J2000) for the North and South fields, respectively. The small and large red dashed circles have sizes of $6\arcmin$ and $12\arcmin$. 
}          
  \label{Err_maps}
\end{figure*}

\begin{figure*}[!h]
   \centering
     \resizebox{19.cm}{!}{\hspace{-0.5cm}
\includegraphics[angle=0]{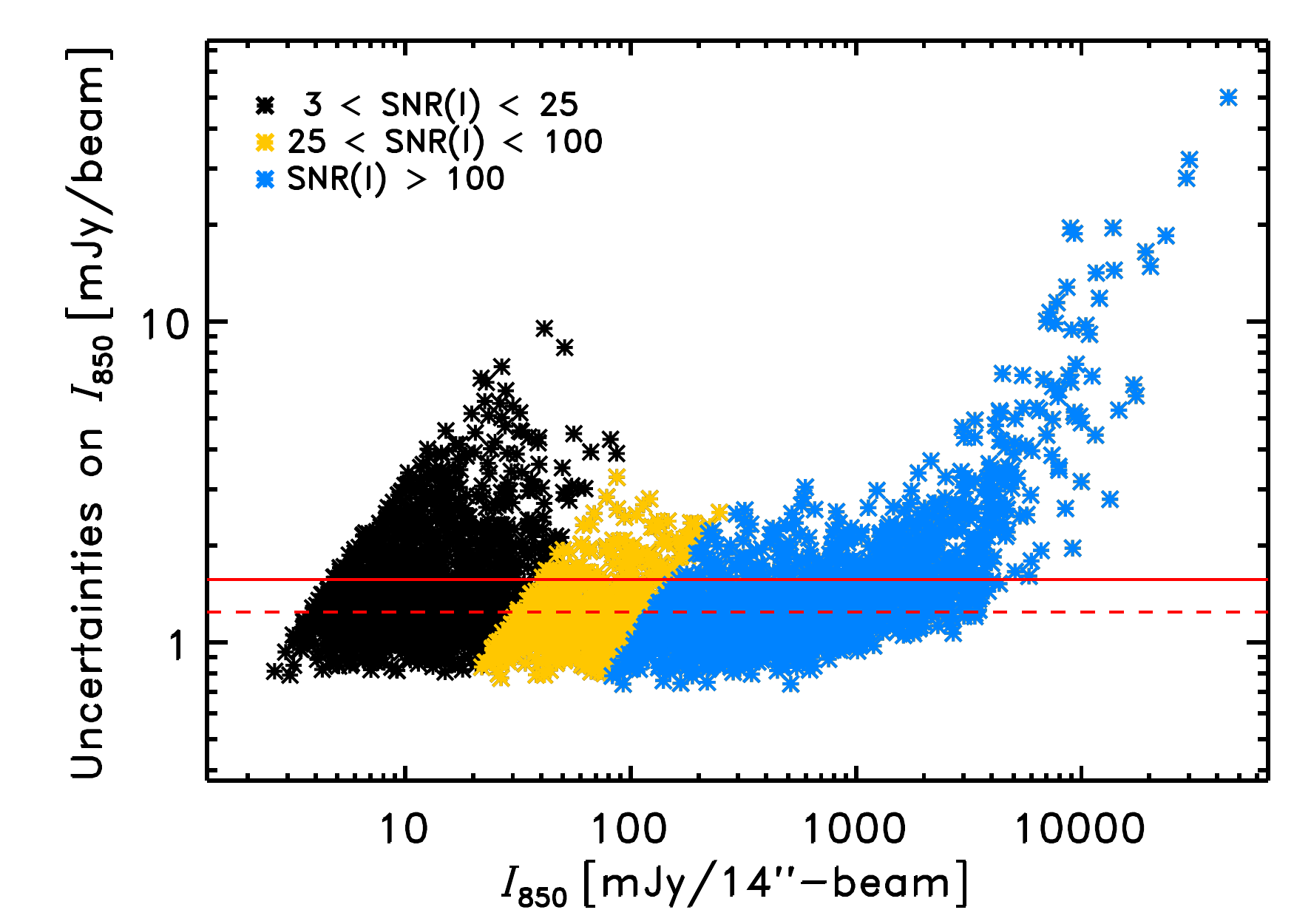}
\includegraphics[angle=0]{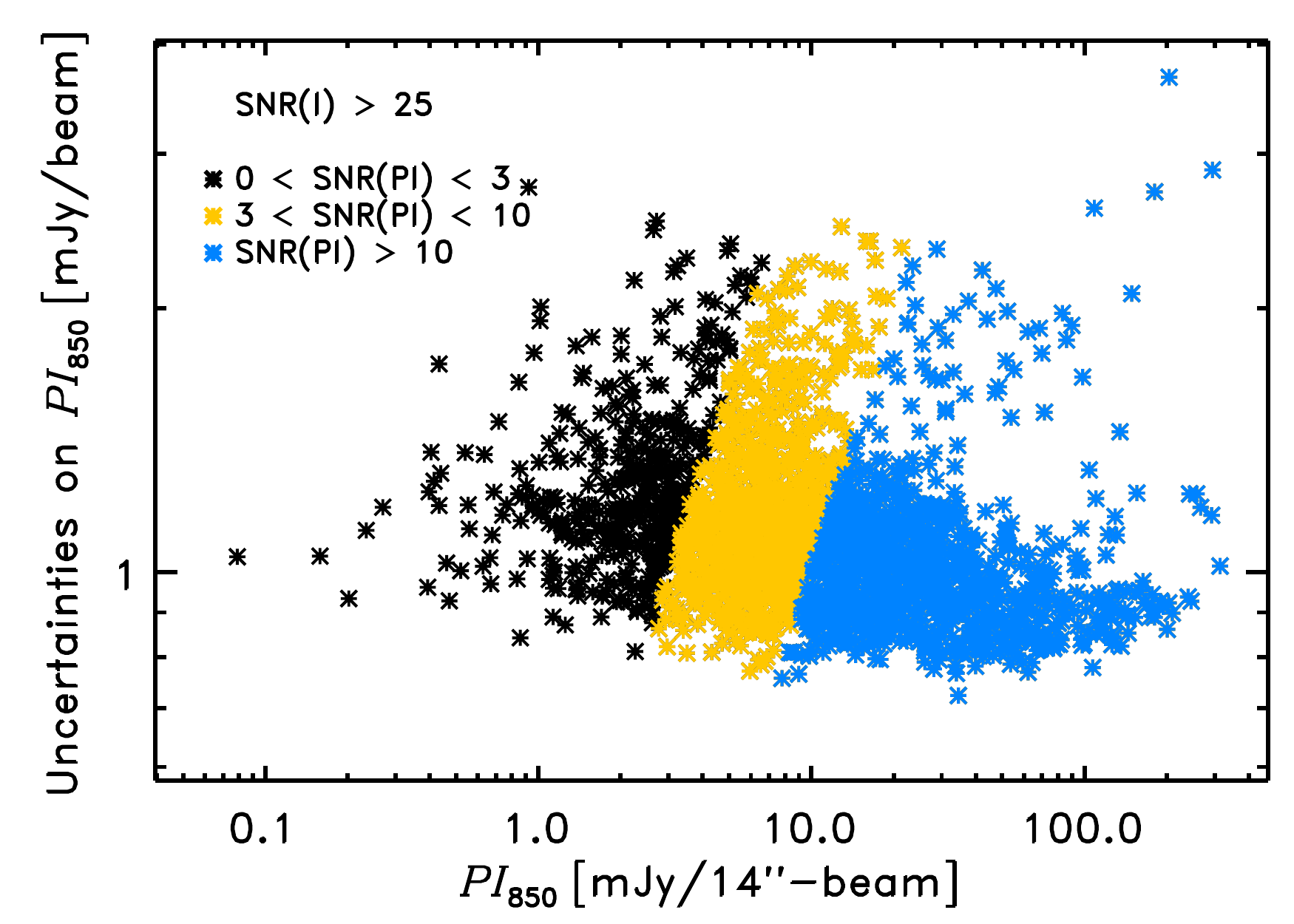}
\includegraphics[angle=0]{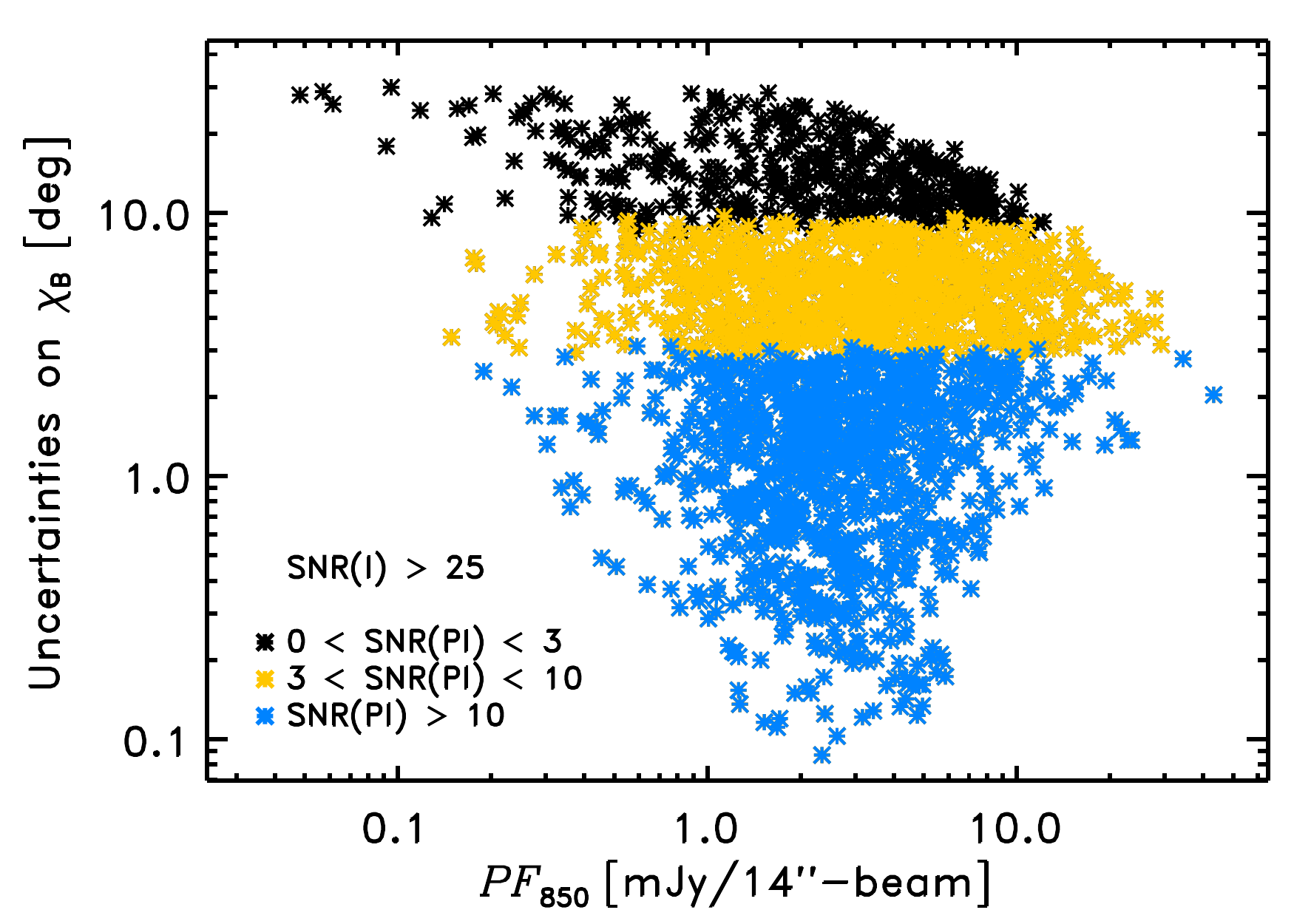}
}
\vspace{-0.2cm}
  \caption{{\it Left:} Measurement uncertainties on Stokes $I$ as a function of the signal on $I$. The black, yellow, and blue data points correspond to 3\,$<\,${\it SNR(I)}$\,<$\,25, 25\,$<\,${\it SNR(I)}$\,<$\,100, and {\it SNR(I)}$>$100, respectively. The dashed and solid red lines show the median (1.24\,mJy/beam) and mean (1.57\,mJy/beam) uncertainties, respectively.  
    {\it Middle:} Uncertainties on  $PI$ as a function of $PI$ for {\it SNR(I)}$>25$. The black, yellow, and blue data points correspond to 0\,$<\,${\it SNR(PI)}$\,<$\,3, 3\,$<\,${\it SNR(PI)}$\,<$\,10, and {\it SNR(PI)}$>$10, respectively. 
   {\it Right:} Uncertainties on the POS B-field angle, $\chi_{B_{\rm POS}}$, as a function of  $PF$ for {\it SNR(I)}$>25$.  
    The colours are the same as in the middle plot.  
   }          
  \label{Err_scatter}
\end{figure*}

 \begin{figure*}[!h]
   \centering
     \resizebox{17.cm}{!}{
\includegraphics[angle=0]{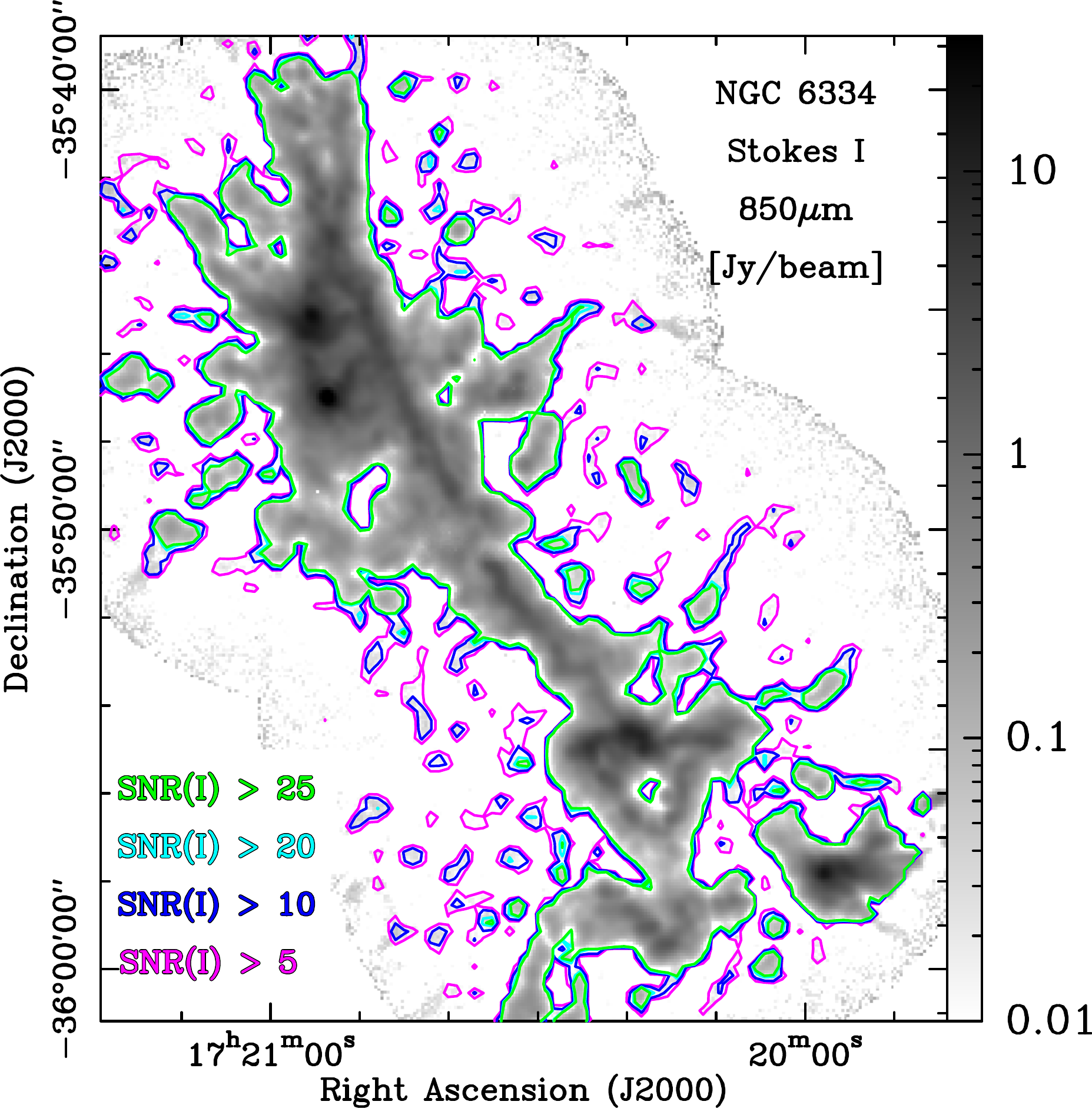}
\hspace{0.5cm}
\includegraphics[angle=0]{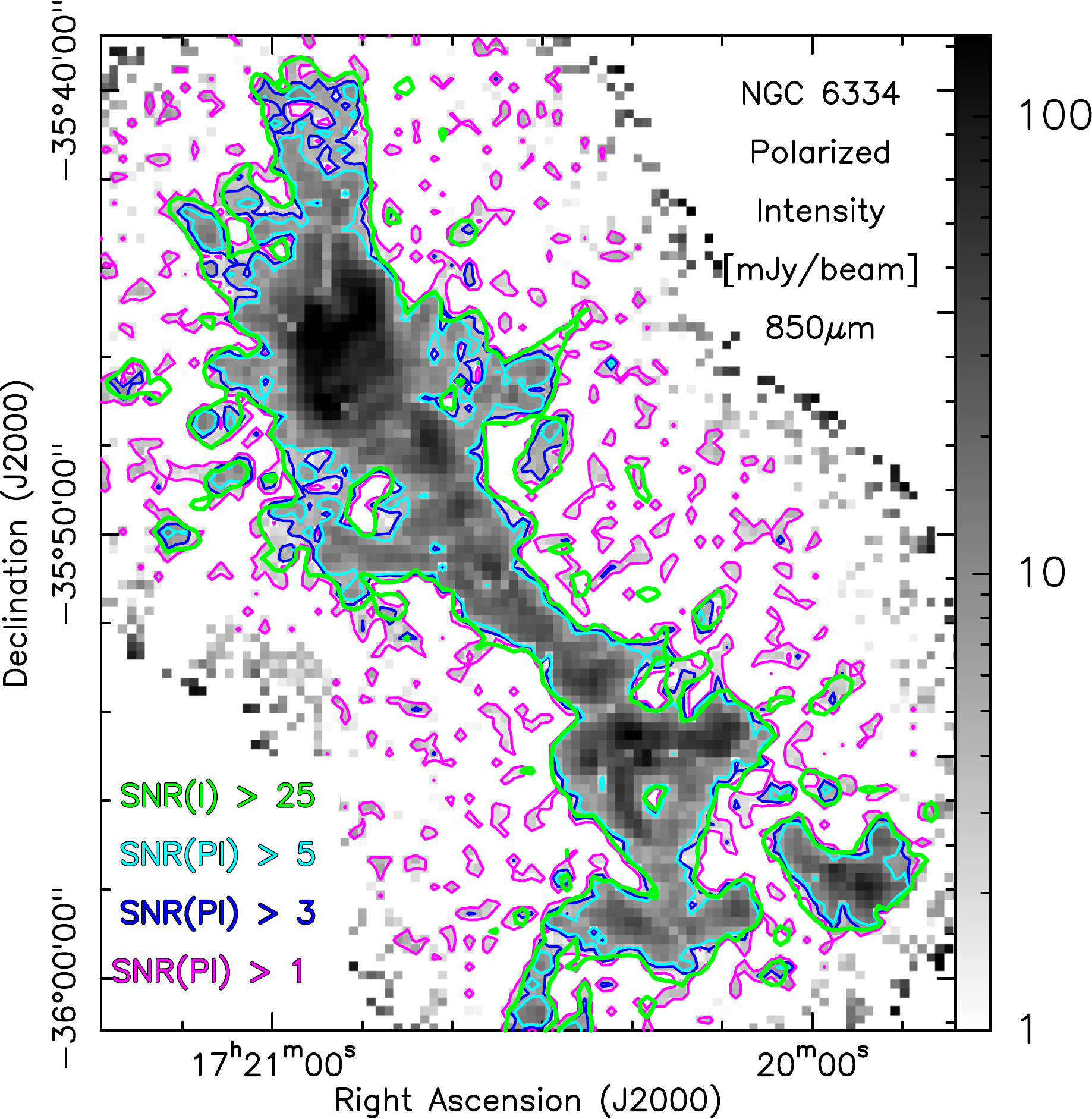}}
\vspace{-0.1cm}
  \caption{ {\it Left:} 
   Stokes  $I$ map at 850\,$\mu$m  observed with the JCMT SCUBA-2/POL-2  towards NGC 6334.
    The HPBW resolution of these maps is $14\arcsec$. The pixel size of the map is $4\arcsec$.
    The magenta, blue, cyan, and green contours are {\it SNR}$(I)>5, 10, 20,$ and 25, respectively. 
    The {\it SNR}$(I)=25$ correspond to $I\sim30\,$mJy\,beam$^{-1}$ or $\nhh\sim1.5\times10^{21}\NHUNIT$.
     {\it Right:} Map of the polarized emission $PI$. The HPBW resolution of these maps is $14\arcsec$. The pixel size of the map is $12\arcsec$. 
       The magenta, blue, and cyan  contours are {\it SNR}$(PI)>1, 3,$ and 5, respectively.
       The green contour corresponds to   {\it SNR}$(I)>25$ and is the same as in the left hand side panel. 
}          
  \label{SNRcontours}
\end{figure*}

\begin{figure*}[!h]
   \centering
 \resizebox{19.cm}{!}{\hspace{-0.5cm}
\includegraphics[angle=0]{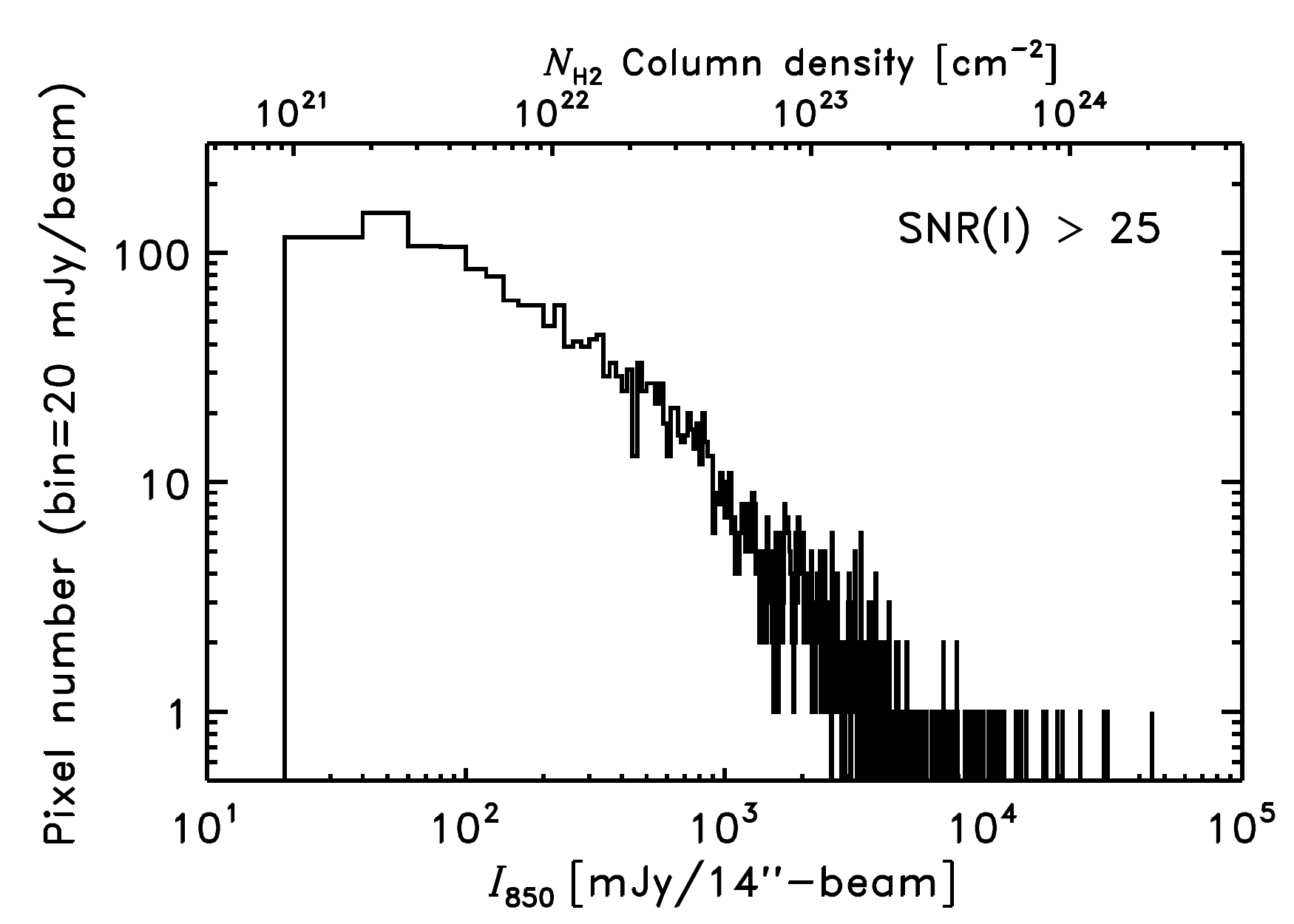}
\includegraphics[angle=0]{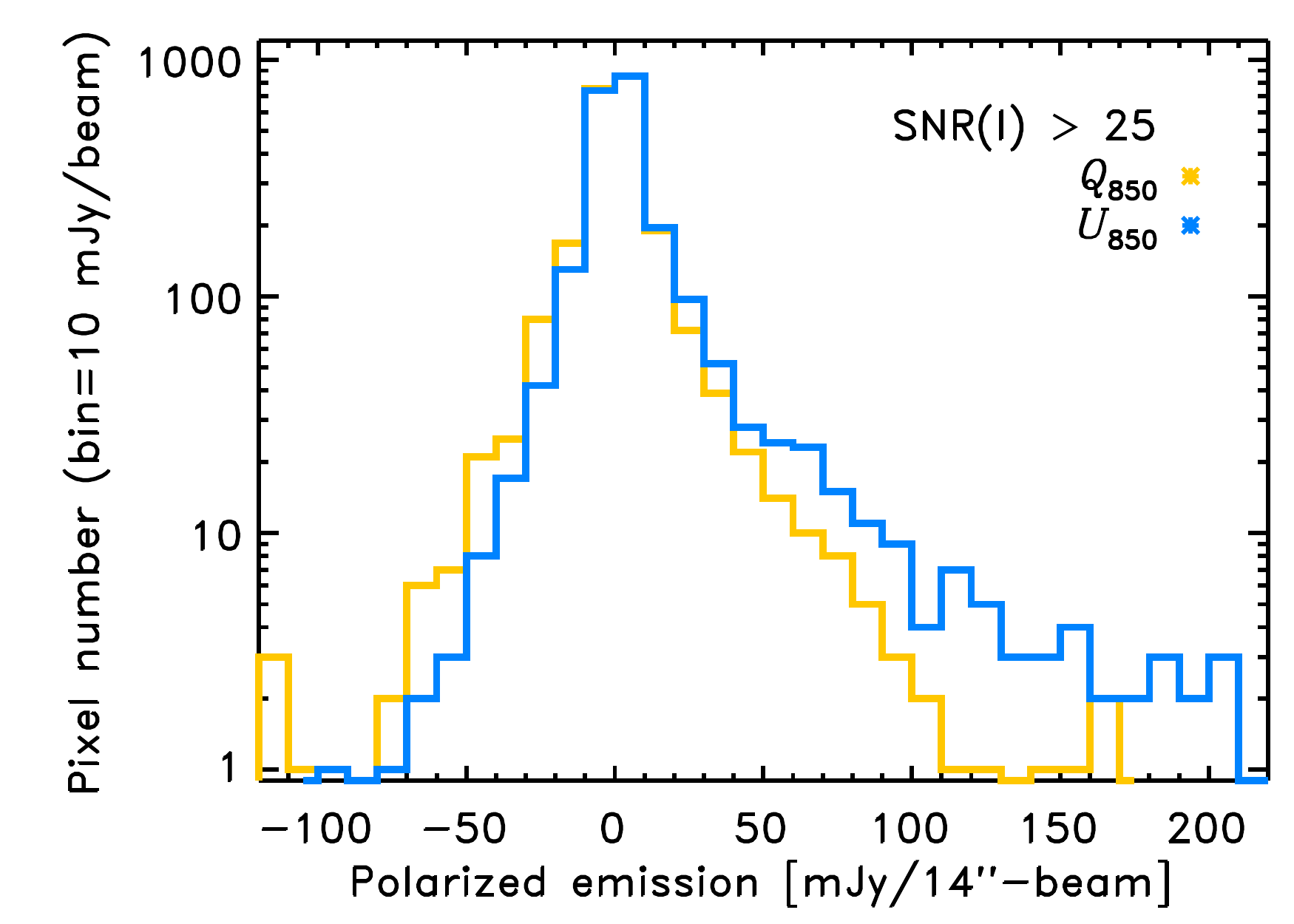}
\includegraphics[angle=0]{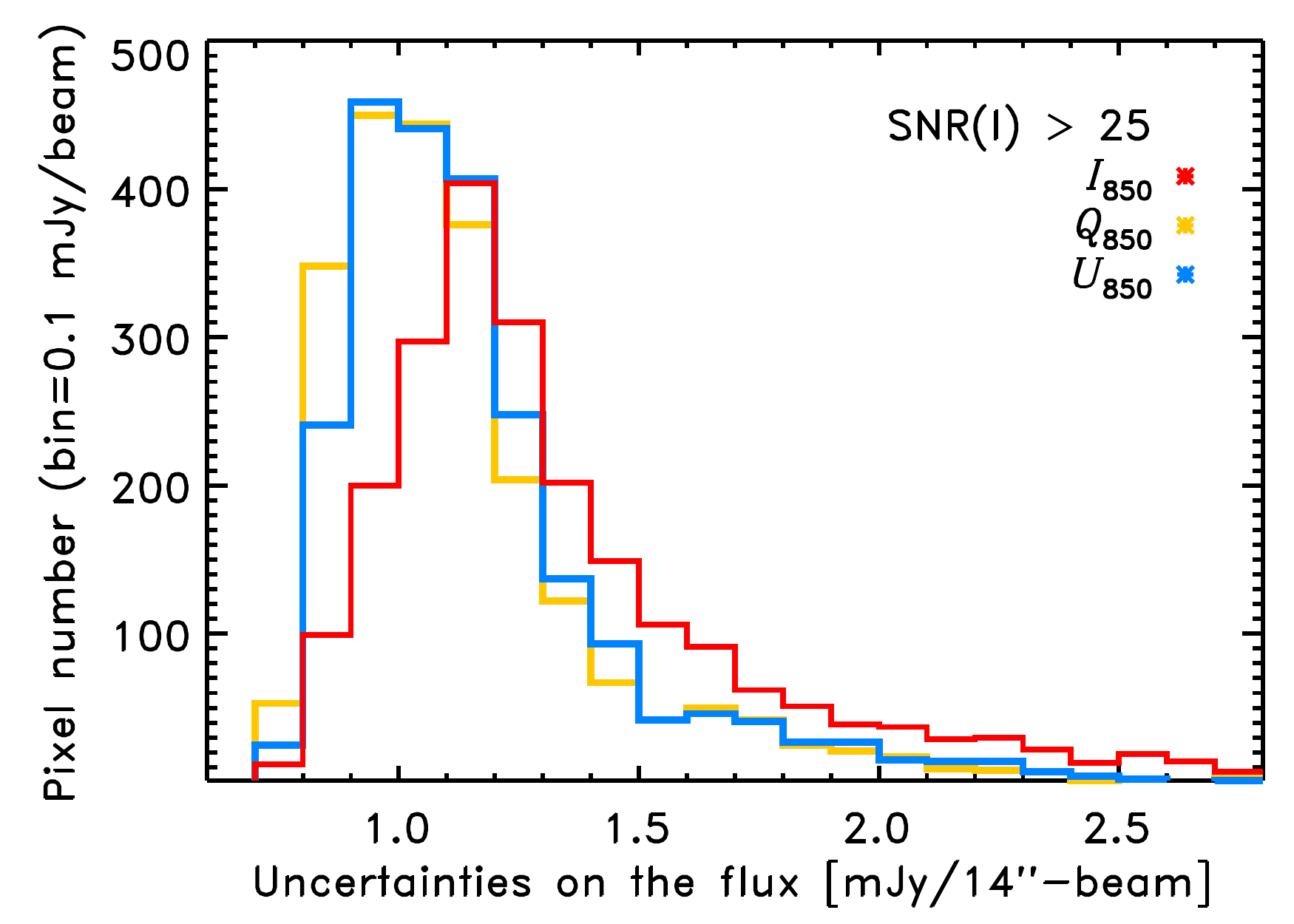}
}
\vspace{-0.2cm}
  \caption{Distribution of the signal of Stokes $I$ (left), $Q$ and $U$ (middle) for  {\it SNR}$(I)>25$. The measured uncertainties on the  three Stokes parameters are shown on the right hand  plot. 
}          
  \label{IQUerr_histo}
\end{figure*}

 \begin{figure*}[!h]
   \centering
     \resizebox{18.cm}{!}{ \hspace{-2cm}
\includegraphics[angle=0]{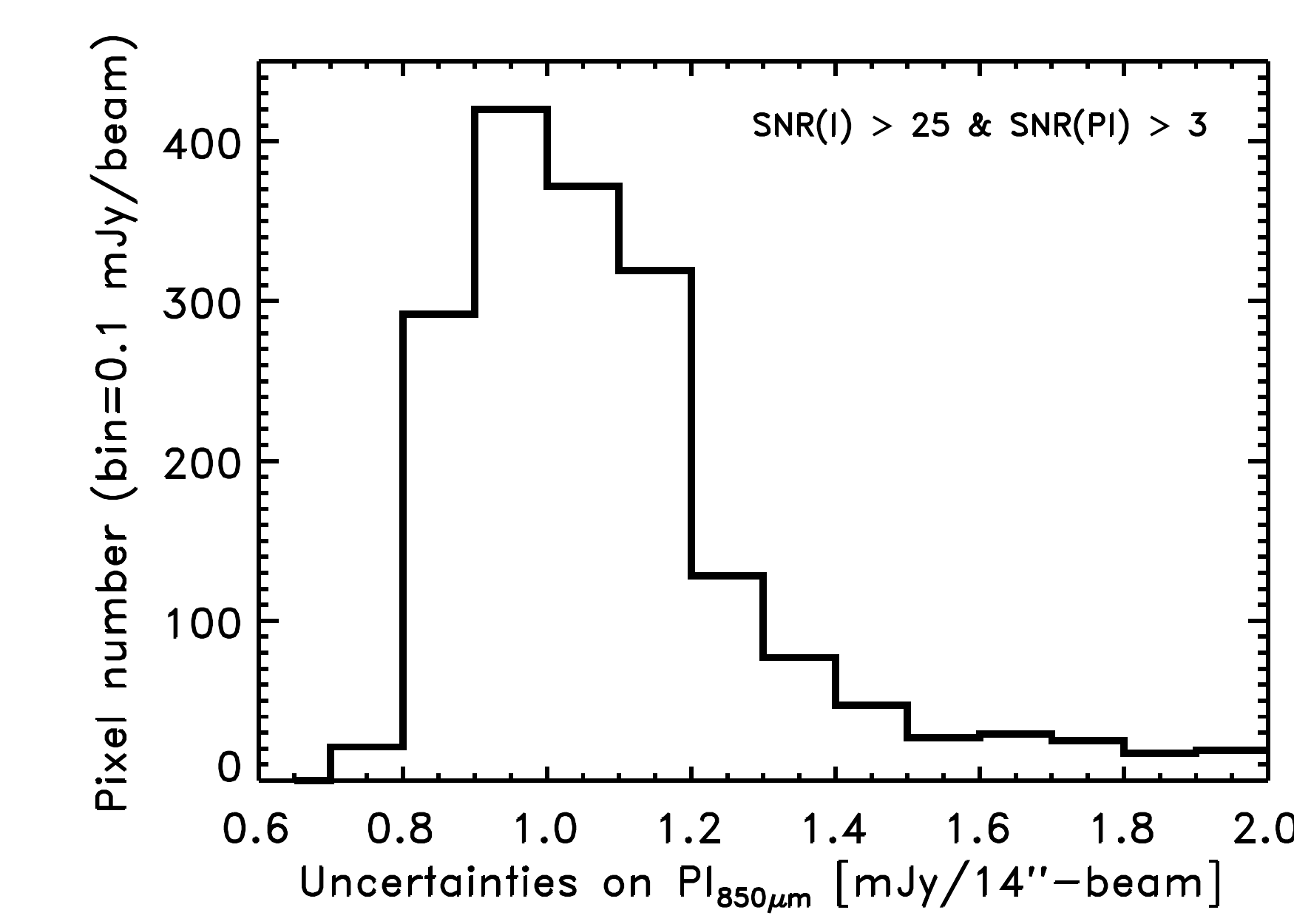}
\includegraphics[angle=0]{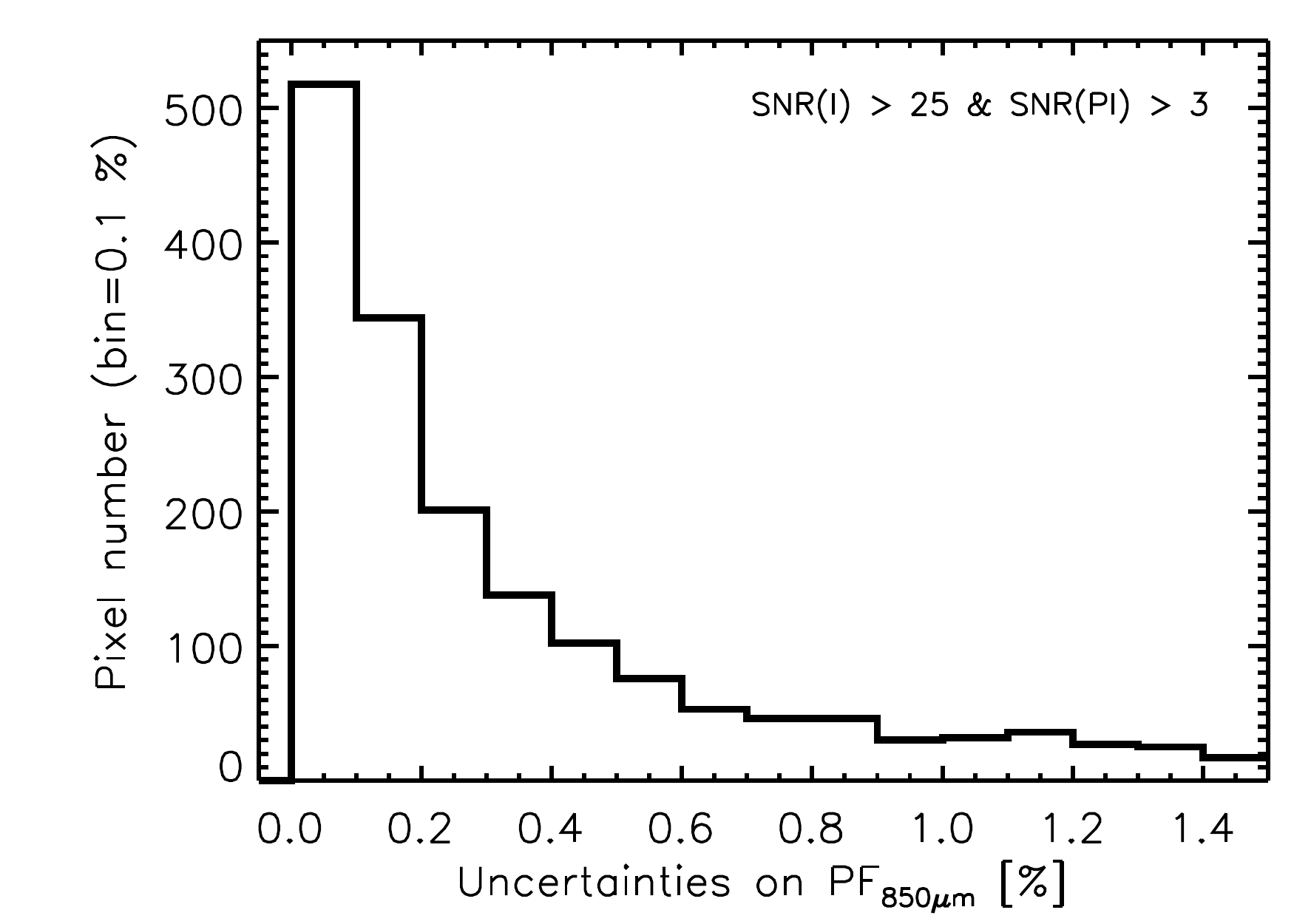}
\includegraphics[angle=0]{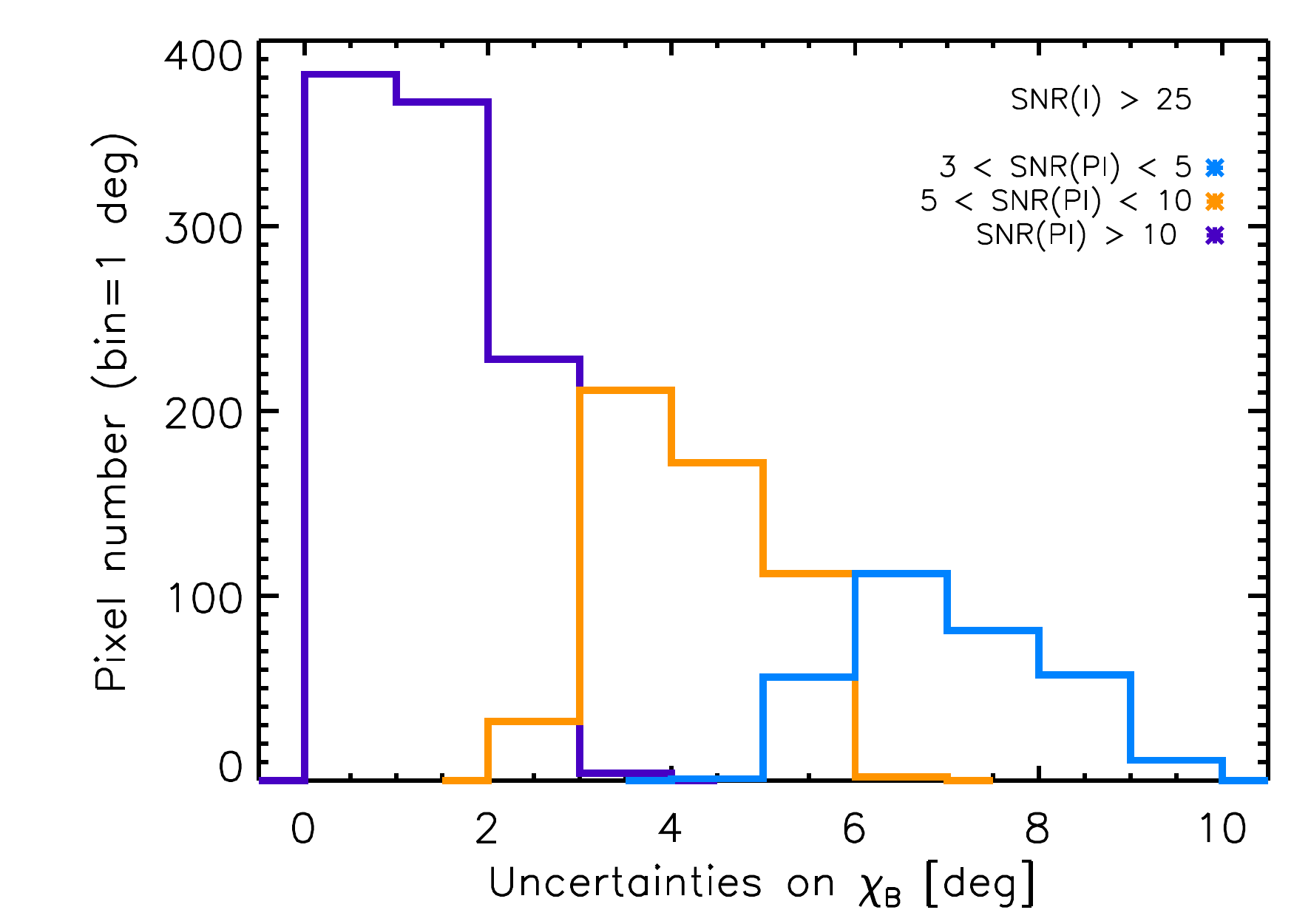}
}\vspace{-0.2cm}
  \caption{  Distributions of the uncertainties on $PI$ (left), $PF$ (middle), and $\chi_{B_{\rm POS}}$ (right) for {\it SNR}$(I)>25$ and {\it SNR}$(PI)>3$ corresponding to the maps shown in Fig.\,\ref{Err_maps}. 
  On the right hand side plot the different colors are as follow: blue for 3\,$<\,${\it SNR(PI)}$\,<$\,5, orange for  5\,$<\,${\it SNR(PI)}$\,<$\,10, and purple for {\it SNR(PI)}$>$10. 
}          
  \label{Err_histo}
    \end{figure*}

To complement the presentation of the data of the JCMT/SCUBA-2/POL-2 BISTRO observations at 850\,$\mu$m towards the NGC 6334 field (see Sect.\,\ref{obs}), we here present maps of the uncertainties of the total intensity $I$, the polarized intensity $PI$, polarization fraction $PF$, and POS B-field angle. 
These maps show the statistical uncertainties derived from the {\it pol2map}  data reduction pipeline reprojected onto 12\arcsec-pixel size grids (cf.,\,\url{http://starlink.eao.hawaii.edu/docs/sc22.htx/sc22.html}).  
Two DAISY maps have been observed towards the North and the South of the field and combined 
into a $\sim 10\,{\rm pc}\times2\,$pc mosaic. The limits of the two sub-fields  are indicated on the top-right hand side panel of Fig.\,\ref{Err_maps}. The $6\arcmin$-diameter circles indicate the area where the observation pattern of the POL2-DAISY mode yields  uniform noise coverage. 
Towards these areas of the map, the data have the lowest uncertainties (see Fig.\,\ref{Err_maps}). The $I$ map shows larger uncertainties towards the bright compact sources due to the increase of pointing and calibration uncertainties towards steep intensity gradients.  We see larger uncertainties towards the edge of the maps and in the regions between the two sub-fields. These larger uncertainties on $I$ ($\delta I$) can be seen on the right-hand side  panel of Fig.\,\ref{Err_scatter}, which  shows a scatter plot of $\delta I$ as a function of $I$. 
 For {\it SNR}$(I)>3$, $\delta I$ is mostly uniform with a mean value of $\sim 1.6\,$mJy/14\arcsec-beam. We see larger  $\delta I$ for the extreme $I$ values $\gtrsim 8\,$Jy/14\arcsec-beam towards the bright compact sources (c.f., Fig.\,\ref{Err_maps}). 
More than $65\%$ and $70\%$ (2293 pixels) of the emission for {\it SNR}$(I)>3$ and {\it SNR}$(I)>5$, respectively,  correspond to {\it SNR}$(I)>25$.
 For {\it SNR}$(I)>25$ (2293 pixels),  $79\%$ of the data points (1822 pixels) have {\it SNR}$(PI)>3$. Only $4\%$ (81 pixels) of the total pixels with {\it SNR}$(PI)>3$ have {\it SNR}$(I)<25$.  
This excellent overall SNR can be also seen on Fig.\,\ref{SNRcontours} showing  SNR contours over-plotted on the $I$ and $PI$ maps. For the analysis presented in this paper, we thus select data points with {\it SNR}$(I)>25$ that correspond also mostly to data points with {\it SNR}$(PI)>3$. The selection criteria on {\it SNR}$(PI)$ is defined by the debiasing method that we use and which is shown to be reliable for {\it SNR}$(PI)>3$ \citep[c.f.,\,\ref{polParam},][]{Vaillancourt2006,Plaszczynski2014}.

         Figure\,\ref{IQUerr_histo} shows the distribution of the observed Stokes parameters for {\it SNR}$(I)>25$. The total intensity spans $\sim3$ orders of magnitude showing the large intensity dynamic range of these observations from $\nhh\sim10^{21}\NHUNIT$ to $\sim10^{24}\NHUNIT$. The Stokes $Q$ and $U$ values have asymmetric  distributions about 0 with different shapes. The $U$ distribution is  skewed towards positive values. This distribution is compatible with a dominant POS B-field mostly at $45^\circ+90^\circ=135^\circ$ with respect to the North direction (a peak in the histogram  of $\chi_{B_{\rm POS}}$ can be seen at $\sim130^\circ$, c.f, Fig.\,\ref{histoPlot}).   The mean and dispersion of the uncertainties of the  $I$, $Q$, and $U$ Stokes parameters are $\delta I=(1.6\pm1.8)$\,mJy/14\arcsec-beam, $\delta Q=(1.13\pm0.29)$\,mJy/14\arcsec-beam, and $\delta U=(1.6\pm0.30)$\,mJy/14\arcsec-beam (see Fig.\,\ref{IQUerr_histo}-right).  
    
Figure\,\ref{Err_histo} shows the distributions of the uncertainties of $PI$, $PF$, and 
$\chi_{B_{\rm POS}}$ for {\it SNR}$(I)>25$ and {\it SNR}$(PI)>3$.  The mean and dispersion of the uncertainties of these distributions  are $\delta PI=(1.1\pm0.3)$\,mJy/14\arcsec-beam, $\delta PF=(0.2\pm0.3)$\,mJy/14\arcsec-beam, and $\delta\chi_{B_{\rm POS}}=(7\pm1)^\circ, (4\pm1)^\circ,$ and $(1.3\pm0.8)^\circ$ for  data points with $3\,<\,${\it SNR(PI)}$\,<$\,5,  $5\,<\,${\it SNR(PI)}$\,<$\,10, and  {\it SNR(PI)}$>$10, respectively. 
 
\section{Column density and dust temperature  maps of the NGC 6334 filament network}\label{App_coldensTdust}

\begin{figure*}[!h]
   \centering
     \resizebox{16.cm}{!}{ 
\includegraphics[angle=0]{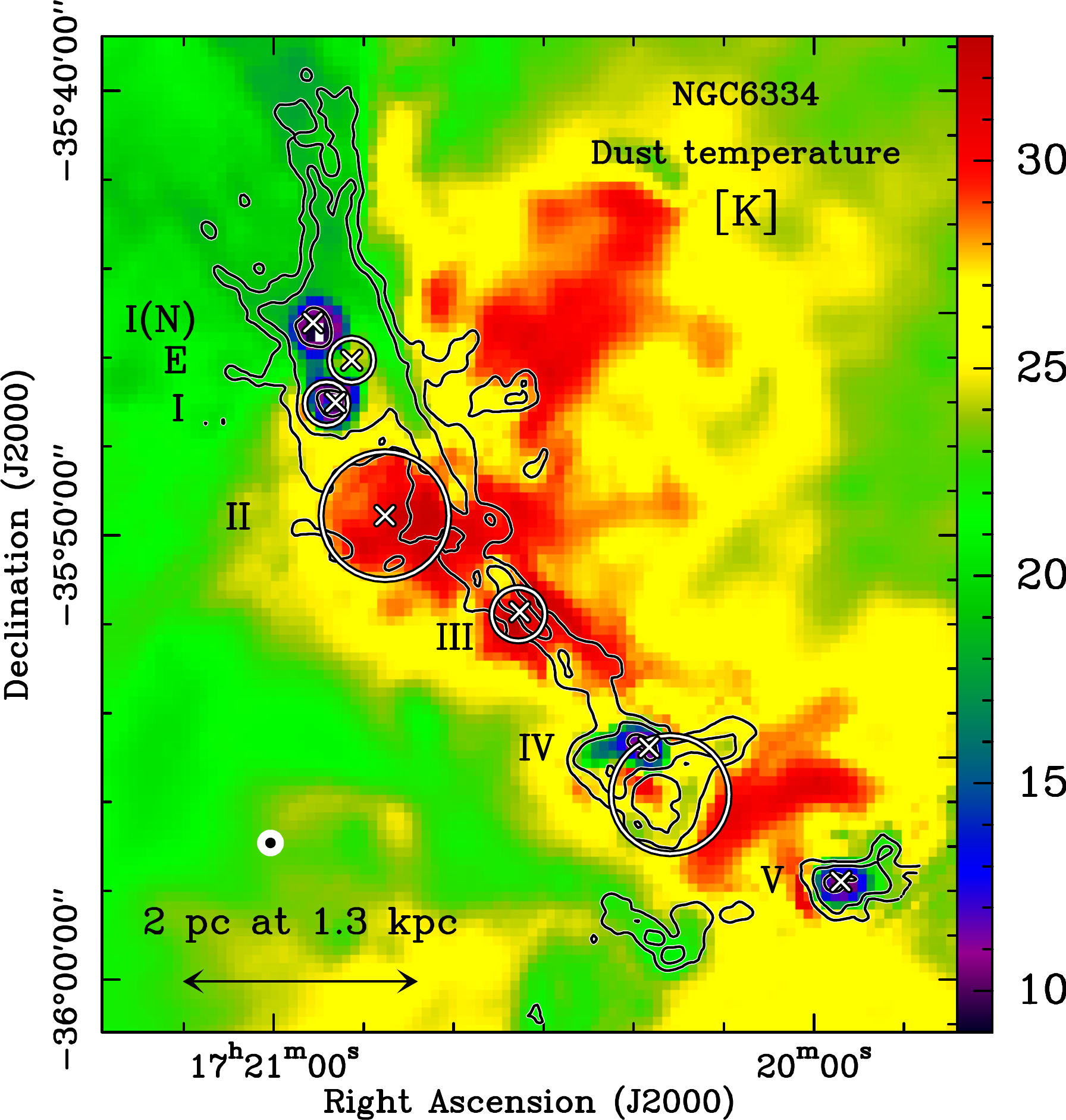}
 \hspace{0.8cm}
\includegraphics[angle=0]{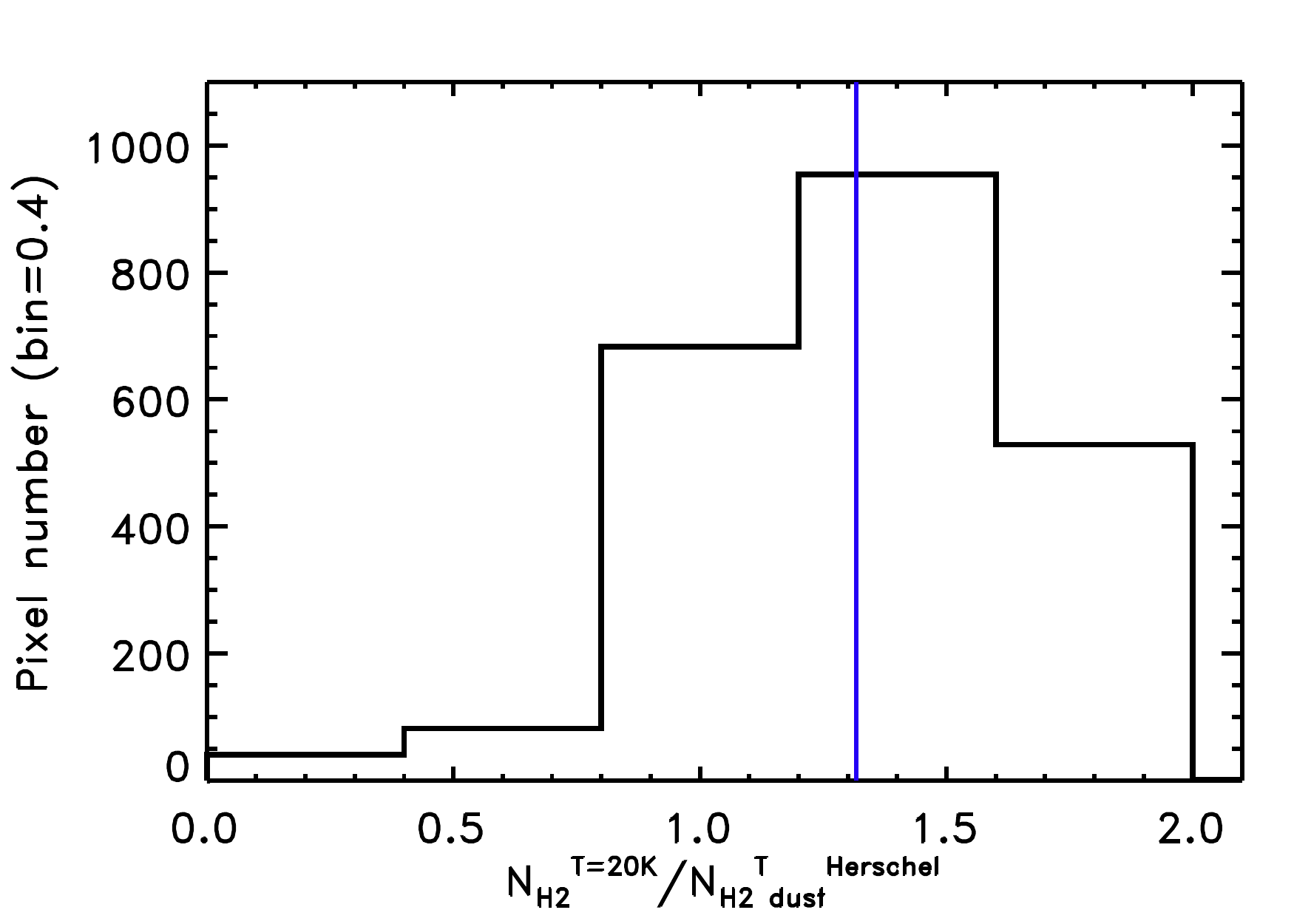}
} \vspace{-.1cm}
  \caption{ 
  {\rev {\it Left:} Dust temperature T$_{ \rm dust}^{Herschel}$ map \citep[][]{Russeil2013} derived from $Herschel$ data observed as part of the HOBYS key program \citep[][]{Motte2010}. This map is at the resolution of $36\arcsec$.
  The  contours (corresponding to 
 $I=0.4, 1.4,$ and $8\,$Jy\,beam$^{-1}$ of the BISTRO data),  the crosses, and the circles are the same as in Fig.\,\ref{IQUmaps}-top-left. 
 The $Herschel$ ($36\arcsec$) and BISTRO ($14\arcsec$) beams are shown on the left bottom of the plot as concentric filled white and black circles, respectively. 
   {\it Right:}  Distribution of the ratio $\nhh^{\rm T=20K}/\nhh^{\rm T_{ \rm dust}^{Herschel}}$ (see Sect.\,\ref{App2a} describing the derivation of the two quantities).         The mean value of this distribution is $1.3$, as indicated by the vertical line. 
}          }
  \label{Bi_Tdust}
    \end{figure*}

{\rev 
\subsection{Column density and dust temperature}\label{App2a}

We derive the  column density (\nhh) 
 map of the NGC 6334 field observed by BISTRO from the total intensity  Stokes $I$ map, 
  with  the relation 
  \begin{equation}
  \nhh=I_{850}/(B_{850}[T]\kappa_{850}\mu_{\rm H_2}m_{\rm H}),
     \end{equation}
  where $\kappa_{\nu}^{850}=0.0182 \,{\rm cm}^2$/g  is the dust opacity per unit mass of dust + gas at 850\,$\mu$m
    \citep[e.g.,][]{Ossenkopf1994}, $\mu_{\rm H_2}=2.8$  is the mean molecular weight per hydrogen molecule \citep[e.g.,][]{Kauffmann2008},  $m_{\rm H}$ is the mass of a hydrogen atom, and $B_{850}[T]$ is the Planck function for a given temperature $T$.

  Figure\,\ref{Bi_Tdust}-$left$, shows the dust temperature map of this region  derived  from grey-body fits to the $Herschel$ five wavelength data presented in \citet[][]{Russeil2013}.     As can be seen in this latter figure, in the northeast section of the NGC 6334 filamentary structure (e.g., crests 4, 5, 6, and 7), which is the least affected by the heating from the multiple \hii\ bubbles present in the field \citep[][]{Russeil2016},  the dust temperature is   T$_{\rm dust}\sim20\,$K. Towards the Southern parts,  
      the dust temperatures derived from the LOS integrated $Herschel$ emission  reach values up to $\sim30\,$K, which  possibly trace the emission of  warmer dust grains located mostly in the outer layers of the cloud. 
   The dust temperatures  ($>25\,$K) derived from  $Herschel$ data in these regions may not provide a good estimate of the gas temperature of the cold and dense molecular cloud traced with the 850\,$\mu$m emission of the BISTRO data. 
   
     We thus derive a column density map, $\nhh^{\rm T=20K}$, for  a constant  $T=20\,$K corresponding to the dominent temperature observed towards the Northern part of the NGC 6334 molecular cloud that is less affected by the warm \hii\ regions. 
 Temperatures of about $20\,$K are also compatible with the kinetic temperature toward star forming high-mass filaments and hubs derived from NH$_3$ GBT observations of the KEYSTONE project \citep[][]{Keown2019}.

  For completeness, we also derive a  column density map, $\nhh^{\rm T_{ \rm dust}^{Herschel}}$, using the $Herschel$ dust temperature map at a resolution of $36\arcsec$ \citep[][and Fig.\,\ref{Bi_Tdust}-$left$]{Russeil2013} to compare with $\nhh^{\rm T=20K}$.  
    The distribution of the $\nhh^{\rm T=20K}/\nhh^{\rm T_{ \rm dust}^{Herschel}}$  ratio of these two maps, towards the field observed by BISTRO with $SNR(I)>25$, is shown in the right panel of Fig.\,\ref{Bi_Tdust}. This distribution is narrow, with a mean of 1.32 and a standard deviation of 0.35. The difference between the two maps, $\nhh^{\rm T=20K}$ and $\nhh^{\rm T_{ \rm dust}^{Herschel}}$, of about $30\%$ on average is mostly smaller than the dispersion of the $\nhh^{\rm T=20K}$ found along the different crests studied in the paper, cf., column (2) of Table\,\ref{tab:paramCritic}.
    
  Thus, for the analysis presented in this paper, we derive the 
    column density %
    using a constant dust temperature of $20\,$K. 
    
 }

\subsection{Comparison of BISTRO data with $Herschel$/SPIRE+ArT\'eMiS data}\label{App2b}

Here we compare the BISTRO observations at 850$\,\mu$m with observations towards the same region obtained by
 $Herschel$/SPIRE \citep{Motte2010} combined with ArT\'eMiS data at 350$\,\mu$m at 8\arcsec resolution  \citep{Andre2016}.  

     To compare the two observations at different wavelengths, we estimate the column density  from the total intensity  Stokes $I$ values  with  the dust opacity per unit mass of dust + gas $\kappa_{\nu}^{850}=0.0182 \,{\rm cm}^2$/g for the BISTRO data at   850\,$\mu$m  and 
  $\kappa_{\nu}^{350}=0.1 \,{\rm cm}^2$/g for the $Herschel$/SPIRE+ArT\'eMiS data at   350\,$\mu$m  \citep{Ossenkopf1994,Kauffmann2008}. For both maps, we use  a mean dust temperature of $T=20\,$K as 
  {\rev explained above and also adopted by \citet{Andre2016}}. 

\begin{figure*}[!h]
   \centering
     \resizebox{19.cm}{!}{ \hspace{-2cm}
\includegraphics[angle=0]{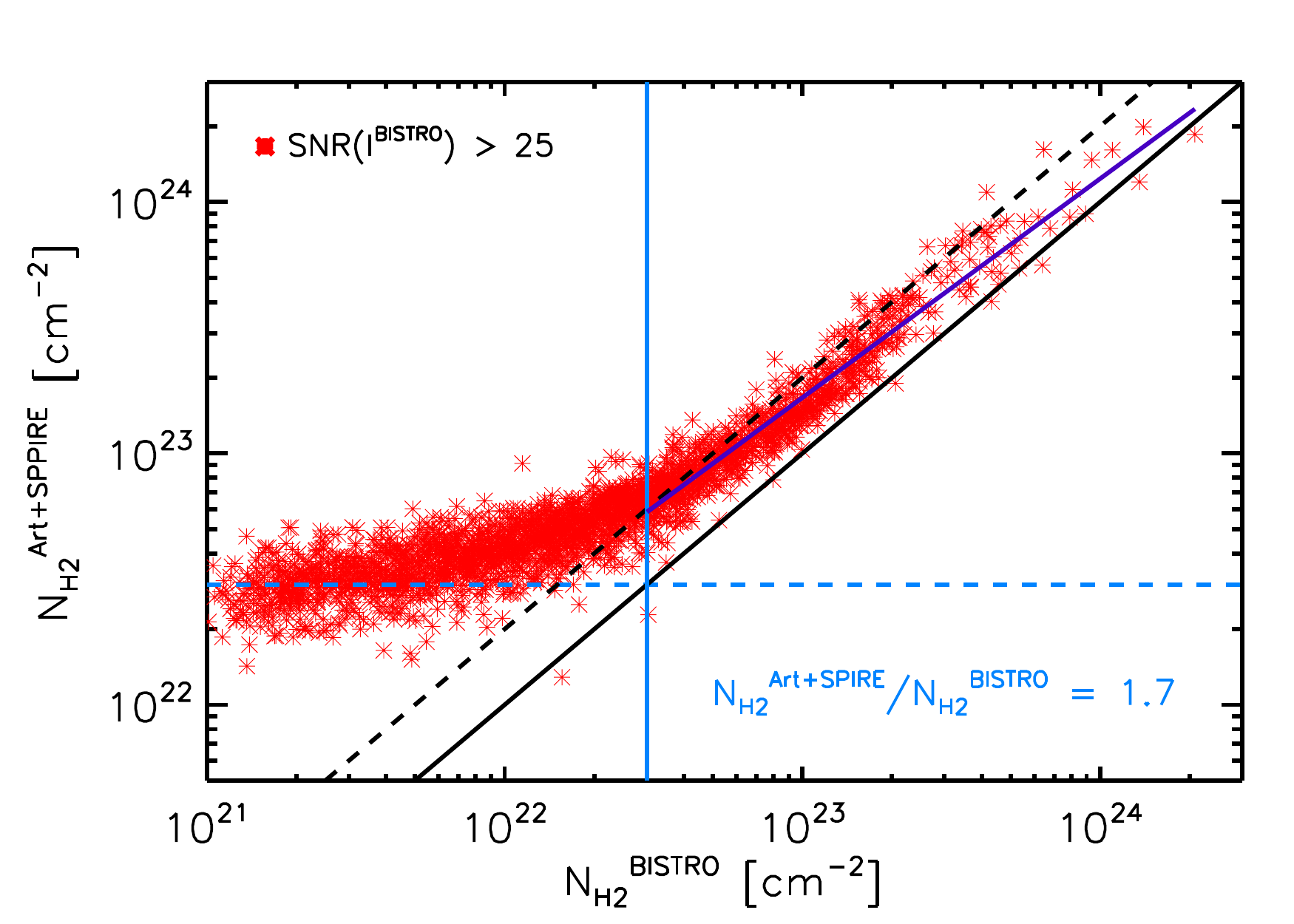}
\includegraphics[angle=0]{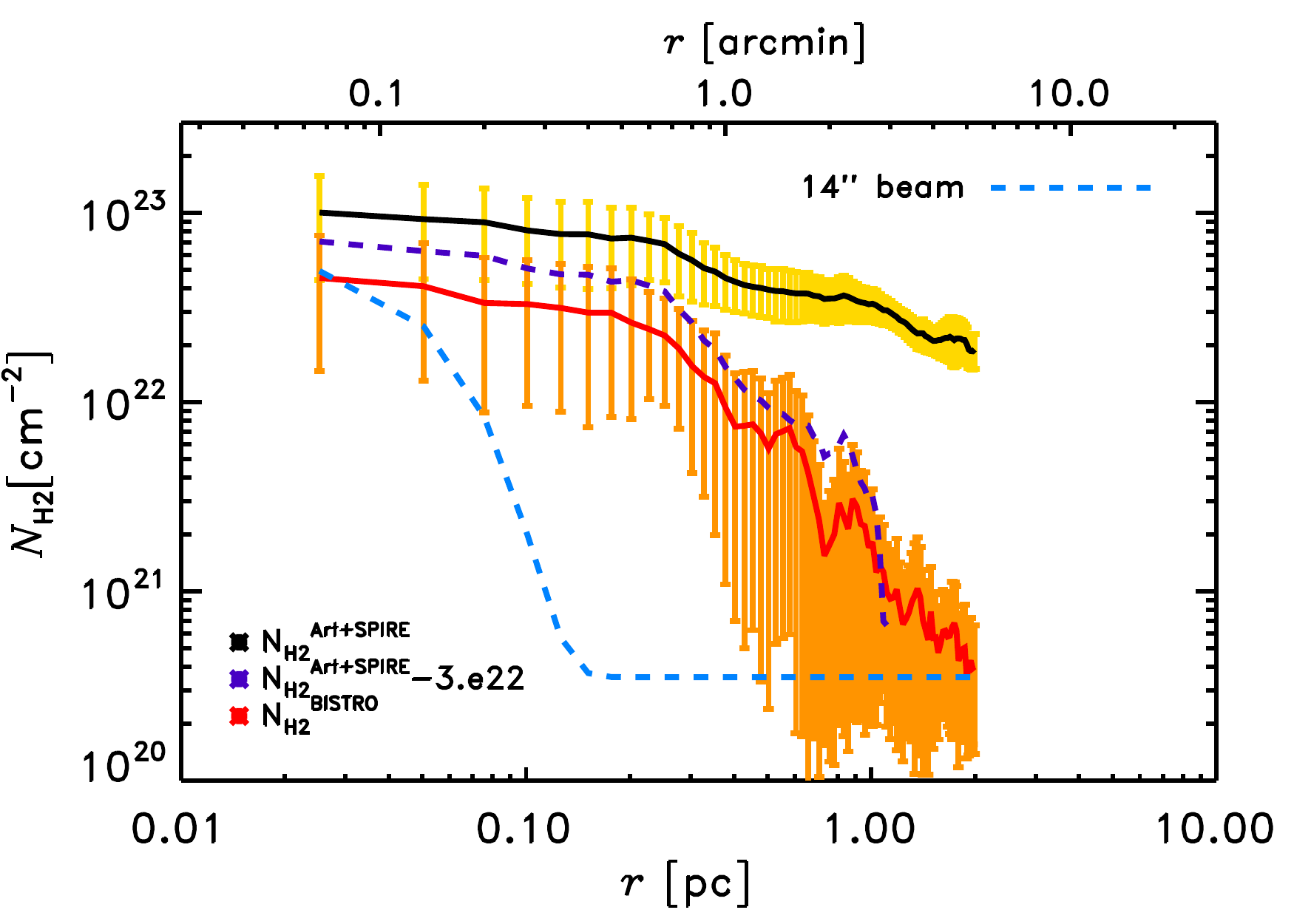}
 \hspace{-1cm}
\includegraphics[angle=0]{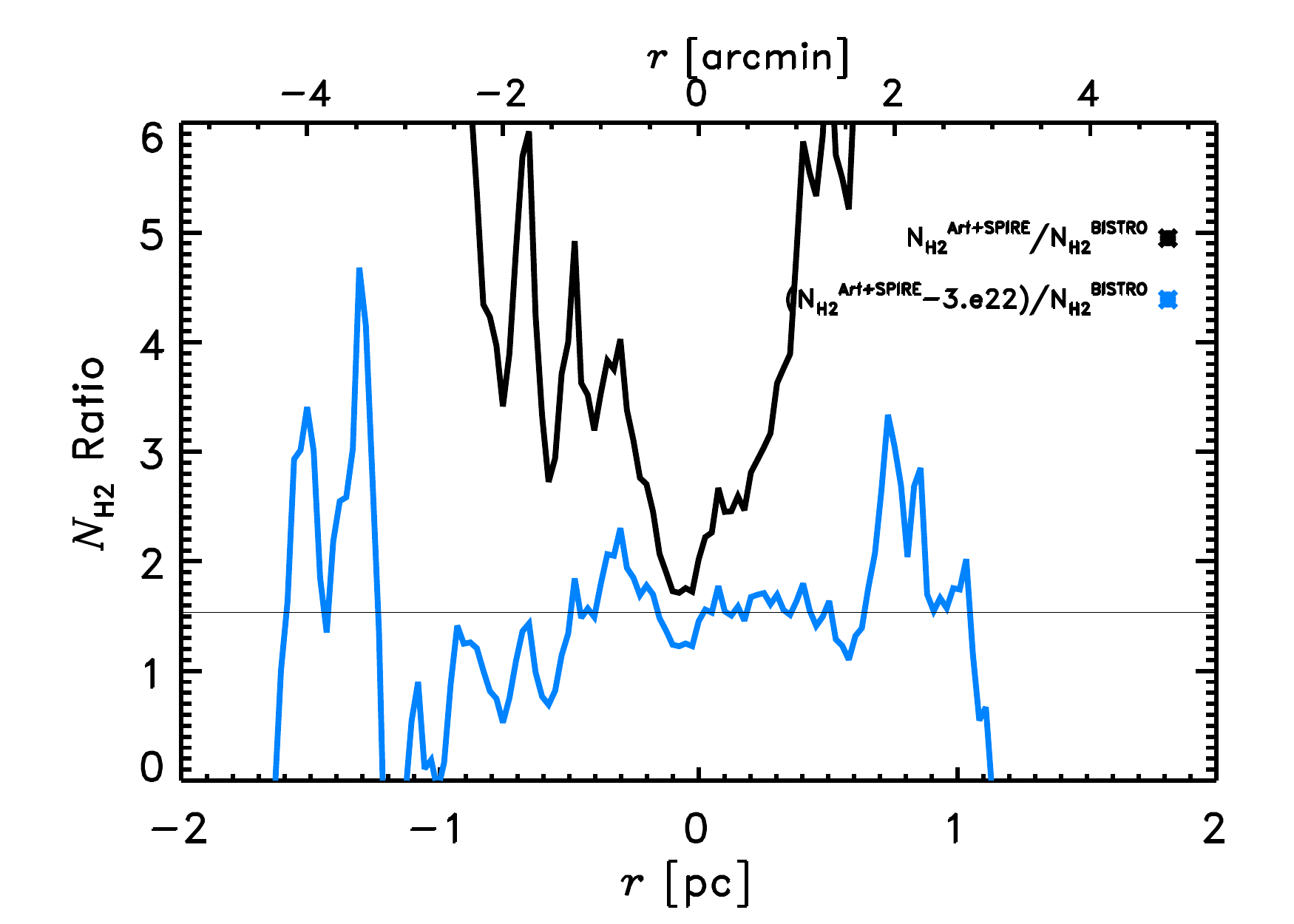}
} \vspace{-.5cm}
  \caption{ 
  {\it Left:} Comparing the column density values derived from ArT\'eMiS+SPIRE ($\nhh^{\rm Art+SPIRE}$ ) at 350$\,\mu$m \citep[from][]{Andre2016} and BISTRO ($\nhh^{\rm BISTRO}$) at 850$\,\mu$m (see Appendix\,\ref{App2b} for the expression used to estimate the column density from the observed total intensity). 
  The vertical blue line indicates  $\nhh=3\times10^{22}\,\NHUNIT$ above which 
   the two maps agree within a factor $<2$. 
   The oblique black solid and dashed lines show the  $y=x$ and $y=2x$ relations. 
  {\it Middle:}  Radial column density profiles (in log scale) perpendicular to and averaged along the 10\,pc  NGC 6334 elongated structure (i.e., crests from 1 to 5 in Fig.\,\ref{SkelFig}). The solid black and red lines correspond to the average profiles derived from ArT\'eMiS+SPIRE  and  BISTRO data, respectively. The purple dashed curve corresponds to $\nhh^{\rm Art+SPIRE}-\nhh^{\rm Galactic}$, where  $\nhh^{\rm Galactic}=3\times10^{22}\,\NHUNIT$ corresponds to 
 the large scale "Galactic emission" filtered out in the BISTRO maps. The angular resolution of the data is indicated by the blue dashed Gaussian function. 
      {\it Right:} The mean profiles (in linear scale) of the ratios $\nhh^{\rm Art+SPIRE}/\nhh^{\rm BISTRO}$  (black) and $(\nhh^{\rm Art+SPIRE}-\nhh^{\rm Galactic})/\nhh^{\rm BISTRO}$ (blue).
        The mean value of this latter ratio is $\sim1.5$ indicated by the horizontal line. 
}          
  \label{BiArtMaps}
    \end{figure*}

Figure\,\ref{BiArtMaps} shows  the column density values derived from BISTRO 850\,$\mu$m data  and those derived from  SPIRE+\Artemis\ data at 350\,$\mu$m data smoothed to the same resolution of 14\arcsec. For column densities $\gtrsim3\times10^{22}\,\NHUNIT$  both maps agree %
 within a factor $<2$ (Fig.\,\ref{BiArtMaps}-left). 

We derive  mean radial profiles perpendicular to the main axis of the elongated filament and averaged along its length. In practice, the maps have first been rotated by $30^\circ$ (corresponding to the mean orientation of NGC 6334 on the plane-of-the-sky) from East to North. Second, horizontal cuts, perpendicular to the main axis of NGC 6334 have been averaged  from South to North along the 10\,pc structure. 
Figure\,\ref{BiArtMaps}-middle shows the mean radial column density profiles perpendicular to the NGC 6334 cloud. 
For $\nhh\gtrsim3\times10^{22}\,\NHUNIT$ the mean ratio between the two profiles is $\sim1.6$. For $\nhh<3\times10^{22}\,\NHUNIT$, corresponding to scales $r\gtrsim$0.3\,pc from the central part of the filament, the ratio $\nhh^{\rm Art+SPIRE}/\nhh^{\rm BISTRO}$ increases up to values of $\sim10$ for  $r\sim$1\,pc or $r\sim$2\parcm6 (Fig.\,\ref{BiArtMaps}-right). 

The drop of the observed emission of the BISTRO data compared to the emission observed by $Herschel$ may be due to the removal of the large scale emission affecting the JCMT ground based data while $Herschel$ data are in principle not affected by this problem. 

Since NGC 6334 is located within 0.6 deg of the Galactic Plane, the emission is affected by large-scale 
"Galactic emission", which corresponds to background and foreground  emission ($\nhh\sim3\times10^{22}\,\NHUNIT$) observed along the LOS  in the surrounding of the NGC 6334 molecular cloud. We estimate here that "Galactic emission" emission to be $\nhh\sim3\times10^{22}\,\NHUNIT$. 

We hence compare the column density derived from BISTRO data with the "Galactic emission" subtracted column density values of the SPIRE+\Artemis\ data. As can be seen in Fig.\,\ref{BiArtMaps}-middle and left, these "Galactic emission" column density subtracted SPIRE+\Artemis\ data are compatible within $\sim50\%$ with the column density derived from BISTRO observations at 850\,$\mu$m down to $\nhh\sim10^{21}\,\NHUNIT$. 
 
We thus conclude that our BISTRO data trace the intrinsic emission of the molecular cloud of NGC 6334 filtering out the 
large-scale emission on the order of $\nhh\sim3\times10^{22}\,\NHUNIT$ corresponding, in this case, {\rev to a combination of  extended emission physically linked to the NGC 6334 molecular complex and} to LOS 
foreground/background Galactic emission.

    \section{Tests on the power spectrum analysis}\label{App3}
    
\begin{figure*}[!h]
   \centering
     \resizebox{16.5cm}{!}{ 
\includegraphics[angle=0]{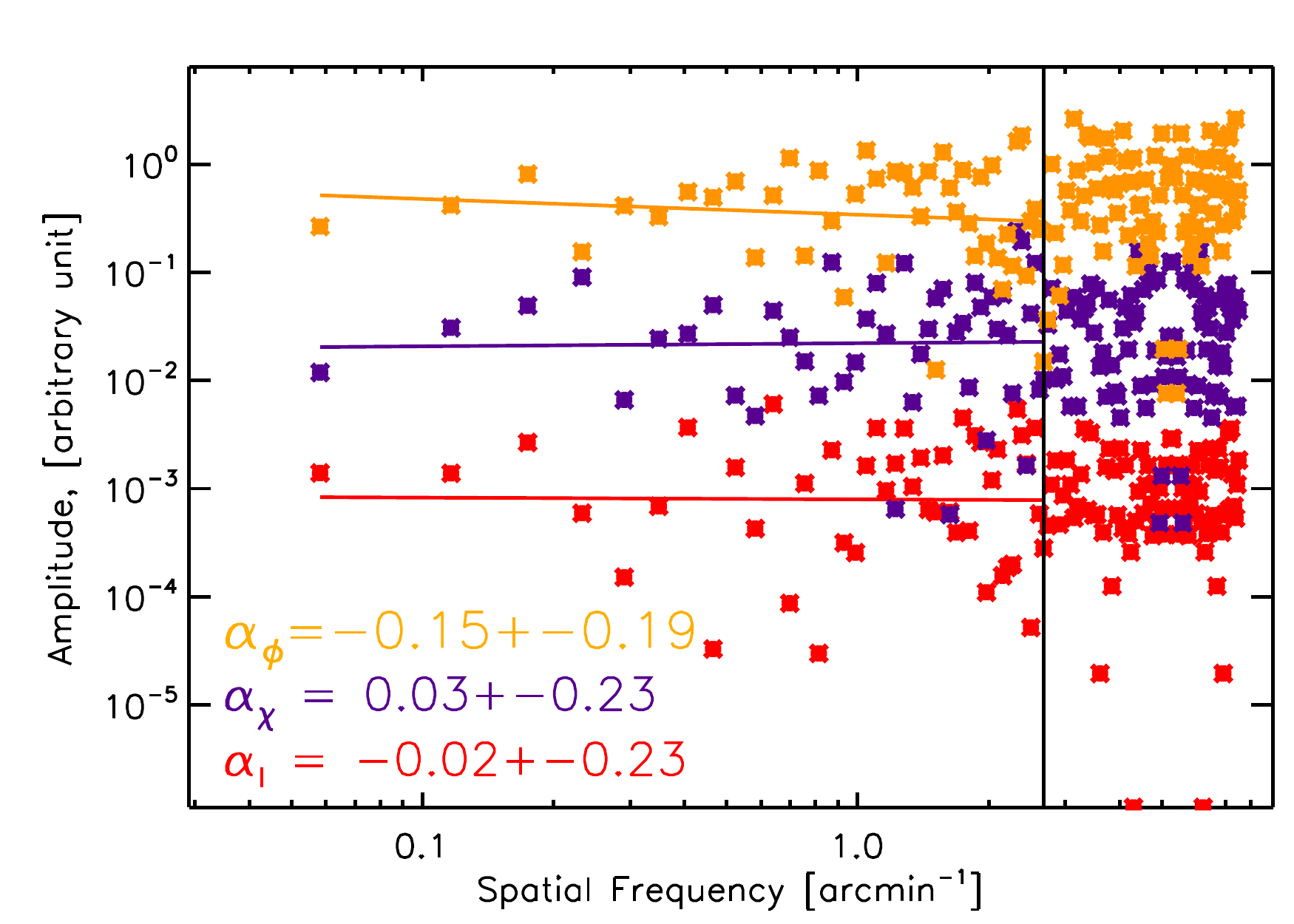}
 \hspace{2cm}
\includegraphics[angle=0]{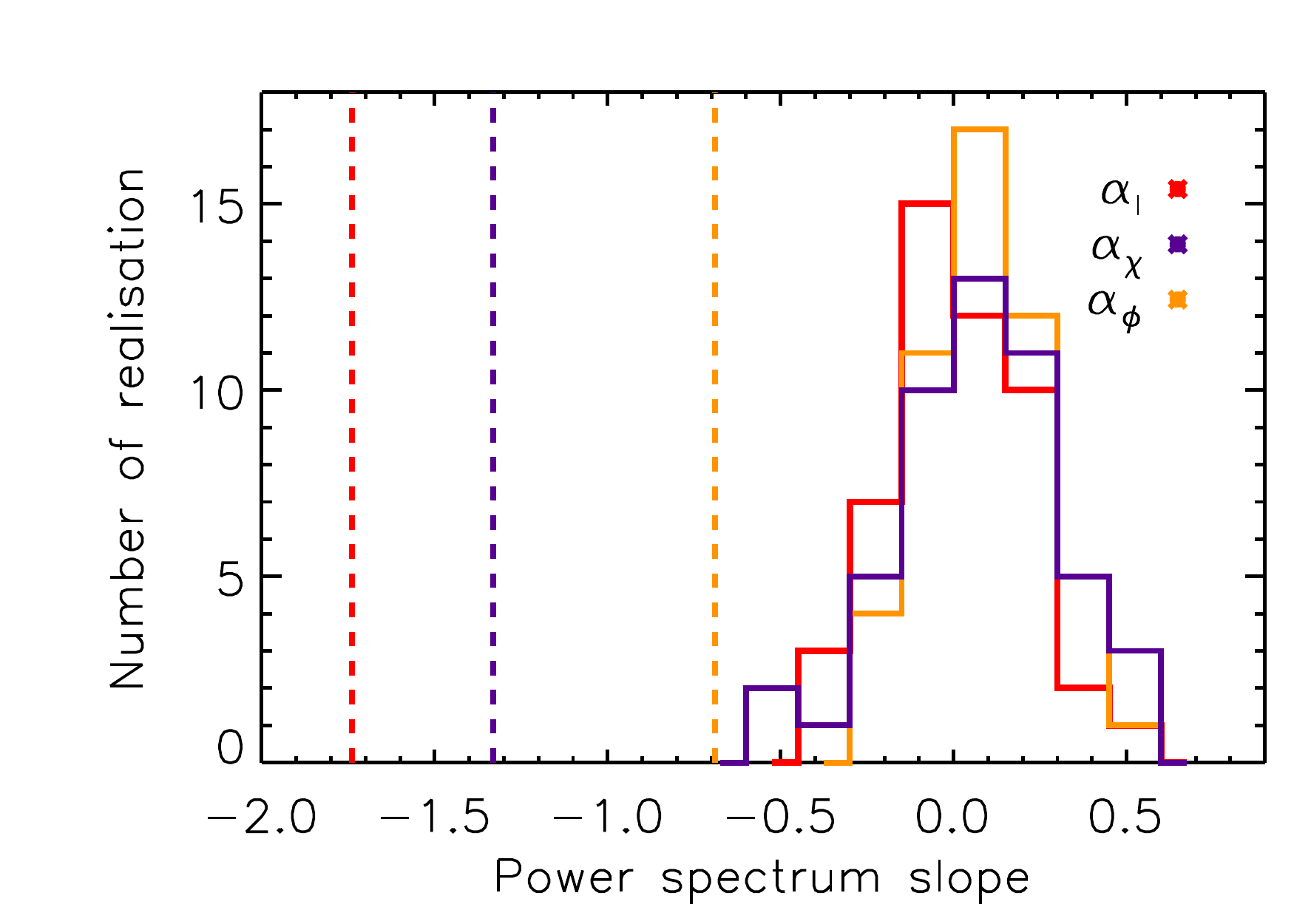}
}\vspace{-.2cm}
  \caption{%
 { \it Left:} A power spectrum  example of the test data (intensity in red, POS B-field angle in purple, and difference angle in yellow), where the observed values along the ridge crest are randomly  reordered. 
 The vertical line shows the spatial resolution of the data. The data points have pixel size of 4\arcsec. 
  The straight coloured lines show the best power-law fits to the power spectra and the derived power-law indexes ($\alpha$) are shown on the bottom left hand side of the plot.
   { \it Right:} Histogram of $\alpha$ for 50 realisations of random reorganization of the observed values. The vertical dashed lines indicate the derived power spectrum slopes from the observed values (see Fig.\,\ref{PS_crest3_5}).  
   }          
  \label{PS_test}
    \end{figure*}

    The observations show a  well defined power law shape  of the power spectra of $I$, $\chi_{B_{\rm POS}}$, and $\phi_{\rm diff}$ (cf.,\,\ref{PS_crest3_5}). To check whether such well-defined power law power spectra could be the result of uncorrelated emission,  we randomly reordered the observed data along the crest of the studied filament and derived the power spectrum for 50 realisations of  test-data. Figure\,\ref{PS_test}-left shows an example of  the power spectrum derived from one of the realisations. These power spectra  are not well represented by power laws.  
    Figure\,\ref{PS_test}-right shows the histogram of the slopes measured by fitting the test power spectra for the 50 realisations. The derived slopes are compatible with a flat spectrum. We hence conclude that the observed power spectra with well defined power law shapes are  a result of the correlated fluctuations of these properties along the filament crest.

\end{appendix}

\end{document}

%% file: bistro_authors2.tex
\author{\small
D. Arzoumanian\inst{1,2}\thanks{\url{Doris.Arzoumanian@astro.up.pt}}
\and
R. Furuya\inst{3,4}
\and
T. Hasegawa\inst{5}
\and
M. Tahani\inst{6}
\and
S. Sadavoy\inst{7}
\and
C.  L.  H. Hull\inst{8,9,10}
\and
D. Johnstone\inst{11,12}
\and
P.  M. Koch\inst{13}
\and
S.-i. Inutsuka\inst{2}
\and
Y. Doi\inst{14}
% ...
}

\institute{
Instituto de Astrof\'isica e Ci{\^e}ncias do Espa\c{c}o, Universidade do Porto, CAUP, Rua das Estrelas, PT4150-762 Porto, Portugal\goodbreak
\and
Department of Physics, Graduate School of Science, Nagoya University, Furo-cho, Chikusa-ku, Nagoya 464-8602, Japan\goodbreak
\and
Tokushima University, Minami Jousanajima-machi 1-1, Tokushima 770-8502, Japan\goodbreak
\and
Institute of Liberal Arts and Sciences Tokushima University, Minami Jousanajima-machi 1-1, Tokushima 770-8502, Japan\goodbreak
\and
National Astronomical Observatory of Japan, Osawa, Mitaka, Tokyo 181-8588, Japan\goodbreak
\and
Dominion Radio Astrophysical Observatory, Herzberg Astronomy and Astrophysics Research Centre, National Research Council Canada, P. O. Box 248, Penticton, BC V2A 6J9 Canada\goodbreak
\and
Department for Physics, Engineering Physics and Astrophysics, Queen's University, Kingston, ON, K7L 3N6, Canada\goodbreak
\and
National Astronomical Observatory of Japan, NAOJ Chile, Alonso de C\'ordova 3788, Office 61B, 7630422, Vitacura, Santiago, Chile\goodbreak
\and
Joint ALMA Observatory, Alonso de C\'ordova 3107, Vitacura, Santiago, Chile\goodbreak
\and
NAOJ Fellow\goodbreak
\and
NRC Herzberg Astronomy and Astrophysics, 5071 West Saanich Road, Victoria, BC V9E 2E7, Canada\goodbreak
\and
Department of Physics and Astronomy, University of Victoria, Victoria, BC V8W 2Y2, Canada\goodbreak
\and
Academia Sinica Institute of Astronomy and Astrophysics, No.1, Sec. 4., Roosevelt Road, Taipei 10617, Taiwan\goodbreak
\and
Department of Earth Science and Astronomy, Graduate School of Arts and Sciences, The University of Tokyo, 3-8-1 Komaba, Meguro, Tokyo 153-8902, Japan\goodbreak
% ...
}